%% file: main.tex
\documentclass[twocolumn,a4paper,preprintnumbers,amsmath,amssymb,floatfix,prb,showpacs]{revtex4}
\pdfoutput=1
\usepackage{amsmath}
\usepackage{amssymb}
\usepackage{graphicx}
\usepackage{color}

\usepackage{hyperref}

\newcommand{\E}{\mathrm{e}}
\newcommand{\D}{\mathrm{d}}
\newcommand{\up}{\uparrow}
\newcommand{\dw}{\downarrow}

\newcommand{\mat}[1]{\mathbf{#1}}
\newcommand{\matgr}[1]{\boldsymbol{#1}} 
\renewcommand{\vec}[1]{\mathbf{#1}}

\newcommand{\ket}[1]{\left| \right. \! #1 \! \left. \right\rangle}
\newcommand{\bra}[1]{\left\langle \right. \! #1 \! \left. \right|}
\newcommand{\thavg}[1]{\left\langle \right. \! #1 \! \left. \right\rangle}

\DeclareMathOperator{\tr}{Tr}

\begin{document}
\title{ A weak coupling  CTQMC  study of the  single impurity and periodic  Anderson models   with s-wave 
        superconducting baths }

\author{David J. Luitz}
\email{dluitz@physik.uni-wuerzburg.de}
\author{Fakher F. Assaad}
\affiliation{Institut f\"ur theoretische Physik und Astrophysik\\Universit\"at W\"urzburg}

\date{\today}
\begin{abstract}

We apply the unbiased weak-coupling continuous time quantum Monte Carlo (CTQMC) method   to review the 
physics of a single magnetic impurity coupled to s-wave superconducting leads described by  the BCS 
reduced Hamiltonian.   As a function of  the superconducting gap $\Delta$, we study the signature of the first
order transition between  the singlet and  doublet (local moment) states on various quantities.  
In particular we concentrate on  the  Josephson current  with $0$ to $\pi$ phase  shift, the 
crossing of the Andreev bound states in the single particle spectral function, as well as the local 
dynamical spin structure factor.   
Within DMFT, this impurity problem provides a link to the periodic Anderson model with
superconducting conduction electrons (BCS-PAM).  The first order transition  observed  in  the impurity model 
is reproduced in the BCS-PAM and is signalized by the crossing of the low energy excitations in the local 
density of states.  The momentum resolved single particle spectral function  in  the singlet state   reveals 
the coherent, Bloch-like, superposition of Andreev bound states. In the doublet or local moment phase the single particle 
spectral function is  characterized by incoherent quasiparticle excitations.
\end{abstract}


\maketitle

\section{Introduction}

Magnetic degrees of freedom in superconducting  environments have attracted considerable interest due to the 
underlying competing effects.  Already  a classical spin oriented along the $z$-axis  
\cite{shiba.spin.supercond,sakurai.supercond.mag.imp}  embedded  in a  superconducting host generates  a localized
state  within  the  superconducting gap. As a function of the interaction strength this excitation
crosses the Fermi energy  thereby triggering a first order transition 
between a ground state with
vanishing total electronic spin and a ground state with nonzero total electronic spin.

For a quantum spin, the Kondo effect sets in. Being a Fermi surface instability, the opening of the superconducting 
gap competes with Kondo screening and ultimately leads to a local moment regime.    This transition is accompanied by 
a $0$ to $\pi$ phase shift in the Josephson current. In the local moment regime  the  $\pi$-shift 
occurs since a  Cooper pair tunneling  through the  junction necessarily  accumulates a phase $\pi$
\cite{josephson.effect,kulik-pi-shift,glazman-matveev,PhysRevB.43.3740}.


The interest in the  problem has been renewed in the last decade by  the rapid progress in nanotechnology 
which made a direct experimental realization of quantum dots
coupled to superconducting leads feasible so that  many experiments have been designed to directly measure
the $0$ to $\pi$ transition of the Josephson current. Experiments using a carbon
nanotube\cite{eichler:161407,wernsdorfer-squid-cnt,0-pi-transition-jorgensen} but also InAs
nanowires\cite{vandam-supercurrent-reversal} as a
quantum dot coupled to superconducting leads were able to observe the sign change of the Josephson
current by increasing the gate
voltage and thus manipulating the number of electrons on the quantum dot.
The effect of the changing electron number on the behavior of such systems has been extensively
studied\cite{1367-2630-9-5-124,rgensen:207003,sand-jespersen:126603,eichler:126602} and the
theoretical expectation of the collapse of the Kondo effect if the superconducting gap $\Delta$
exceeds the Kondo temperature $T_K$ has been confirmed by experiments of Buitelaar et al.\cite{PhysRevLett.89.256801}.

From the  numerical point of view,  a combination of algorithmic development and computational power 
has allowed  for a more detailed study of the problem using  the numerical renormalization
group\cite{PhysRevB.70.020502,JPSJ.73.2494,0953-8984-19-48-486211,0953-8984-20-27-275213},
quantum Monte Carlo
simulations\cite{PhysRevLett.93.047002,PhysRevLett.94.039902,PhysRevLett.94.229702} as well as
functional renormalization group calculations\cite{karrasch:024517}.
Most numerical works present in the literature only present either the study of the Josephson current
\cite{karrasch:024517,PhysRevLett.93.047002,PhysRevLett.94.039902,PhysRevB.70.020502} or the study
of the spectral properties of the Quantum dot \cite{0953-8984-19-48-486211}.
One of the goals of this article is to use the  weak coupling  CTQMC method \cite{rubtsov:035122}
to compute the  Josephson current as well as the spectral functions for the same parameter set in order to present 
a comprehensive study of the $0$ to $\pi$ transition of a Josephson quantum dot. Our numerically exact data 
clearly confirms the picture of a first order phase transition from a singlet phase linked to the 
$0$-junction regime of the Josephson current to a doublet phase corresponding to the $\pi$-junction regime.

In addition to numerical efforts, many analytical approximations have been introduced to tackle
different aspects of the physics of the problem. The non crossing approximation has been used to
show that Andreev bound states crossing the Fermi energy are connected to the $0$ to $\pi$
transition of the Josephson current\cite{PhysRevB.61.9109}. 
Perturbative methods as well as mean field theory have brought a quite complete understanding of the
phase diagram featuring the $0$ and $\pi$ phases as well as the intermediate phases $0'$ and
$\pi'$\cite{PhysRevB.62.6687,PhysRevB.68.035105,meng:224521}. 
Another method employed by several authors is the introduction of different analytically solvable effective models, 
which are valid in different limits \cite{PhysRevB.68.035105,meng:224521,0953-8984-19-48-486211}. These models 
are very useful to acquire  an  intuitive understanding of the physics.  
We will present the study of an effective Hamiltonian for the limit of a superconducting gap
$\Delta$ much larger than the bandwidth to support the interpretation of the CTQMC data.

Another motivation of this paper, is to  study within  dynamical mean field theory (DMFT) \cite{RevModPhys.68.13} 
the periodic Anderson model with an s-wave  BCS-conduction band (BCS-PAM).  Within this approximation, the BCS-PAM 
maps onto the  single impurity Anderson  model with superconducting baths  supplemented with a self-consistency 
condition.   We will show that the physics  of the impurity model can be taken over to the lattice case. 
In particular  the first order transition  observed  in  the impurity model 
is reproduced in the BCS-PAM and is signalized by the crossing of the low energy excitations in the local 
density of states.  The momentum resolved single particle spectral function  in  the singlet  phase  reveals 
the coherent, Bloch-like, superposition of Andreev bound states.  In the doublet or local moment phase the single 
particle spectral function is 
characterized by incoherent quasiparticle excitations.  We  provide an understanding of  this  in terms of 
models of disorder. 
  

The paper is organized as  follows. After introducing  the model  in Sec.  \ref{sec:impurity-model}, 
we  discuss in Sec. \ref{sec:toy-model} an effective toy  model valid in the limit of a 
superconducting gap, $\Delta$,  much larger than the bandwidth $W$.   This simple toy model goes a good way 
at understanding certain aspects of the underlying physics. 
A brief outline of the employed CTQMC result including the proof of Wick's theorem for each
configuration in the Monte Carlo simulation will be presented in Sec. \ref{sec:CTQMC}.
The results of the toy model are then compared to the results of the
CTQMC simulation, which are discussed in detail in Sec. \ref{sec:NumericalResults}.
Sec. \ref{sec:DMFT} is dedicated to the study of the BCS-PAM within DMFT.
We include an  appendix \ref{sec:proof_of_det_identity} featuring the proof of a general
determinant identity needed for the proof of Wick's theorem for every configuration in the CTQMC.

\section{Model}
\label{sec:impurity-model}
The physics of a  quantum dot coupled to two superconducting leads (L=left, R=right) via a hybridization
term is captured  by the single impurity Anderson model with  the leads described
by the BCS mean-field Hamiltonian:
\begin{equation}
 \label{eq:qd_supercond_hamiltonian}
\tilde{H} = \sum _ {\alpha=L} ^{R} \tilde{H}_{0, \alpha} + \tilde{H}_d + \tilde{H}_V,
\end{equation}
with 
\begin{equation}
\begin{split}
&\tilde{H}_{0,\alpha} = \sum_{k,\sigma} \xi_k \tilde{c}^\dagger_{k,\sigma, \alpha} \tilde{c}_{k,
\sigma, \alpha} \\ & \quad \quad- \sum_{k} \left( {\Delta} \E^{i \phi_{\alpha}} \tilde{c}^\dagger_{k,\up,\alpha}
\tilde{c}^\dagger_{-k,\dw,\alpha} + \text{h.c.} \right),\\
&\tilde{H}_d=\sum \limits_\sigma \xi_d \tilde{d}^\dagger_\sigma \tilde{d}_\sigma + U \left(
\tilde{d}^\dagger_\up \tilde{d}_\up - \frac{1}{2} \right)\left( \tilde{d}^\dagger_\dw \tilde{d}_\dw
- \frac{1}{2} \right), \\
& \tilde{H}_V= -\frac{V}{\sqrt{N}} \sum \limits_{\alpha=L}^R \sum \limits_{\sigma, k} \left(
\tilde{c}^\dagger_{k,\sigma,\alpha } \tilde{d}_\sigma + \tilde{d}^\dagger_\sigma
\tilde{c}_{k,\sigma,\alpha } \right).
\end{split}
\end{equation}
The operators $\tilde{c}^\dagger_{k,\sigma,\alpha}$ are creation operators for electrons with a
$z$-component of the spin
$\sigma$ and momentum $k$ in lead $\alpha$, $\tilde{d}^\dagger_{\sigma}$ is a creation operator of
an electron with a $z$-component of the spin $\sigma$ on the quantum dot. $\xi_k=\epsilon(k) - \mu = -2t \cos(k) -\mu$ is
the dispersion relation for the electrons in the leads, where we assume, that the dispersion is
independent of the lead index $\alpha$, and $\xi_d=\epsilon_d-\mu$ is the position of the dot level.
Throughout this paper, we will express all quantities in units of $t=1$.
The superconducting order parameter has a modulus $\Delta$ and a phase $\phi_\alpha$. 
The parameter $V$ characterizes the
strength of the hybridization, and $U$ corresponds to the Coulomb blockade.

Since the Hamiltonian does not conserve the electron number as a consequence of the BCS-term, we use the standard trick of rewriting the Hamiltonian in terms of creation and annihilation operators of quasiparticles, which for spin up are identical to the electrons, but correspond to holes in the spin down sector. This can also be expressed as a canonical transformation:
\begin{equation}
\label{eq:canonical_transformation}
 \tilde{d}^\dagger_\up \rightarrow d^\dagger_\up, \,\, \tilde{d}^\dagger_\dw \rightarrow d_\dw,\,\, \tilde{c}^\dagger_{k,\up,\alpha} \rightarrow c^\dagger_{k,\up,\alpha}, \,\, \tilde{c}^\dagger_{-k,\dw,\alpha} \rightarrow c_{k,\dw,\alpha}.
\end{equation}

Using the new operators, the Hamiltonian can be written in a Nambu notation:
\begin{equation}
\label{eq:qd_bcs_hamiltonian_nambu}
\begin{split}
 &H = H_0+H_U = \sum_{k,\alpha} \vec{c}_{k,\alpha}^\dagger \mat{E}_\alpha(k) \vec{c}_{k,\alpha} +\vec{d}^\dagger \matgr{\epsilon}_d \vec{d} \\&- \frac{V}{\sqrt{N}} \sum \limits_{k,\alpha} \left( \vec{c}_{k,\alpha}^\dagger \matgr{\sigma}_z \vec{d} + \vec{d}^\dagger \matgr{\sigma}_z \vec{c}_{k,\alpha} \right) + H_U 
\end{split}
\end{equation}
with $H_U=-U(d^\dagger_\up d_\up - \frac{1}{2})(d^\dagger_\dw d_\dw - \frac{1}{2})$, the Nambu spinors
\begin{equation}
 \vec{d}=\begin{pmatrix}
                   d_\up\\
		d_\dw\\
                  \end{pmatrix},\quad \vec{c}_{k,\alpha}=\begin{pmatrix}
                   c_{k,\up,\alpha}\\
		c_{k,\dw,\alpha}\\
                  \end{pmatrix},
\end{equation}
the matrices
\begin{equation}
 \mat{E}_\alpha(k) = \begin{pmatrix}
                      \xi_k & - \Delta  \E^{i \phi_\alpha} \\
			- \Delta \E^{-i \phi_\alpha} & -\xi_k 
                     \end{pmatrix}, \quad 
\matgr{\epsilon}_d = \begin{pmatrix}
                      \xi_d & 0\\
			0 & -\xi_d\\
                     \end{pmatrix}
\end{equation}
and the Pauli matrix
\begin{equation}
\matgr{\sigma}_z=\begin{pmatrix}
                  1  & 0 \\
		0 & -1 \\
                 \end{pmatrix}.
\end{equation}

For practical reasons, we use the following definition for the single particle Green's function
throughout Sec. \ref{sec:impurity-model} to Sec. \ref{sec:NumericalResults}:
\begin{equation}
  G_{dd}^{\sigma \sigma'} ( i \omega_m)= \int \limits_0 ^\beta \D \tau \exp(i \omega_m \tau)
  \thavg{T d^\dagger_\sigma (\tau) d_{\sigma'}  }.
  \label{eq:Green-function-definition}
\end{equation}

With this definition, the resolvent operator $ \mat{G}^0(i \omega_m) = \left( - i \omega_m \mat{1} -
\mat{H}_0^T \right)^{-1}$ can be used to obtain the Green's function of the noninteracting system:

\begin{equation}
\label{eq:greens_function_Gdd}
\begin{split}
\mat{G}_{dd}^0(i\omega_n)^{-1} &=  (-i\omega_n \mat{1} - \matgr{\epsilon}_d) \\
&+ \frac{V^2}{N} \sum \limits_{\alpha, k}  \matgr{\sigma}_z  \left( i \omega_n \mat{1} +
\mat{E}_\alpha^T(k) \right)^{-1} \matgr{\sigma}_z .
\end{split}
\end{equation}

\section{Effective Hamiltonian in the Limit $\Delta/W \rightarrow \infty$}

\label{sec:toy-model}

To gain a deeper understanding of the physics on the quantum dot, it is useful to search for analytically solvable toy models. We will study an effective model, which reproduces the physics of the Hamiltonian (\ref{eq:qd_supercond_hamiltonian}) in the limit $\Delta/W \rightarrow \infty$, where $W$ is the band width.
To derive the effective model, we look at the limit $\Delta\rightarrow \infty$ of the Green's
function in Eq. (\ref{eq:greens_function_Gdd}). The superconducting order parameter $\Delta$ appears only in the matrix $\mat{E}_\alpha(k)$, thus we examine the behavior of this matrix for large values of $\Delta$. This can easily be done by diagonalizing $\mat{E}_\alpha(k)$ for $\phi_\alpha=0$:
\begin{equation}
\label{eq:diagonalization_Ek}
 \mat{E}_\alpha(k) = \mat{U}_\Delta^{-1} \begin{pmatrix}
                                   -\sqrt{\Delta^2 +\xi_k^2} & 0 \\ 0 & \sqrt{\Delta^2 +\xi_k^2}
                                  \end{pmatrix} \mat{U}_\Delta.
\end{equation}
Let us first look at the limit $\Delta \rightarrow \infty$ of the transformation matrix $U_\Delta$,
which for brevity is not a unitary matrix.
\begin{equation}
 \mat{U}_\Delta = \begin{pmatrix}
      -\frac{\xi_k - \sqrt{\Delta^2 + \xi_k^2}}{\Delta} & 1 \\
	-\frac{\xi_k + \sqrt{\Delta^2 + \xi_k^2}}{\Delta} & 1 
     \end{pmatrix}  \Rightarrow
\mat{U}_\infty = \begin{pmatrix}
                  1 & 1 \\ -1 & 1
                 \end{pmatrix}.
\end{equation}
The diagonal matrix in Eq. (\ref{eq:diagonalization_Ek}) can be considered in a similar manner and we obtain for $\lim \limits_{\Delta \rightarrow \infty}\mat{E}_\alpha(k)= \mat{E}_\infty$:
\begin{equation}
 \mat{E}_\infty = \mat{U}_\infty^{-1} \begin{pmatrix}
                                       -\Delta & 0 \\ 0 & \Delta
                                      \end{pmatrix} \mat{U}_\infty = 
\begin{pmatrix}
 0 & -\Delta \\ - \Delta & 0
\end{pmatrix}.
\end{equation}
Using this result, for large values of $\Delta$ the sum over $k$ and $\alpha$ in Eq. (\ref{eq:greens_function_Gdd}) can be carried out yielding
\begin{equation}
\label{eq:greens_function_Gdd_eff}
 \mat{G}_{dd}^{0,\infty}(i\omega_n)^{-1} =  (-i\omega_n \mat{1} - \epsilon_d) + 2 V^2  \matgr{\sigma}_z  \left( i \omega_n \mat{1} + \mat{E}_\infty \right)^{-1} \matgr{\sigma}_z.
\end{equation}
This is exactly the free Green's function obtained from a Hamiltonian of the form:
\begin{equation}
\label{eq:Heff}
H_\text{eff} = -\sqrt{2} V ( \vec{c}^\dagger \matgr{\sigma}_z \vec{d} + \vec{d}^\dagger \matgr{\sigma}_z \vec{c} ) + \vec{c}^\dagger \mat{E}_\infty \vec{c} + \vec{d}^\dagger \matgr{\epsilon}_d \vec{d} +H_U.
\end{equation}
$H_\text{eff}$ describes a system consisting of one bath site $c$ connected by a hybridization term to the correlated quantum dot $d$. The dispersion of the bath has completely vanished, as the superconducting band gap becomes much larger than the bandwidth.

We chose a basis of the 16 dimensional Hilbert space and write the Hamiltonian
as a matrix, which subsequently can be diagonalized.  As we have restricted the
parameter space for the Monte Carlo simulations to $\epsilon_d = 0$ and $\mu=0$
in the original Hamiltonian of Eq. (\ref{eq:qd_supercond_hamiltonian}),
we will use the same parameters for the exact diagonalization results.

\begin{figure}
 \resizebox{\columnwidth}{!}{\input{tex-Heff-eigenvalues-U.tex}}
 \vspace{-0.7cm}
\caption{\label{fig:level_crossing_eigenenergies} (Color online) Eigenenergies of the effective Hamiltonian (\ref{eq:Heff}) for varying $U$. The fixed parameters are given by $V=0.5$ and $\Delta=1$. The crossing of the two lowest levels is clearly seen at $U \approx 1.7$. The ground state for $U < 1.7$ is a singlet state. For larger values of $U$, the twofold degenerate doublet state becomes energetically more favorable. }
\end{figure}
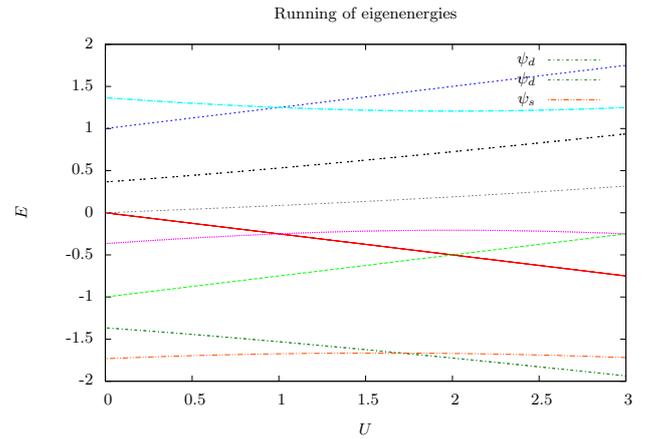

\begin{figure}
 \resizebox{\columnwidth}{!}{\input{tex-weightrunning.tex}} 
 \vspace{-0.7cm}
 \caption{\label{fig:weightrunning} (Color online) For
 $\Delta < 1.412$ the ground state is the singlet state from Eq.
 (\ref{eq:singlet-ground-state}). If $\Delta$ is increased, the weight $\alpha$ of the single
 occupied states $\ket{\tilde \up, \tilde \dw}$ and $\ket{ \tilde \dw,\tilde \up}$ decreases in
 favor of the states with a double occupied quantum dot, corresponding to the weights $\beta$ and
 $\gamma$. At $\Delta = 1.412$ the ground state changes to the twofold degenerate doublet state given
 in (\ref{eq:doublet-ground-state}) and the weight of the states with a single occupied quantum dot
 $b$ increases with $\Delta$. The parameters in this plot are $V=0.5$ and $U=1.0$. }
\end{figure}
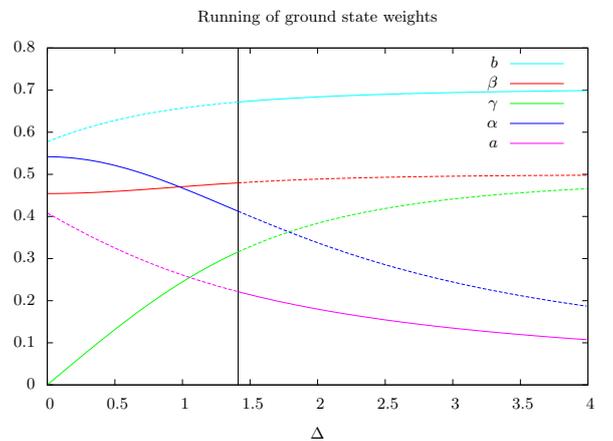

\subsection{Ground state of the effective model}
The ground state of the system (\ref{eq:Heff}) can be determined by diagonalizing the Hamiltonian $H_\text{eff}$. As
depicted in Fig. \ref{fig:level_crossing_eigenenergies}, the energy levels cross at a critical
value of $U=U_c$ and a similar behavior can be observed by varying $\Delta$ with a corresponding
critical value $\Delta_c$. For $U<U_c$ and $\Delta<\Delta_c$, the ground state is given by
$\ket{\psi_s}= -\alpha \left( \ket{\up\dw ,0} - \ket{0,\up\dw} \right) - \beta \left( \ket{\up,\dw} +
\ket{\dw,\up} \right) - \gamma \left( \ket{\dw,\dw} + \ket{\up,\up} \right)$, with the notation
$c_\sigma^\dagger \ket{0,0} = \ket{\sigma,0}$ and $d_\sigma^\dagger \ket{0,0} = \ket{0,\sigma}$.
Note, that we are using the unphysical basis introduced in Eq. 
(\ref{eq:canonical_transformation}).  To interpret this ground state it is better to return to the
physical basis by inverting the canonical transformation in Eq.
(\ref{eq:canonical_transformation}) and transforming the vacuum state $\ket{0,0} \rightarrow
\ket{\tilde{\dw},\tilde{\dw}}$.  The ground state can then be rewritten in the physical basis as:
\begin{equation}
\label{eq:singlet-ground-state}
\begin{split}
\ket{\psi_s}&=\alpha \left( \ket{\tilde \dw, \tilde \up} - \ket{\tilde \up,\tilde \dw} \right) \\ &+
\beta \left( \ket{ \tilde 0,\tilde \up \tilde \dw} + \ket{ \tilde \up \tilde \dw, \tilde 0} \right)
+ \gamma \left( \ket{\tilde 0, \tilde 0} + \ket{ \tilde \up \tilde \dw, \tilde \up \tilde \dw}
\right).
\end{split}
\end{equation}

This state is clearly a singlet state, corresponding to a Kondo singlet between the quantum dot and
the bath with the dominant weight $\alpha$. The states representing a pairing on the quantum dot or
in the bath have the suppressed weights $\beta$ and $\gamma$ for small values of $\Delta$ but grow
more important if $\Delta$ is increased as is shown in Fig. \ref{fig:weightrunning}.

At  $U > U_c$, the ground state changes and we get the twofold degenerate ground states
$\ket{\psi_{d,\up}}=a \left( \ket{\up \dw,\up } - \ket{\up\dw,\dw} \right) + b \left(
\ket{\up,\up\dw} + \ket{\dw,\up\dw} \right)  $ and $\ket{\psi_{d,\dw}} = a \left( \ket{ 0,\up} -
\ket{0,\dw} \right) + b \left( \ket{\dw, 0} + \ket{\up ,0} \right)$, rewritten in the physical basis:
\begin{equation}
\label{eq:doublet-ground-state}
\begin{split}
 \ket{\psi_{d,\up}} &=a \left( \ket{\tilde \up ,\tilde 0} - \ket{ \tilde \up,\tilde{\up\dw}} \right)
 + b \left( \ket{\tilde 0, \tilde \up } + \ket{ \tilde {\up\dw},\tilde \up} \right) \\
 \ket{\psi_{d,\dw}} &=a \left( \ket{ \tilde \dw , \tilde 0} - \ket{ \tilde \dw, \tilde {\up \dw}}
 \right) + b \left( \ket{ \tilde 0, \tilde \dw } + \ket {\tilde {\up\dw}, \tilde \dw} \right).
\end{split}
\end{equation}
This two-fold degenerate ground state  has a $z$-component of  the total spin  $\pm 1/2$  and hence corresponds to a 
local moment.

\subsection{Phase diagram}

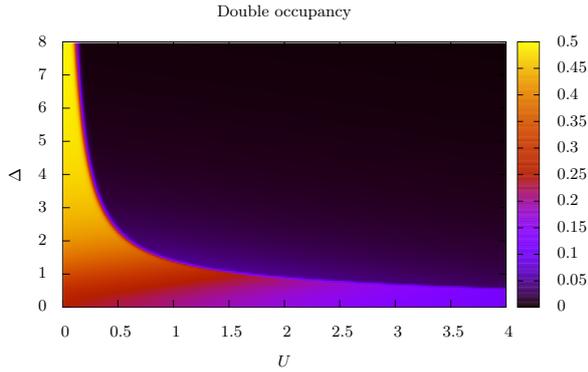
\begin{figure}
\resizebox{\columnwidth}{!}{\input{tex-double-occ-delta-U-B200-3d.tex}}
\vspace{-0.7cm}
\caption{ \label{fig:double_occupancy} (Color online) Double occupancy $\thavg{\tilde d_\up ^\dagger \tilde d_\up \tilde d_\dw^\dagger \tilde d_\dw}$ of the quantum dot in the effective model at $\beta=200$ and $V=0.5$. This plot can be understood as a phase diagram of the effective model, as the phase boundary is accompanied by a sharp decay of the double occupancy.  }
\end{figure}

To further illustrate the phase transition between the singlet state $\ket{\psi_s}$ and the doublet
states $\ket{\psi_{d,\up \dw}}$, the double occupancy $\thavg{\tilde d_\up ^\dagger \tilde d_\up
\tilde d_\dw^\dagger \tilde d_\dw}$ of the quantum dot in the effective model is shown in Fig. 
\ref{fig:double_occupancy}. At low temperature a very sharp drop of the double occupancy on the
phase boundary can be observed, which evolves to a jump at $T=0$.
Here the larger values of the double occupancy are connected to the
singlet phase, while the lower values belong to the doublet phase, where single occupancy is
favored. This can be understood by studying the expectation value of the double occupancy in the
ground state. In the singlet phase, we obtain
\begin{equation}
 \bra{\psi_s} \tilde{d}^\dagger _{\up} \tilde d_\up \tilde d_\dw^\dagger \tilde d_\dw \ket{\psi_s} =
 \left| \beta \right|^2 + \left| \gamma \right|^2,
\end{equation}
and for the doublet phase:
\begin{equation}
 \bra{\psi_{d,\up \dw}} \tilde{d}^\dagger _{\up} \tilde d_\up \tilde d_\dw^\dagger \tilde d_\dw \ket{\psi_{d,\up\dw}} = \left| a \right|^2.
\end{equation}
From the behavior of the weights $\beta$, $\gamma$ and $a$ shown in Fig. \ref{fig:weightrunning}
it is clear that the double occupancy increases with $\Delta$ in the singlet phase and decreases in
the doublet phase.

Note, that many of the results presented in this paper can be observed either at fixed $U$ or
$\Delta$ as can be conjectured from Fig. \ref{fig:double_occupancy}.

\subsection{Proximity effect}
\label{subsec:proximity_effect_eff}

To gain further insight in the sign change of the local pair correlations
$\thavg{\tilde{d}_\up^\dagger \tilde{d}_\dw^\dagger}$
\cite{PhysRevB.70.020502,PhysRevB.55.12648,balatsky:373}, we calculate the ground state expectation value of the local pair
correlations in the effective model (\ref{eq:Heff}). For the singlet phase, we obtain
\begin{equation}
  \begin{split}
	\bra{\psi_s} \tilde{d}_\up^\dagger \tilde{d}_\dw^\dagger \ket{\psi_s} &= \bra{\psi_s} \left(
	\beta \ket{\tilde{\up}\tilde{\dw},\tilde{\up}\tilde{\dw}} + \gamma \ket{\tilde{\up}
	\tilde{\dw},\tilde{0}} \right) \\ &= 2 \text{Re} (\beta^* \gamma) \geq 0.
      \end{split}
	\label{eq:eff_pair_corr_singlet}
\end{equation}
Clearly, only terms describing the pairing on the quantum dot contribute to the pair correlations, whereas
the Kondo singlet of electrons on the quantum dot and in the bath does not. From Fig. 
\ref{fig:weightrunning}, it is obvious that the resulting pairing
correlation is positive and increases with $\Delta$. This illustrates the proximity effect, as a
pair field in the bath induces a pair field on the quantum dot.

On the other hand, in the doublet phase, we obtain 
\begin{equation}
	\bra{\psi_{d,\dw}} \tilde{d}_\up^\dagger \tilde{d}_\dw^\dagger \ket{\psi_{d,\dw}} = 	
	\bra{\psi_{d,\dw}} a \ket{\tilde{\dw}, \tilde{\up\dw}} = - \left| a \right|^2 <0.
	\label{eq:eff_pair_corr_doublet}
\end{equation}
As in the singlet phase, only the states corresponding to a pairing on the quantum dot contribute to
the pair correlations. The local moment part of the ground state does not generate pair
correlations. As the weight $a$ in the doublet phase ground state is positive and decreases with
$\Delta$ (see Fig. \ref{fig:weightrunning}), the local pair correlations have a negative sign in
contrast to the positive sign in the singlet phase and decrease with $\Delta$.

\subsection{Spectral function}
\label{subsec:spectral_function_eff}
Using the Lehmann representation, the spectral function $A_{\up\up}(\omega)$ of the effective model is easily calculated. It is defined by
\begin{equation}
 A_{\up \up}(\omega) =  \frac{\pi}{Z} \sum \limits_{n,m} 
M_{nm}
 \left( \E^{-\beta E_m} \! + \! \E^{-\beta E_n} \right) \delta(\omega \! + \! E_n \! - \! E_m),
\end{equation}

with the matrix elements $ M_{nm} = \left|\bra{n} \tilde d_\up^\dagger \ket{m} \right|^2 $.
\begin{figure}
  \vspace{-0.5cm}
\resizebox{\columnwidth}{!}{\input{tex-spectralfunction-B200-3d.tex}}
\vspace{-1.5cm}
\caption{ \label{fig:spectral-effective} (Color online) Spectral function $A_{\up \up}(\omega)$ of the effective model for different values of $\Delta$ at $\beta=200$, $U=1$ and $V=0.5$. The $\delta$-peaks have been broadened by a Gaussian function of width $\sigma=0.04$ for better visibility. }
\end{figure}
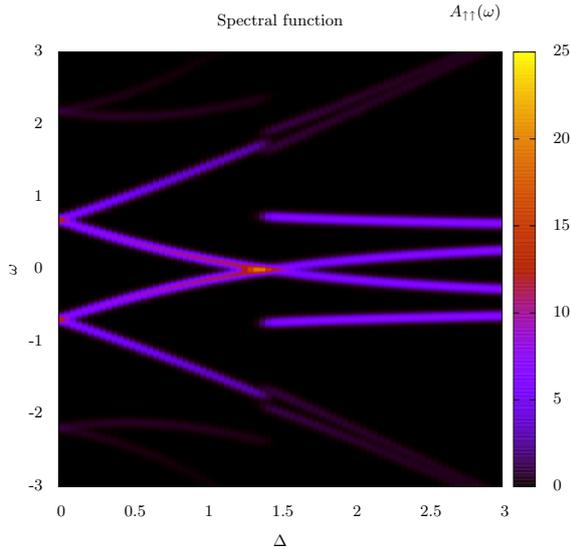
The spectral function is shown in Fig. \ref{fig:spectral-effective}. 
Comparing this plot to the numerical solution of the full model as depicted in Fig. \ref{fig:Aom-of-Delta}, we observe, that the simple model already shows the important feature of an excitation at the position $\omega=0$ at the critical value of $\Delta$. 
Even though for very small values of $\Delta$, the Kondo resonance at $\omega=0$ can not be seen in
the simple model, we see a precursor of the Kondo resonance as a pole of the Green's function, which
develops into a resonance if we increase the number of sites in the bath\cite{hewson}.

A careful analysis reveals, that the low frequency signature of the spectral function reflects the excitation between the two lowest lying states of the spectrum. 
These states are the ground states of the singlet and the doublet phase and therefore, the position
$\omega$ of the excitation marks precisely the energy difference of the two ground states.
At the critical value of $\Delta=1.412$, the level crossing occurs and leads to a vanishing energy
difference of the two ground states, meaning that the excitation between the two states lies now
precisely at $\omega=0$.

\subsection{Dynamical spin structure}
\label{subsec:dynamical_spin_structure_eff}

Like the spectral function, the dynamical spin structure factor $S(\omega)$ can be calculated using the Lehmann representation:
\begin{equation}
	\label{eq:spinstructure-lehmann_2}
	S(\omega) = \frac{\pi}{Z} \sum_{n,m} \E^{-\beta E_n} \left| \bra{n} \tilde{S}_+ \ket{m}
	\right|^2 \delta( \omega + E_n - E_m).
\end{equation}
In the Monte Carlo simulation, a numerically more stable quantity is obtained by replacing $S_+$ by $S_z$ in the
above equation. This quantity is completely equivalent to $S(\omega)$, as we only make use of the
$SU(2)$-symmetry of the problem, and is therefore used in the following.

In the representation (\ref{eq:spinstructure-lehmann_2}) of $S(\omega)$, it is clear that the dynamical spin structure factor will show excitations
at frequencies corresponding to the energy needed to flip the spin on the quantum dot. Therefore,
the dynamical spin structure factor is very well suited to determine whether the system is in the
singlet or in the doublet regime.

\begin{figure}
 \resizebox{\columnwidth}{!}{\input{tex-spinstructure-eff-3d.tex}}
 \vspace{-0.7cm}
\caption{\label{fig:spinstructure-eff}
(Color online)
 Dynamical spin structure factor $S(\omega)$ of the effective model at $\beta=200$. The phase transition from the singlet-phase to the doublet-phase for $U=1$ and $V=0.5$ occurs at $\Delta \approx 1.412$. At this point a transition from a gapped excitation to a peak at $\omega=0$ corresponding to a local magnetic moment in the doublet phase is observed. To visualize the $\delta$-functions, a Gaussian broadening of width $\sigma=0.05$ has been applied. }
\end{figure}
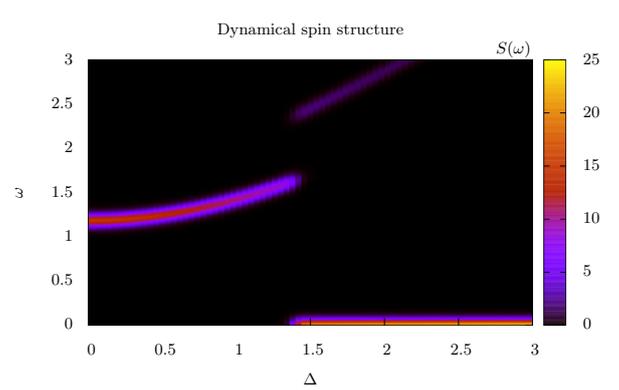

In Fig. \ref{fig:spinstructure-eff} the phase transition from the singlet phase to the doublet phase is reflected by the fact, that in the singlet phase, a gapped excitation can be observed, whereas in the doublet phase, a peak at $\omega=0$ emerges, which corresponds to a local magnetic moment.

\subsection{Dynamical charge structure}
\label{subsec:dynamical_charge_structure_eff}

The dynamical charge structure factor $N(\omega)$ can be defined by the Lehman representation 
\begin{equation}
	N(\omega) = - \frac{\pi}{Z} \sum_{n,m} \left| \bra{n} \tilde{n} - \delta_{n,m} \ket{m}
	\right|^2 \E^{-\beta E_m} \delta(\omega + E_n - E_m).
	\label{eq:chargestructure-lehmann}
\end{equation}

\begin{figure}
 \resizebox{\columnwidth}{!}{\input{tex-chargestructure-b200-eff-3d.tex}}
 \vspace{-0.7cm}
 \caption{ \label{fig:chargestructure-eff} 
 (Color online)
 Dynamical charge structure factor $N(\omega)$ of the effective model at $\beta=200$. We have used
 the same parameters as for Fig. \ref{fig:spectral-effective} }

\end{figure}
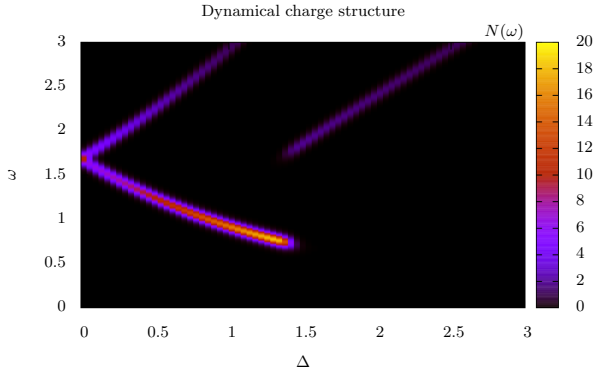

As for the other spectral functions, the charge structure factor $N(\omega)$ shown in Fig. 
\ref{fig:chargestructure-eff}, exhibits a sharp change of its behavior at the phase transition for
the critical value of the superconducting gap $\Delta$. We observe, that the charge structure shows a
finite gap for all values of $\Delta$ and that for large values of $\Delta$, the gap increases in a
slightly nonlinear manner.

A more detailed study of the matrix elements contributing to the charge structure factor reveals,
that because of correlation we have completely different excitations than for the spectral function.
In fact, the most prominent excitations are excitations from the respective ground states in the two
different phases to higher energy states with structure similar  to that of the  ground states.

\section{CTQMC}
\label{sec:CTQMC}
\subsection{Basic outline of the algorithm}

For the numerically exact solution of the BCS-Anderson-model, we used the weak coupling CTQMC-method
\cite{rubtsov:035122}, which is based on a perturbation expansion around the limit of $U=0$.
Following the presentation of the CTQMC-algorithm in \cite{assaad:035116}, we will shortly outline
the basic principles of the method.

As pointed out in \cite{rubtsov:035122,assaad:035116} the interacting Hamiltonian $H_U$ in Eq. (\ref{eq:qd_bcs_hamiltonian_nambu}) can up to a constant be rewritten as 
\begin{equation}
 H_U = -\frac{U}{2} \sum_{s=\pm 1} \left( d_\up^\dagger d_\up - \alpha_\up^s \right) \left( d_\dw^\dagger d_\dw - \alpha_\dw^s \right)
\end{equation}
introducing the parameters $\alpha_\sigma^s$ to minimize the sign problem. For the present case, a choice of $\alpha_\up^s = \alpha_\dw^s = \frac{1}{2} + s \delta$ with $\delta = \frac{1}{2} + 0^+$ was found to completely eliminate the sign problem at half filling, even after the complex phase factors $\exp(i\phi_\alpha)$ in the Hamiltonian were introduced.

Using perturbation theory, the partition function $Z$ of the full Hamiltonian (\ref{eq:qd_bcs_hamiltonian_nambu}) can be written as:
\begin{equation}
\begin{split}
 \frac{Z}{Z_0} &= \thavg{ T \E^{- \int_0^\beta \D \tau H_U(\tau)  } }_0 =\\
&= \sum_{n=0}^\infty \left( \frac{U}{2} \right)^n \int _0 ^\beta \D \tau_1 \dots \int_0 ^{\tau_{n-1}} \D \tau_n \sum_{s_1,\dots, s_n} \times \\
&\times \thavg{T \left(\hat{n}_\up(\tau_1) - \alpha_\up^{s_1} \right) \dots
\left(\hat{n}_\dw(\tau_n) - \alpha_\dw^{s_n} \right) }_0.
\end{split}
\end{equation}
with the number operators $\hat{n}_\sigma = d^\dagger_\sigma d_\sigma$ and the thermal expectation value  $\thavg{ \bullet }_0 = \frac{1}{Z_0} \tr \left[ \E^{-\beta H_0} \bullet \right]$. 
As $H_0$ is a noninteracting Hamiltonian, Wick's theorem holds, and the expectation value $\thavg{T (\hat{n}_\up(\tau_1) - \alpha_\up^1 ) \dots (\hat{n}_\dw(\tau_n) - \alpha_\up^n ) }_0$ can be cast in a determinant of a matrix $\mat{M}_{C_n}$ of size $2n\times 2n$, where $C_n$ is a configuration of vertices $\{\tau_i,s_i\}$. In contrast to the formulation for the Hubbard model given in \cite{assaad:035116}, we do not need to include an index for the lattice site as we only have one correlated site, the impurity. The Matrix $\mat{M}_{C_n}$ is not block diagonal for the two spin sectors in the case $\Delta \neq 0$, so we cannot factor the determinant in two determinants of $n\times n$ matrices. Finally, the partition function of the model is given by
\begin{equation}
 \frac{Z}{Z_0} = \sum_{C_n} \left( \frac{U}{2} \right)^n \det \mat{M}_{C_n},
\end{equation}
where the sum runs over all possible configurations $C_n$ of vertices as in \cite{assaad:035116}. The matrix $\mat{M}_{C_n}$ is defined by
\begin{equation}
 \mat{M}_{C_n} = 
\begin{pmatrix}
 \mat G_{dd}^0(\tau_1,\tau_1) - \matgr \alpha_1 & \dots & \mat G_{dd}^0(\tau_n,\tau_1) \\
\vdots & \ddots & \vdots \\
 \mat G_{dd}^0(\tau_1,\tau_n) & \dots & \mat G_{dd}^0(\tau_n,\tau_n)  - \matgr \alpha_n \\
\end{pmatrix}
\end{equation}
using the $2\times 2$ Green's function matrices $\mat{G}_{dd}^0(\tau,\tau') = \left(\begin{smallmatrix}
                                                                  \thavg{T d^\dagger_\up(\tau) d_\up(\tau')}_0 & \thavg{T d^\dagger_\dw(\tau) d_\up(\tau')}_0 \\
								  \thavg{T d^\dagger_\up(\tau) d_\dw(\tau')}_0 & \thavg{T d^\dagger_\dw(\tau) d_\dw(\tau')}_0 \\
                                                                 \end{smallmatrix}\right)$
								 and with $\matgr{\alpha}_i= \left( \begin{smallmatrix} \alpha_\up^i & 0 \\ 0 & \alpha_\dw^i \end{smallmatrix}\right)$.

A similar reasoning yields an expression for the thermal expectation value $\thavg{O(\tau)} = \frac{1}{Z} \tr \left[ \E^{-\beta H} O(\tau) \right]$ of the full model:
\begin{equation}
\label{eq:ddqmc_thermal_expectation_value}
\thavg{O(\tau)} = \frac{  \sum_{C_n} \left(\frac{U}{2} \right)^n \det \mat{M}_{C_n} \langle \langle O(\tau) \rangle \rangle_{C_n}  }{ \sum_{C_n} \left(\frac{U}{2} \right)^n \det \mat{M}_{C_n}  }.
\end{equation}
Here $\langle \langle O(\tau) \rangle \rangle_{C_n}$ is the contribution of the configuration $C_n$ to the observable $O(\tau)$, which is given by
\begin{equation}
\label{eq:config_contrib}
 \langle \langle O(\tau) \rangle \rangle_{C_n} = \frac{\thavg{T (\hat{n}_\up(\tau_1) - \alpha_\up^1 ) \dots (\hat{n}_\dw(\tau_n) - \alpha_\dw^n ) O(\tau) }_0}{ \thavg{T (\hat{n}_\up(\tau_1) - \alpha_\up^1 ) \dots (\hat{n}_\dw(\tau_n) - \alpha_\dw^n ) }_0 }.
\end{equation}
Both, the numerator and the denominator of the above Eq. (\ref{eq:config_contrib}) can be written as determinants of matrices using Wick's theorem.
Eq. (\ref{eq:ddqmc_thermal_expectation_value}) is the central relation of the CTQMC algorithm,
because starting from this equation, the Metropolis-Hastings-Algorithm can be employed to generate a
Markov chain of configurations $C_n$.
At this point, we have to interpret $\left(\frac{U}{2} \right)^n \det \mat{M}_{C_n}$ as the
statistical weight of a given configuration $C_n$ what in general is impossible, as $\det
\mat{M}_{C_n}$ is a complex number. Therefore, we have to replace $\left(\frac{U}{2} \right)^n \det
\mat{M}_{C_n}$ by its modulus and account for the phase in the measurement of the observables.
Fortunately, in the present case, the statistical weights are always real and nonnegative, so that  we
can simply calculate the contribution to the observable $O(\tau)$ for a given configuration $C_n$ in
the Markov chain as $\langle \langle O(\tau) \rangle \rangle_{C_n}$.

\subsection{ Wick's theorem for each configuration }

For the measurement of higher Green's functions of the form $\thavg{ T \gamma_1^\dagger \gamma_{1'}
\dots \gamma_m^\dagger \gamma_{m'}}$, where $\gamma_i^\dagger$ stands for
$d^\dagger_{\sigma_i}(\tau_{i,\text{meas}})$ or
$c^\dagger_{k_i,\sigma_i,\alpha_i}(\tau_{i,\text{meas}})$ depending on the quantity of interest, the
calculation of the contribution $\langle \langle T \gamma_1^\dagger \gamma_{1'} \dots
\gamma_m^\dagger \gamma_{m'} \rangle \rangle_{C_n}$ is tedious and time consuming. Luckily for every
configuration $C_n$ a relation similar to Wick's theorem can be found, which greatly simplifies the
calculation of higher Green's functions. It is closely connected to the determinant identity
(\ref{eq:determinant_identity}) proven in appendix \ref{sec:proof_of_det_identity}. The application
of the ordinary Wick's theorem to the denominator and the numerator of Eq. (\ref{eq:config_contrib}) yields
\begin{equation}
 \langle \langle T \gamma_1^\dagger \gamma_{1'} \dots \gamma_m^\dagger \gamma_{m'} \rangle \rangle_{C_n} = 
\frac{ \det \mat{B}_{C_n} }{ \det \mat{M}_{C_n} },
\end{equation}
where we have defined the matrix $\mat B_{C_n} \in \mathbb{C}^{(2n+m) \times (2n+m)} $ as
\begin{widetext}
\begin{equation}
\mat B_{C_n} =  
\begin{pmatrix}
    & & & \thavg{T \gamma_1^\dagger \vec d(\tau_1)}_0 & \dots & \thavg{T \gamma_m^\dagger \vec d(\tau_1)}_0 \\
    & \mat{M}_{C_n} & & \vdots & \ddots & \vdots \\
    & & & \thavg{T \gamma_1^\dagger \vec d(\tau_n)}_0 & \dots & \thavg{T \gamma_m^\dagger \vec d(\tau_n)}_0 \\
   \thavg{T \vec d^\dagger(\tau_1) \gamma_{1'}}_0 & \dots & \thavg{T \vec d^\dagger(\tau_n) \gamma_{1'}}_0 & \thavg{T \gamma_1^\dagger \gamma_{1'}}_0  &  \dots  & \thavg{T \gamma_m^\dagger \gamma_{1'}}_0  \\
  \vdots & \ddots & \vdots & \vdots & \ddots & \vdots \\
\thavg{T \vec d^\dagger(\tau_1) \gamma_{m'}}_0 & \dots & \thavg{T \vec d^\dagger(\tau_n) \gamma_{m'}}_0 & \thavg{T \gamma_1^\dagger \gamma_{m'}}_0  & \dots  & \thavg{T \gamma_m^\dagger \gamma_{m'}}_0  \\
 \end{pmatrix}.
\end{equation}
\end{widetext}
Defining the matrices $\mat {B}^{\mathbf{ij} }_{C_n} \in \mathbb{C}^{(2n+1)\times(2n+1)}$, we can make use of the determinant identity (\ref{eq:determinant_identity})
\begin{equation}
 \! \mat {B}^{ \mathbf{ij} }_{C_n} = \!\!
\begin{pmatrix}
  & & & \!\! \thavg{T \gamma_j^\dagger \vec d(\tau_1)}_0 \\
 & \!\!\!\! \mat{M}_{C_n} \!\!\!\! & & \vdots\\
& & & \!\! \thavg{T \gamma_j^\dagger \vec d(\tau_n)}_0 \\
 \thavg{T \vec d^\dagger(\tau_1) \gamma_{i'}}_0 \! & \! \dots \! & \! \thavg{T \vec d^\dagger(\tau_n) \gamma_{i'}}_0 \!\! & \thavg{T \gamma_j^\dagger \gamma_{i'}}_0 
\end{pmatrix}\!\!,
\end{equation}
yielding
\begin{equation}
 \frac{ \det \mat B_{C_n} }{ \det \mat M_{C_n} } = \frac{1}{(\det \mat M_{C_n})^n } 
\det \begin{pmatrix}
      \det \mat {B}^{ \mathbf{11} }_{C_n} & \dots & \det \mat {B}^{ \mathbf{1m} }_{C_n} \\
 	\vdots & \ddots & \vdots \\
      \det \mat {B}^{ \mathbf{m1} }_{C_n} & \dots & \det \mat {B}^{ \mathbf{mm} }_{C_n} \\
     \end{pmatrix}\!\!.
\end{equation}
From Eq. \ref{eq:config_contrib} it is obvious, that $\det \mat {B}^{ \mathbf{ij} }_{C_n} / \det \mat M_{C_n} $ is identical to the contribution of the configuration $C_n$ to the one particle Green's function $\thavg{ T \gamma_j^\dagger \gamma_{i'} }$. Hence, Wick's theorem holds for every configuration $C_n$ and is given by
\begin{equation}
\begin{split}
  &\langle \langle T \gamma_1^\dagger \gamma_{1'} \dots \gamma_m^\dagger \gamma_{m'} \rangle \rangle_{C_n} =\\ &\det 
\begin{pmatrix}
 \langle \langle T \gamma_1^\dagger \gamma_{1'} \rangle \rangle_{C_n} & \dots & \langle \langle T \gamma_m^\dagger \gamma_{1'} \rangle \rangle_{C_n} \\
\vdots & \ddots & \vdots \\
 \langle \langle T \gamma_1^\dagger \gamma_{m'} \rangle \rangle_{C_n} & \dots & \langle \langle T \gamma_m^\dagger \gamma_{m'} \rangle \rangle_{C_n} \\
\end{pmatrix}.
\end{split}
\end{equation}
This relation is particularly useful in a simulation measuring multiple physical observables as measurements of single particle Green's functions can be reused in an economic way.

\section{Numerical Results}
\label{sec:NumericalResults}

In this section, we present the results obtained by CTQMC simulations for the model
(\ref{eq:qd_supercond_hamiltonian}). We restrict ourselves to the case of half filling, 
$\epsilon_d=0$ and $\mu=0$. In the first part of this section, we will discuss the results for
static quantities  including  the Josephson current, double occupancy  and  pair
correlations on the quantum dot.  We then  proceed to dynamical quantities such as  the single particle
spectral function and the dynamical spin structure factor.

\subsection{Josephson current}
\label{sec:Josephson}
The Josephson current flowing through the Quantum dot can be calculated directly within the CTQMC method, as it is given by an equal time Green's function:

\begin{equation}
	\thavg{j_\alpha} = i \frac{V}{\sqrt{N}} \sum_{k,\sigma} \thavg{ \tilde{c}_{k,\sigma,\alpha}^\dagger \tilde{d}_\sigma - \tilde{d}_\sigma^\dagger \tilde{c}_{k,\sigma,\alpha}   }	
	\label{eq:josephson_current}
\end{equation}

\begin{figure}[h]
	\begin{center}
 \resizebox{\columnwidth}{!}{\input{tex-josephson_pi_shift_1.tex}}
	\end{center}
	\vspace{-0.7cm}
	\caption{(Color online) Josephson current in the $0$ junction regime}
	\label{fig:josephson-0-junction}
\end{figure}
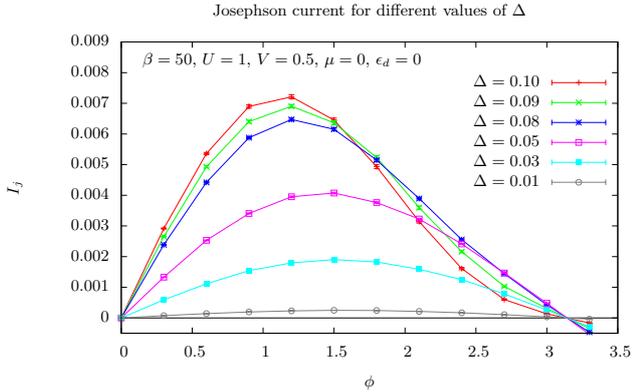

We show here our results for the Josephson current at an inverse temperature of $\beta=50$ as a function of the superconducting gap $\Delta$. 
For small values of $\Delta$, we observe a sinusoidal form of the Josephson current as a function of
the phase difference $\phi$ with increasing amplitude, as $\Delta$ increases (see Fig. 
\ref{fig:josephson-0-junction}).

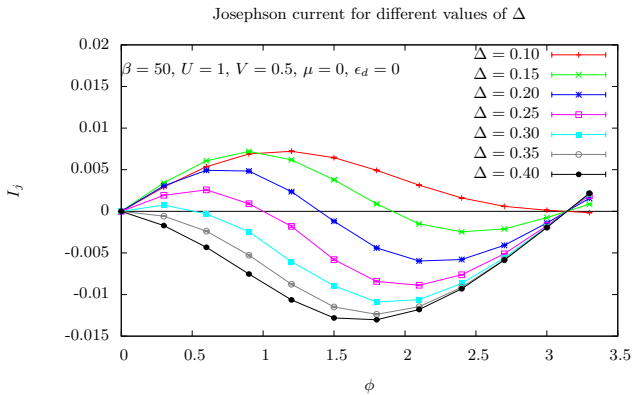
\begin{figure}[h]
	\begin{center}
 \resizebox{\columnwidth}{!}{\input{tex-josephson_pi_shift_2.tex}}
	\end{center}
	\vspace{-0.7cm}
	\caption{(Color online) Josephson current in the $0'$ and $\pi'$ junction regime.}
	\label{fig:josephson-transition}
\end{figure}

This parameter regime is known as the 0-Junction regime, because the Josephson current $I_j(\phi) =
\frac{\partial \Omega}{\partial \phi}$ has a zero with positive slope at $\phi=0$, corresponding to
a minimum in the grand potential $\Omega$ at $\phi=0$ (see Fig. 5 in reference
\cite{karrasch:024517}). 

If the value of $\Delta$ is further increased, the behavior of the Josephson current changes, as in
the region $\Delta \approx 0.15 \dots 0.35$ the Josephson current shows a zero between $\phi=0$ and
$\phi=\pi$. (see Fig. \ref{fig:josephson-transition}).
This leads to a minimum in the grand potential at $\pi$ and the parameter regime is called $0'$ or
$\pi'$ regime depending on which minimum of the grand potential is the global one
\cite{PhysRevLett.82.2788}.
The behavior of the Josephson current is in accordance with the behavior of the double occupancy
seen in Fig. \ref{fig:double-occ-of-Delta}, as in the same parameter region, where we observe the $0'$ to $\pi'$ transition, the drop of the double occupancy as a function of $\phi$ can be observed, which is linked to the change of the curvature of the current-phase relation of the Josephson current.

\begin{figure}[h]
	\begin{center}
 \resizebox{\columnwidth}{!}{\input{tex-josephson_pi_shift_3.tex}}
	\end{center}
	\vspace{-0.7cm}
	\caption{(Color online) Josephson current in the $\pi$ junction regime.}
	\label{fig:josephson-pi-junction}
\end{figure}
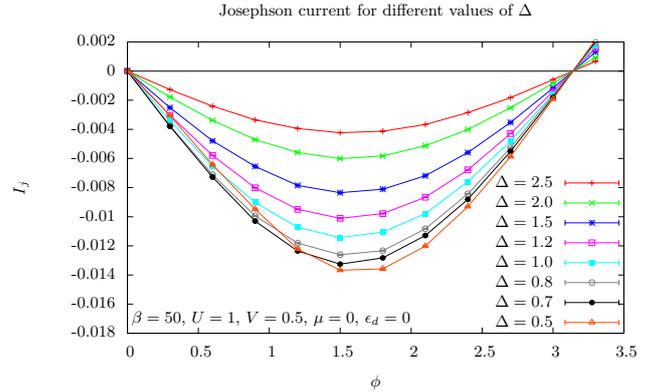

For larger values of $\Delta$, the sign of the Josephson current changes and the grand potential
shows now a single minimum at $\phi=\pi$, this regime is therefore called the $\pi$ regime.
(see Fig. \ref{fig:josephson-pi-junction}).

The picture for the behavior of the grand potential as a function of $\phi$ that we get from the
current phase relation of the Josephson current agrees very nicely with the results presented by
Benjamin et al.\cite{controllable.pi.junction.benjamin}.

The current phase relations for the different phases presented here were also extensively studied by
Karrasch et al. using the fRG and NRG methods \cite{karrasch:024517}, Choi et al. using the NRG
method \cite{PhysRevB.70.020502}, as well as by Siano and Egger using
the Hirsch-Fye QMC method \cite{PhysRevLett.93.047002,PhysRevLett.94.039902,PhysRevLett.94.229702}. 
Even though the numerical exactness of certain results has been debated, the results of all
numerical works show very good qualitative agreement and are confirmed by the present results.

\begin{figure}
\resizebox{\columnwidth}{!}{\input{tex-josephson-curr-temp.tex}}
\vspace{-0.7cm}
\caption{ \label{fig:jos-curr-temperature} (Color online) Josephson current for different temperatures. The current
phase relations do not intersect at one single point as suggested by the NRG results of Karrasch et
al.\cite{karrasch:024517}.
}
\end{figure}
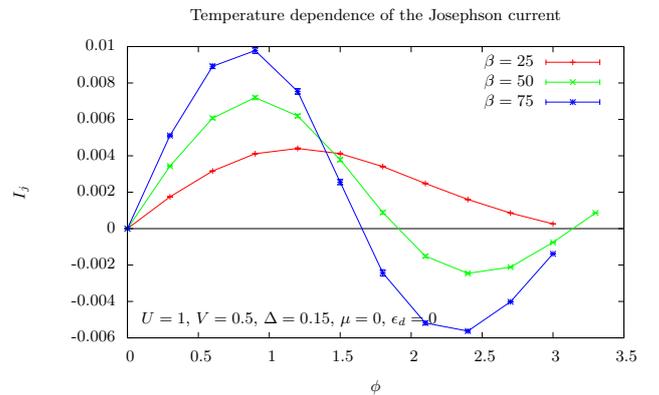

In the literature\cite{PhysRevLett.94.229702,karrasch:024517}, the temperature dependence of the
current phase relation of the Josephson current has been discussed. We show CTQMC results in Fig. 
\ref{fig:jos-curr-temperature} which look
very similar to the Siano and Egger result\cite{PhysRevLett.94.229702}. As CTQMC is numerically
exact, our result suggests that the crossing of all curves in one single point\cite{karrasch:024517}
at $I_j=0$ found in the approximate finite temperature NRG is not universal.

\subsection{Double occupancy}

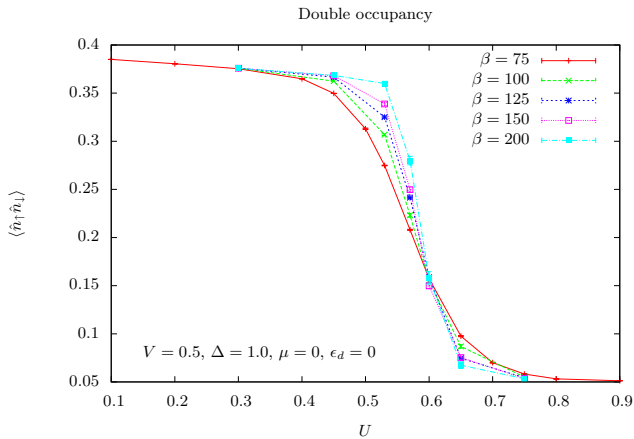
\begin{figure}[h]
 \resizebox{\columnwidth}{!}{\input{tex-double-occ-measurements-D1.tex}}
 \vspace{-0.7cm}
\caption{ \label{fig:double_occupancy_D1.0} (Color online) Double occupancy $\thavg{\hat{n}_\up \hat{n}_\dw}$ of
the quantum dot at $\Delta=1.0$. The data shows a jump in the double occupancy becoming sharper with
decreasing temperature.
}
\end{figure}

We learned from the toy model described in Sec. \ref{sec:toy-model} that the system exhibits  
a phase transition from the singlet phase to the doublet phase as  $U$ is increased. This picture is 
consistent with the NRG results of Bauer et al. \cite{0953-8984-19-48-486211}.
The phase transition can be observed in the double occupancy $\thavg{\hat{n}_\up \hat{n}_\dw}$ of
the quantum dot, which is proportional to $\frac{\partial \Omega}{\partial U}$, where $\Omega$ is
the grand potential.
At $T=0$, a sharp step function of the double occupancy is expected. 
While the $T=0$ regime is not directly accessible to quantum Monte Carlo calculations, we calculated the double occupancy for different temperatures using the CTQMC-method. 
The results are shown in Fig.  
\ref{fig:double_occupancy_D1.0}. From the data, it is obvious that with decreasing temperature the
curves converge to the step function of the limit $T=0$, which is a clear sign for a first order
phase transition, reflecting a level crossing of the two ground states. This is in complete
accordance with the results for the toy model.

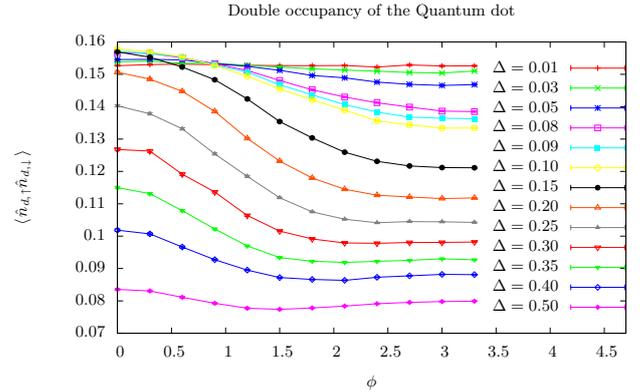
\begin{figure}[h]
	\begin{center}
 \resizebox{\columnwidth}{!}{\input{tex-josephson_double_occ1.tex}}
	\end{center}
	\vspace{-0.7cm}
	\caption{(Color online) Double occupancy of the quantum dot as a function of the phase difference $\phi=\phi_L-\phi_R$ for different values of $\Delta$.}
	\label{fig:double-occ-of-Delta}
\end{figure}

It is interesting to correlate the Josephson current as a function of the phase difference 
$\phi=\phi_L-\phi_R$ for various values of $\Delta$ (see Sec. \ref{sec:Josephson}), with  the double occupancy on 
the  dot.
As depicted in Fig. \ref{fig:double-occ-of-Delta}, for very small values of $\Delta$ as well as 
for $\Delta >\approx 0.4$, we see that the double occupancy is a constant function of $\phi$. 
This corresponds to a current-phase-relation for the Josephson current fixed in either 
the $\pi$- or the $0$-junction regime.
For intermediate values of $\Delta$, we observe a far more interesting behavior of the double
occupancy: At a certain value of $\phi$, the double occupancy drops to a smaller value. This drop is
of course smeared out by the finite temperature, but can be understood as a way to drive the phase
transition from the $0$- to the $\pi$-junction regime by the phase difference~$\phi$.

\subsection{Pair correlation}

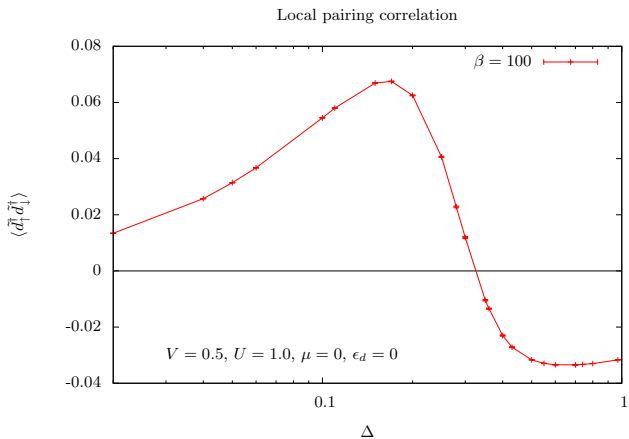
\begin{figure}
 \resizebox{\columnwidth}{!}{\input{tex-local_pair_correlation.tex}}
 \vspace{-0.7cm}
 \caption{ \label{fig:local_pair_correlation} (Color online) Local pair correlation $\Delta_d=\langle \tilde{d}_\up^\dagger
 \tilde{d}_\dw^\dagger \rangle$ as a function of $\Delta$. We observe the same behavior as Choi et
 al. \cite{PhysRevB.70.020502}, which is also in very good agreement with the pair correlation
 expected for the effective model discussed in \ref{subsec:proximity_effect_eff}.
}
\end{figure}

In agreement with the NRG result of Choi et al. \cite{PhysRevB.70.020502} as well as with
the mean field results by Salkola et. al. \cite{PhysRevB.55.12648}, we obtain the
local pair correlation on the quantum dot shown in Fig. \ref{fig:local_pair_correlation}. 
For small $\Delta$, the local pair correlation increases because of the proximity effect, as an
increasing magnitude of the pair field $\Delta$ in the leads induces a growing pair correlation on
the quantum dot.
The sharp sign change at the critical value of $\Delta$
observed at zero temperature is smeared out at finite temperatures, but the qualitative behavior
is exactly the same as for the effective model discussed in Sec.  
\ref{subsec:proximity_effect_eff}. We therefore conclude, that the sign change of the pair
correlation is due to residual pairing on the quantum dot in the doublet phase which decreases with
$\Delta$.

The same qualitative behavior of the local pair correlation is also observed, if $U$ is changed
instead of $\Delta$ as discussed in \cite{0953-8984-19-48-486211,PhysRevB.55.12648}.
The sign change of the local pair correlation $\Delta_d$ is traditionally expressed as a $\pi$-phase shift in
$\Delta_d$.

\subsection{Spectral function}
\label{subsec:impurity-spectral}

All  quantities studied so far  suggest  that a first order phase transition occurs when
we tune the system from the $0$-Junction to the $\pi$-Junction regime. This can be confirmed by
studying dynamical quantities such as the spectral function.

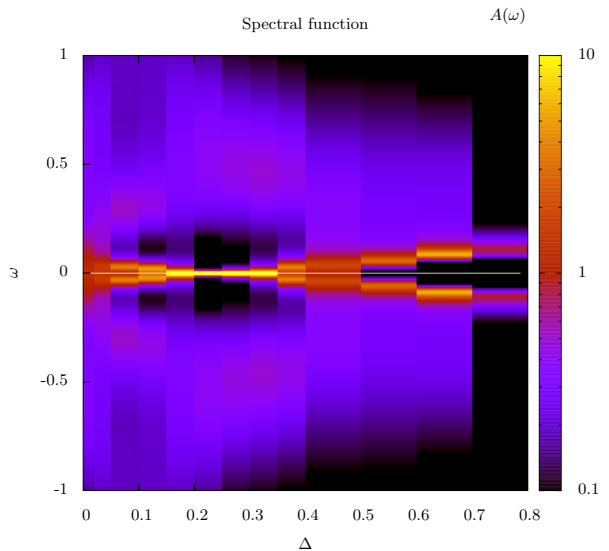
\begin{figure}[h]
  \vspace{-0.6cm}
 \resizebox{\columnwidth}{!}{\input{tex-impurity-Aom-of-Delta-3d.tex}}
 \vspace{-1.5cm}
 \caption{ \label{fig:Aom-of-Delta} (Color online) Spectral function $A(\omega)$ as a function of $\Delta$ for the parameters $\beta=100$, $U=1.0$ and $V=0.5$ at half filling and zero phase difference between the two superconductors.}
\end{figure}

In Fig. \ref{fig:Aom-of-Delta} we show the spectral function $A(\omega)$ of the quantum dot as a function of $\Delta$.
The data has been calculated from the CTQMC data for the Green's function $G_{dd}^{\up\up}(\tau)$
using stochastic analytic continuation\cite{PhysRevB.57.10287,beach-2004}.
This method works especially well for the low energy spectrum and sharp excitations while the high
energy spectrum and excitation continua are more difficult to resolve.
Inside the gap, the formation of Andreev bound states can be seen very well.

In the region of $\Delta\approx0$ we see the Kondo-resonance. As a function of growing values of $\Delta$ 
and as a consequence  of the opening of the quasiparticle gap at the Fermi level, the Kondo resonance  evolves to 
Andreev bound state. Note that at the mean-field level, the Kondo resonance merely corresponds to a 
virtual bound state.   Opening a quasiparticle gap at the Fermi level drives the lifetime  of the  this virtual 
bound state to infinity. In the parameter region which corresponds to the $0$-Junction regime of the Josephson current 
($\Delta \approx 0\dots0.1$), we observe Andreev bound states with  excitation energies approaching $\omega=0$. This corresponds to the crossing point in Fig. \ref{fig:Aom-of-Delta}
and  has also been observed by Bauer et al. for fixed $\Delta$ and increasing $U$ in \cite{0953-8984-19-48-486211}.

The comparison of the Quantum Monte Carlo data shown in Fig. \ref{fig:Aom-of-Delta} with the
result obtained from the effective model discussed in Sec. \ref{subsec:spectral_function_eff} is
particularly insightful. The spectral signature is very similar except for the lack of the Kondo resonance due
to the finite size of the effective model. In the  effective model, the Andreev
bound state excitation corresponds to the energy difference between the ground states of the singlet
and the doublet phase. The position $\Delta$ at which the Andreev bound states cross at $\omega=0$
has been identified as a clear sign for the crossing of the ground states of the singlet and doublet
phases.  Hence, we interpret the crossing of the Andreev bound states in the CTQMC data as a
very strong sign for a level crossing and hence a first order phase transition from the singlet
to the doublet phase in the full model.

\subsection{Dynamical spin structure factor}

In addition to the spectral function, the dynamical spin structure factor $S(\omega)$ defined in
Eq. \ref{eq:spinstructure-lehmann_2}, provides a way of characterizing the phases of the system. 
For $\Delta=0$, we clearly see a suppressed spectral weight at $\omega=0$ and a peak which
corresponds to the characteristic energy scale of the Kondo temperature $T_K$. From the peak
position, we obtain a rough estimate for the Kondo temperature of $T_K\approx 0.06$.

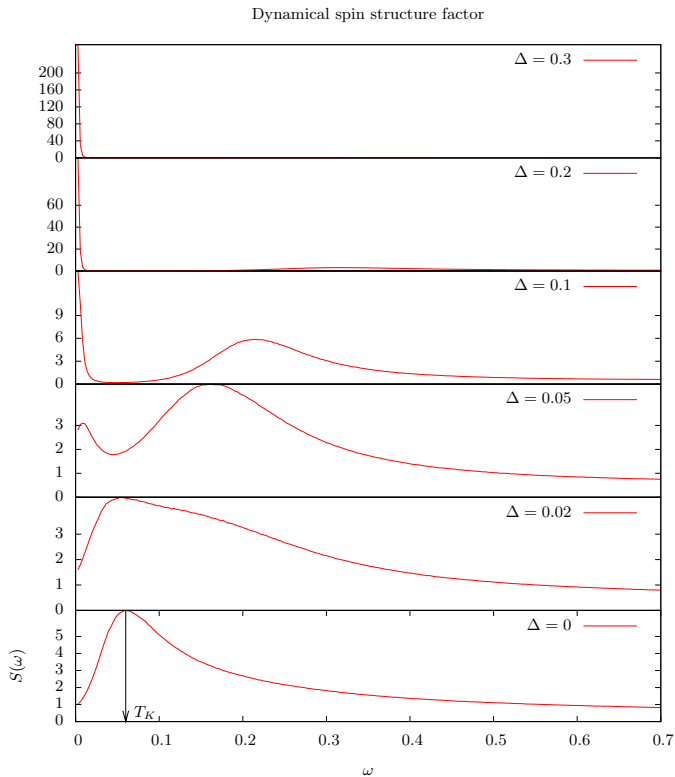
\begin{figure}[h]
  \begin{center}
    \resizebox{\columnwidth}{!}{\input{tex-impurity-Som-of-Delta.tex}   }
 \end{center}
 \caption{ \label{fig:Som-of-Delta} (Color online) Dynamical spin structure factor $S(\omega)$ as a function of $\Delta$ for the parameters $\beta=100$, $U=1.0$ and $V=0.5$ at half filling and zero phase difference between the two superconductors.
 For $\Delta=0$ we can roughly estimate the Kondo-Temperature $T_K\approx 0.06$ from the peak position of $S(\omega)$.
 }
\end{figure}

From  $\Delta\approx0.05$ onwards,  spectral weight is accumulated at $\omega=0$  ultimately forming  a  pronounced
sharp local moment peak for large values of  $\Delta$.  As the Kondo temperature is a measure for the energy 
required to break the Kondo singlet, we expect the Kondo effect to break down at a value of $\Delta\approx T_K$.
This  is indeed observed in Fig. \ref{fig:Som-of-Delta}.

The signature of the breakdown of the Kondo resonance also shows up in the spectral function  plotted  Fig. \ref{fig:Aom-of-Delta}.
Since  the Kondo resonance stems from a screening of the magnetic moment by conduction electrons in an energy window 
$T_K$ around the Fermi level, the opening of a single particle gap of order $T_K$ destroys the Kondo resonance giving 
way to an  Andreev bound state. 

The breakdown of the Kondo resonance  is
accompanied by a change of the curvature in the current-phase-relation of the Josephson current
which is a precursor for the transition to the $0'$ phase
(see the  curves for $\Delta=0.05$ and $\Delta=0.08$ in Fig. \ref{fig:josephson-0-junction}). 
We also observe that after the transition from the $\pi'$- to the $\pi$- regime has occurred
(see the current-phase-relation of the Josephson current of Fig. \ref{fig:josephson-transition}) 
 the peak at finite $\omega$ vanishes  and all the spectral weight is accumulated in the very sharp local moment peak at $\omega=0$.

\subsection{Charge gap}

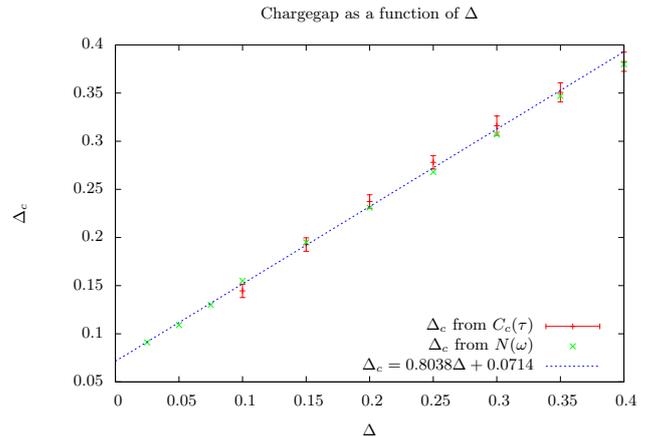
\begin{figure}[h]
	\begin{center}
		\resizebox{\columnwidth}{!}{\input{tex-chargegap.tex}}
	\end{center}
	\vspace{-0.7cm}
	\caption{(Color online) Charge gap $\Delta_c$ as a function of $\Delta$. We calculated the dynamical charge
	structure factor from the charge-charge correlation function $C_c(\tau)$ using 
	stochastic analytic continuation and extracted the charge gap using two different methods.
	First, we read off the charge gap directly from the stochastic analytic continuation data, secondly, we calculated the
	charge gap from the charge-charge correlation function. The straight line is a linear fit
	through the numerical data.}
	
	\label{fig:chargegap}
\end{figure}

From the dynamical charge structure factor, we can determine the  gap $\Delta_c$ to local charge fluctuations on 
the dot  with two different methods
 \footnote{The dynamical charge structure factor itself can in principle be calculated from the
CTQMC result for the charge correlation function $C_c(\tau)=\thavg{\tilde{n}(\tau)\tilde{n}} -
\thavg{\tilde{n}} \thavg{\tilde{n}}$ using 
stochastic analytic continuation. This, however is numerically demanding and requires a very high
quality of data. In the present case, we were unable to extract more than the lowest lying
excitation of the dynamical charge structure factor, which is directly connected to the charge gap.
The higher energy spectrum showed an extremely complex structure which is difficult to capture with
stochastic analytic continuation.}.
 One way to extract the charge gap is to read off the peak position of
 the lowest lying excitation in the dynamical charge structure factor obtained from the charge
 correlation function $C_c(\tau) = \thavg{\tilde{n}(\tau)\tilde{n}} -\thavg{\tilde{n}} \thavg{\tilde{n}}$
 via stochastic analytic continuation.
 The other way of extracting the charge gap from $C_c(\tau)$ is based on the fact, that the
 charge structure factor $N(\omega)$ is linked to $C_c(\tau)$ via
 \begin{equation}
	 C_c(\tau) \propto \int \limits_{-\infty}^{\infty} \mathrm{d} \omega \, \E^{-\tau \omega} N(\omega).
	 \label{eq:link_cc_Nom}
 \end{equation}
If $N(\omega)$ is sharply peaked around a certain value $\omega_p$, we can approximate $N(\omega)$
by $N(\omega) \approx \delta(\omega - \omega_p)$. This corresponds to $C_c(\tau) \approx \E^{-\tau
\omega_p}$. Therefore, a least squares fit of an exponential function $\E^{-\tau \omega_p}$ to
$C_c(\tau)$ in a region where only one single mode dominates, can reveal the frequency $\omega_p$ at which
$N(\omega)$ is peaked. The applicability of the method can be seen in the half logarithmic plot of
$C_c(\tau)$, where a sharply peaked charge structure factor $N(\omega)$ is reflected by a region, in
which $C_c(\tau)$ can be well approximated by a straight line.

The data obtained using these methods is shown in Fig. \ref{fig:chargegap}. In the context of the
effective model discussed in Sec. \ref{subsec:dynamical_charge_structure_eff}, we observe, that
the behavior of the charge gap of the full model clearly differs from that of the effective model.
Especially, we do not see any signature of the phase transition in the behavior of the charge gap.

The charge gap opens approximately linearly with $\Delta$. It is very
hard to extract the charge gap from the numerical data at small $\Delta$, therefore we can only
extrapolate to $\Delta=0$. Here, it appears, that we have a finite charge gap even in the absence of
superconductivity.

The fact that the local charge fluctuations remain gaped  confirms the   picture that  the  0 to $\pi$ transition
occurs  only in the spin sector.

\section{DMFT}
\input{dmft-results}

\section{Conclusion}

We have shown that the weak-coupling CTQMC algorithm is an extremely powerful unbiased  tool to compute 
thermodynamic as well as dynamical quantities of impurity models in superconducting environments. The method 
can cope very well with a complex phase of the superconducting order parameter thereby allowing for the 
calculation of the Josephson current.  
Our detailed  results for the impurity problem confirm the  picture  of a first order  phase
transition  between  a single and doublet state. It is accompanied by a $\pi$ 
phase shift in the  Josephson current.  Being completely unbiased, our approach provides the first
numerically exact 
results for this model Hamiltonian. 

Within DMFT, the physics of the BCS-PAM is mapped onto the  single impurity Anderson  model supplemented by a
self-consistency loop.  We have shown that within this approximation, the physics of the impurity model 
can be carried over to the lattice. In particular at fixed  superconducting order parameter $\Delta$  the first order 
transition between a singlet and local moment state as a function of growing values of $U$  shows up  in a hysteresis behavior of
the double occupancy.   Furthermore,  the low energy features of the local $f$-spectral function
are reminiscent  of the 
Andreev bound states with vanishing excitation energy (i.e. a crossing point) at the critical coupling.    Within the DMFT 
approximation, we can look into the single particle spectral function.  In the  singlet phase, the low energy features  can 
be interpreted in terms of a dispersion relation of Andreev bound states.  This state is continuously  linked to the $U = 0$ limit. 
In the doublet state or local moment regime, the low energy features of the spectral functions are incoherent.  We propose to 
understand  this in terms of models of disorder.   In particular  in this state, the spin dynamics
of the $f$-electron is frozen  and 
since we are considering paramagnetic states   it points in a random  different direction  in each unit cell. A  simple model of 
disorder following this picture accounts very well for the observed incoherent spectral function.

\section{Acknowledgments}
We thank Julia Wernsdorfer for interesting discussions and for bringing up the subject in her
diploma thesis. We also wish to thank Volker Meden for fruitful discussion and advice. Part of the
calculations were carried out at the Leibniz Rechenzentrum in Munich on HLRB2. We thank this
institution for allocation of CPU time.

DJL also thanks Jutta Ortloff and Manuel Schmidt for many valuable
discussions as well as Burkhard Ritter for critical reading of the manuscript. 
FFA would like to thank the KITP, where part of this work 
was carried out, for hospitality (NFS Grant  PHY05-51164).
We thank the DFG for financial support. 

\appendix

\section{Proof of the determinant identity}
\label{sec:proof_of_det_identity}

In this section a general determinant identity is proven, which can be used to derive Wick's theorem for contributions of a configuration $C_n$ to physical observables. Let us define the vectors $\vec{u_i},\, \vec{v_i} \in \mathbb{C}^m$ and the numbers $\alpha_{ij} \in \mathbb{C}$. Further, let $A\in \mathbb{C}^{m\times m}$ be a matrix of rank $m$. We define the non-singular matrices $\mat {M_n} \in \mathbb{C}^{(m+n) \times (m+n)}$ and $\mat {A_{ij}} \in \mathbb{C}^{(m+1)\times (m+1)}$ by:
\begin{equation}
\label{eq:m_aij_def}
 \mat{M_n}=\begin{pmatrix}
          \mat{A} & \vec{u_1} &  \dots & \vec{u_n} \\
	  \vec{v_1}^T & \alpha_{11}  & \dots & \alpha_{1n}\\
	  \vdots & \vdots &  \ddots & \vdots \\
	  \vec{v_n}^T & \alpha_{n1}  & \dots & \alpha_{nn}\\
         \end{pmatrix}, \quad
 \mat{A_{ij}}=\begin{pmatrix}
               \mat{A} & \vec{u_j}\\
		\vec{v_i}^T & \alpha_{ij}
              \end{pmatrix}.
\end{equation}

With these definitions, the following determinant identity holds:

\begin{equation}
\label{eq:determinant_identity}
\det{\mat{M_n}} (\det{\mat{A}})^{n-1} = \det{\begin{pmatrix}
                                      \det \mat{A_{11}} &  \dots & \det \mat{A_{1n}}\\
					\vdots &  \ddots & \vdots \\
				      \det \mat{A_{n1}}  & \dots & \det \mat{A_{nn}} \\
                                     \end{pmatrix}}.
\end{equation}

The identity can be proven by induction in $n$. It is trivial for $n=1$, so we have to start with $n=2$, where we have to show
\begin{equation}
\label{eq:matrix_identity_n2}
  \frac{\det{\mat{M_2}}}{\det{\mat{A}}} = \frac{\det{\mat{A_{11}}}}{\det{\mat{A}}} \frac{\det{\mat{A_{22}}}}{\det{\mat{A}}} - \frac{\det{\mat{A_{12}}}}{\det{\mat{A}}} \frac{\det{\mat{A_{21}}}}{\det{\mat{A}}}.
\end{equation}
For the following calculations, we introduce several vectors:
\begin{equation}
 \vec{u_{ij}^1}=\begin{pmatrix}
                 \vec{u_j}\\
		 \alpha_{ij}-1
                \end{pmatrix} \!,\,
\vec{v_{ij}^2}=\begin{pmatrix}
                 \vec{v_i}\\
		 0
                \end{pmatrix} \!,\,
\vec{u^2}=\vec{v^1}=\begin{pmatrix}
			\vec 0 \\   1
                    \end{pmatrix} \in \mathbb{C}^{m+1 }.
\end{equation}

\begin{equation}
 \vec{u_M^1}= \!\begin{pmatrix}
           \vec u_2\\
	   \alpha_{12}\\
		\alpha_{22} - \! 1
             \end{pmatrix} \! \!,
\vec{v_M^2}=\begin{pmatrix}
             \vec{v}_2\\
	     \alpha_{21}\\
		0
            \end{pmatrix} \! \!,
\vec{u_M^2}=\vec{v_M^1}=\begin{pmatrix}
			 \vec 0\\
			 1
                        \end{pmatrix} \! \!  \in \! \mathbb{C}^{m+2} \!.
\end{equation}
Let us define the \emph{expanded} matrix $\mat C_\text{ex}$ of a square matrix $\mat C$ as the matrix $C$ expanded by one row and one column containing a unit vector:
\begin{equation}
 \mat C_\text{ex} = 
\begin{pmatrix}
 \mat C & \vec 0\\
\vec 0^T & 1
\end{pmatrix}.
\end{equation}
As a last definition, we introduce the abbreviation $b_{ij}=\vec {v_i}^T \mat A^{-1} \vec {u_j}$.
Using these notations, we can write the matrices $\mat {A_{ij}}$ as 
\begin{equation}
\label{eq:aij_expansion}
 \mat{A_{ij}} = \mat{A}_{\rm ex} + \vec{u_{ij}^1} \vec{v^1}^T + \vec{u^2} \vec{v_{ij}^2}^T.
\end{equation}
To calculate the determinant $\det \mat{A_{ij}}$, we use the matrix determinant lemma $\det(\mat A + \vec u \vec v^T) = (1+\vec v^T \mat A^{-1} \vec u) \det \mat A$, yielding
\begin{equation}
	\frac{\det{\mat{A_{ij}}} }{\det \mat{A}_\text{ex}} \! = \! \left[1 \! + \vec{v_{ij}^2}^T \! (\mat{A}_{\rm ex} + \vec{u_{ij}^1} \vec{v^1}^T )^{-1} \!  \vec{u^2} \right] ( 1+\vec{v^1}^T \mat{A}_{\rm ex}^{-1} \vec{u_{ij}^1} ).
\end{equation}
The inverse matrix of $(\mat{A}_{\rm ex} + \vec{u_{ij}^1} \vec{v^1}^T )$ can be obtained from the Sherman-Morrison formula and a tedious calculation making use of the special form of the vectors and matrices gives the result
\begin{equation}
\label{eq:det_aij_over_det_a}
 \frac{\det{\mat{A_{ij}}} }{\det \mat{A}} = \alpha_{ij} -  b_{ij}.
\end{equation}
From this, the right hand side of Eq. (\ref{eq:matrix_identity_n2}) can be easily obtained. For
the left hand side, we have to perform an analogous calculation using the decomposition of the matrix $\mat {M_2}$:
\begin{equation}
 \mat {M_2} = \mat{A_{11}}_{\rm ex} + \vec{u_M^1} \vec{v_M^1}^T + \vec{u_M^2} \vec{v_M^2}^T.
\end{equation}
Again, we apply the matrix determinant lemma two times and insert the Sherman-Morrison formula to calculate the inverse matrix of $ (\mat{A_{11}}_{\rm ex} + \vec{u_M^1} \vec{v_M^1}^T ) $. Simplifying the result as far as possible, we finally arrive at
\begin{equation}
\label{eq:det_m_over_det_a}
  \frac{\det \mat{M_2}}{\det \mat{A}} =  \left( \alpha_{11} - b_{11} \right)\left( \alpha_{22} - b_{22} \right) - \left( \alpha_{12} - b_{12} \right)\left( \alpha_{21} - b_{21} \right).
\end{equation}
If we compare (\ref{eq:det_m_over_det_a}) with (\ref{eq:det_aij_over_det_a}), it is clear, that Eq. (\ref{eq:matrix_identity_n2}) holds.

We now assume that for a certain value $n\in
\mathbb{N}$ Eq. (\ref{eq:determinant_identity}) holds. For $n+1$, we can cast the matrix $\mat
{M_{n+1}}$ in a form, where we can make use of Eq. (\ref{eq:determinant_identity}) holding for $n$:
\begin{equation}
 \mat{M_{n+1}} =\begin{pmatrix}
                  \mat{\tilde{A}} & \vec{\tilde{u}_2} & \dots & \vec{\tilde{u}_{n+1}} \\
		  \vec{\tilde{v}_2}^T & \alpha_{2,2} & \dots & \alpha_{2, n+1}\\
		  \vdots & \vdots & \ddots & \vdots \\
		  \vec{\tilde{v}_n}^T & \alpha_{n,2} & \dots &  \alpha_{n,n+1}\\
		  \vec{\tilde{v}_{n+1}}^T & \alpha_{n+1 ,2} & \dots & \alpha_{n+1, n+1}
                 \end{pmatrix},
\end{equation}
where we have introduced the new matrix $\mat{\tilde A}$ and the vectors $\vec{\tilde u_{i}}$ and $\vec{\tilde u_{j}}$ with:
\begin{equation}
 \mat{\tilde A} = \begin{pmatrix}
                   \mat A & \vec {u_1} \\
		   \vec {v_1}^T & \alpha_{11}
                  \end{pmatrix}\!, \, \, \,
\vec{\tilde u_i} = \begin{pmatrix}
                    \vec {u_i}\\
			\alpha_{1i}
                   \end{pmatrix}\!, \, \, \,
\vec{\tilde v_i} = \begin{pmatrix}
                    \vec {v_i}\\
			\alpha_{i1}
                   \end{pmatrix} \! .
\end{equation}
Further, we need the matrices $\mat{\tilde A_{ij}}$ defined analogously to (\ref{eq:m_aij_def}):
\begin{equation}
\label{eq:aijtildedef}
 \mat{\tilde{A}_{ij}} = \begin{pmatrix}
                         \mat{\tilde{A}}& \vec{\tilde{u}_j} \\
			 \vec{\tilde{v}_i}^T & \alpha_{ij}
                        \end{pmatrix}=
\begin{pmatrix}
 \mat{A} & \vec{u_1} & \vec{u_j}\\
\vec{v_1}^T & \alpha_{11} & \alpha_{1j}\\
\vec{v_i}^T & \alpha_{i1} & \alpha_{ij}\\
\end{pmatrix}
\end{equation}
With these definitions, and with the abbreviations $a_{ij} = \det \mat{A_{ij}}$ and $\tilde a_{ij} =
\det \mat{\tilde A_{ij}}$, we are now able to apply Eq. (\ref{eq:determinant_identity}) holding for $n$:
\begin{equation}
\label{eq:Mnp1_eq1}
 \det \mat{M_{n+1}} (\det \mat{\tilde{A}})^{(n-1)} = \det \begin{pmatrix}
                                                     \tilde{a}_{2,2} &  \dots & \tilde{a}_{2, n+1}\\
						     \vdots &  \ddots & \vdots \\
						     \tilde{a}_{n+1, 2} &  \dots & \tilde{a}_{n+1, n+1}\\
                                                    \end{pmatrix}.
\end{equation}
For $\tilde a_{ij} $, we make use of Eq. (\ref{eq:determinant_identity}) with $n=2$, which we have proved above:
\begin{equation}
 \tilde{a}_{ij} = \frac{1}{\det \mat{A}} \left( a_{11}a_{ij} -  a_{i1}a_{1j} \right).
\end{equation}
Inserting this result in (\ref{eq:Mnp1_eq1}) yields a determinant with entries of the form $
a_{11}a_{ij} -  a_{i1}a_{1j}$. We make use of the multi linearity of the determinant to decompose
this expression and we obtain a sum of determinants with prefactors of the form $a_{ij}$.
Eliminating zero contributions, the resulting expression corresponds precisely to the
Laplace-expansion of a larger determinant, and we finally obtain
\begin{equation}
 \det \mat{M_{n+1}} \det \mat{A}^n = \det
\begin{pmatrix}
 a_{1,1} & a_{1,2} & \dots & a_{1, n+1}\\
 a_{2,1} & a_{2,2} & \dots & a_{2, n+1}\\
 \vdots & \vdots & \ddots & \vdots \\
 a_{n+1, 1} & a_{n+1, 2} & \dots & a_{n+1, n+1}
\end{pmatrix}.
\end{equation}
This is the identity (\ref{eq:determinant_identity}) for $n+1$. Hence we have derived the determinant identity for $n+1$ using only the identity for $n$ and $n=2$. By induction, the identity (\ref{eq:determinant_identity}) therefore holds for every $n\in \mathbb{N}$, as it is trivial for $n=1$.

\bibliography{main}

\end{document}

%% file: tex-Heff-eigenvalues-U.tex
\begingroup
  \makeatletter
  \providecommand\color[2][]{%
    \GenericError{(gnuplot) \space\space\space\@spaces}{%
      Package color not loaded in conjunction with
      terminal option `colourtext'%
    }{See the gnuplot documentation for explanation.%
    }{Either use 'blacktext' in gnuplot or load the package
      color.sty in LaTeX.}%
    \renewcommand\color[2][]{}%
  }%
  \providecommand\includegraphics[2][]{%
    \GenericError{(gnuplot) \space\space\space\@spaces}{%
      Package graphicx or graphics not loaded%
    }{See the gnuplot documentation for explanation.%
    }{The gnuplot epslatex terminal needs graphicx.sty or graphics.sty.}%
    \renewcommand\includegraphics[2][]{}%
  }%
  \providecommand\rotatebox[2]{#2}%
  \@ifundefined{ifGPcolor}{%
    \newif\ifGPcolor
    \GPcolortrue
  }{}%
  \@ifundefined{ifGPblacktext}{%
    \newif\ifGPblacktext
    \GPblacktexttrue
  }{}%
  \let\gplgaddtomacro\g@addto@macro
  \gdef\gplbacktext{}%
  \gdef\gplfronttext{}%
  \makeatother
  \ifGPblacktext
    \def\colorrgb#1{}%
    \def\colorgray#1{}%
  \else
    \ifGPcolor
      \def\colorrgb#1{\color[rgb]{#1}}%
      \def\colorgray#1{\color[gray]{#1}}%
      \expandafter\def\csname LTw\endcsname{\color{white}}%
      \expandafter\def\csname LTb\endcsname{\color{black}}%
      \expandafter\def\csname LTa\endcsname{\color{black}}%
      \expandafter\def\csname LT0\endcsname{\color[rgb]{1,0,0}}%
      \expandafter\def\csname LT1\endcsname{\color[rgb]{0,1,0}}%
      \expandafter\def\csname LT2\endcsname{\color[rgb]{0,0,1}}%
      \expandafter\def\csname LT3\endcsname{\color[rgb]{1,0,1}}%
      \expandafter\def\csname LT4\endcsname{\color[rgb]{0,1,1}}%
      \expandafter\def\csname LT5\endcsname{\color[rgb]{1,1,0}}%
      \expandafter\def\csname LT6\endcsname{\color[rgb]{0,0,0}}%
      \expandafter\def\csname LT7\endcsname{\color[rgb]{1,0.3,0}}%
      \expandafter\def\csname LT8\endcsname{\color[rgb]{0.5,0.5,0.5}}%
    \else
      \def\colorrgb#1{\color{black}}%
      \def\colorgray#1{\color[gray]{#1}}%
      \expandafter\def\csname LTw\endcsname{\color{white}}%
      \expandafter\def\csname LTb\endcsname{\color{black}}%
      \expandafter\def\csname LTa\endcsname{\color{black}}%
      \expandafter\def\csname LT0\endcsname{\color{black}}%
      \expandafter\def\csname LT1\endcsname{\color{black}}%
      \expandafter\def\csname LT2\endcsname{\color{black}}%
      \expandafter\def\csname LT3\endcsname{\color{black}}%
      \expandafter\def\csname LT4\endcsname{\color{black}}%
      \expandafter\def\csname LT5\endcsname{\color{black}}%
      \expandafter\def\csname LT6\endcsname{\color{black}}%
      \expandafter\def\csname LT7\endcsname{\color{black}}%
      \expandafter\def\csname LT8\endcsname{\color{black}}%
    \fi
  \fi
  \setlength{\unitlength}{0.0500bp}%
  \begin{picture}(7200.00,5040.00)%
    \gplgaddtomacro\gplbacktext{%
      \csname LTb\endcsname%
      \put(990,660){\makebox(0,0)[r]{\strut{}-2}}%
      \put(990,1125){\makebox(0,0)[r]{\strut{}-1.5}}%
      \put(990,1590){\makebox(0,0)[r]{\strut{}-1}}%
      \put(990,2055){\makebox(0,0)[r]{\strut{}-0.5}}%
      \put(990,2520){\makebox(0,0)[r]{\strut{} 0}}%
      \put(990,2985){\makebox(0,0)[r]{\strut{} 0.5}}%
      \put(990,3450){\makebox(0,0)[r]{\strut{} 1}}%
      \put(990,3915){\makebox(0,0)[r]{\strut{} 1.5}}%
      \put(990,4380){\makebox(0,0)[r]{\strut{} 2}}%
      \put(1122,440){\makebox(0,0){\strut{} 0}}%
      \put(2073,440){\makebox(0,0){\strut{} 0.5}}%
      \put(3023,440){\makebox(0,0){\strut{} 1}}%
      \put(3974,440){\makebox(0,0){\strut{} 1.5}}%
      \put(4925,440){\makebox(0,0){\strut{} 2}}%
      \put(5875,440){\makebox(0,0){\strut{} 2.5}}%
      \put(6826,440){\makebox(0,0){\strut{} 3}}%
      \put(220,2520){\rotatebox{90}{\makebox(0,0){\strut{}$E$}}}%
      \put(3974,110){\makebox(0,0){\strut{}$U$}}%
      \put(3974,4710){\makebox(0,0){\strut{}Running of eigenenergies}}%
    }%
    \gplgaddtomacro\gplfronttext{%
      \csname LTb\endcsname%
      \put(5839,4207){\makebox(0,0)[r]{\strut{}$\psi_{d}$}}%
      \csname LTb\endcsname%
      \put(5839,3987){\makebox(0,0)[r]{\strut{}$\psi_{d}$}}%
      \csname LTb\endcsname%
      \put(5839,3767){\makebox(0,0)[r]{\strut{}$\psi_s$}}%
    }%
    \gplbacktext
    \put(0,0){\includegraphics{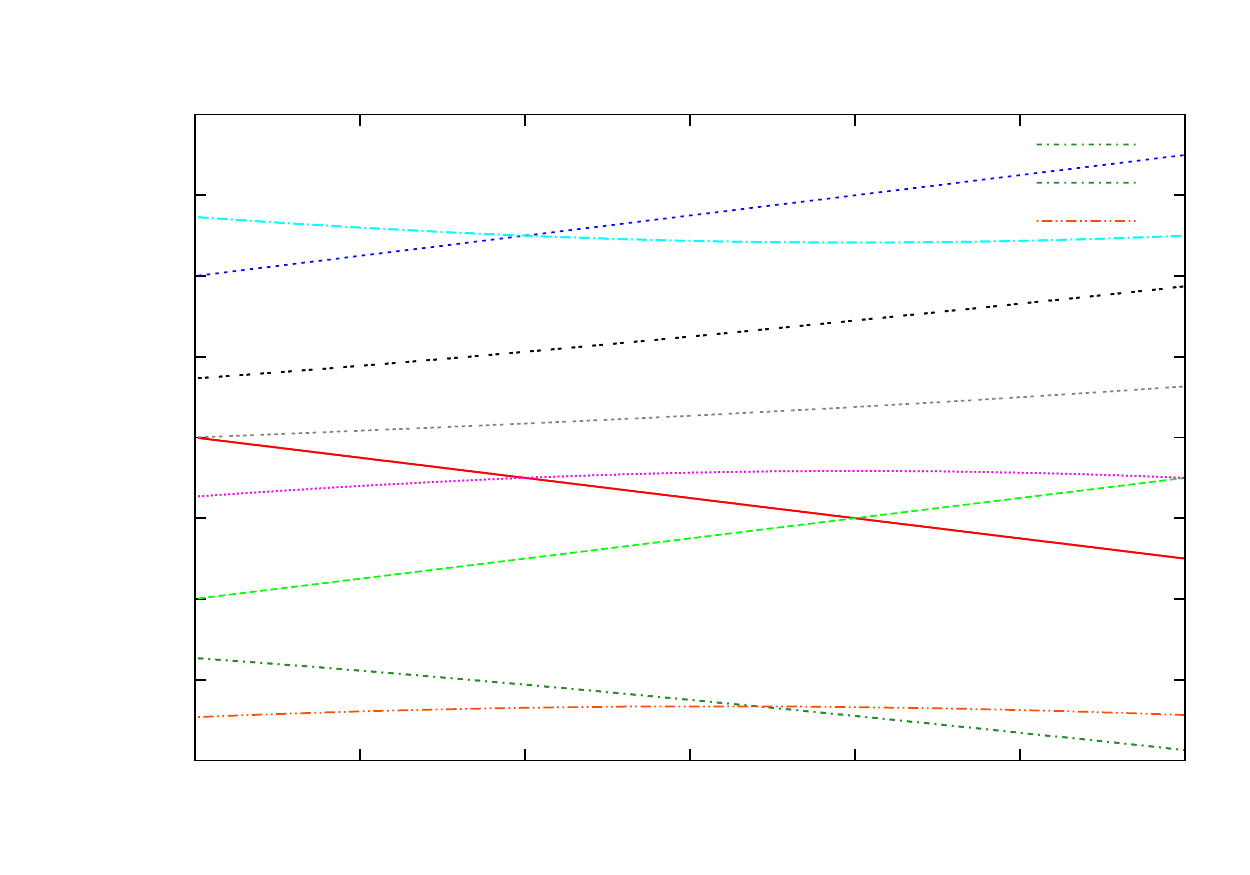}}%
    \gplfronttext
  \end{picture}%
\endgroup

%% file: tex-weightrunning.tex
\begingroup
  \makeatletter
  \providecommand\color[2][]{%
    \GenericError{(gnuplot) \space\space\space\@spaces}{%
      Package color not loaded in conjunction with
      terminal option `colourtext'%
    }{See the gnuplot documentation for explanation.%
    }{Either use 'blacktext' in gnuplot or load the package
      color.sty in LaTeX.}%
    \renewcommand\color[2][]{}%
  }%
  \providecommand\includegraphics[2][]{%
    \GenericError{(gnuplot) \space\space\space\@spaces}{%
      Package graphicx or graphics not loaded%
    }{See the gnuplot documentation for explanation.%
    }{The gnuplot epslatex terminal needs graphicx.sty or graphics.sty.}%
    \renewcommand\includegraphics[2][]{}%
  }%
  \providecommand\rotatebox[2]{#2}%
  \@ifundefined{ifGPcolor}{%
    \newif\ifGPcolor
    \GPcolortrue
  }{}%
  \@ifundefined{ifGPblacktext}{%
    \newif\ifGPblacktext
    \GPblacktexttrue
  }{}%
  \let\gplgaddtomacro\g@addto@macro
  \gdef\gplbacktext{}%
  \gdef\gplfronttext{}%
  \makeatother
  \ifGPblacktext
    \def\colorrgb#1{}%
    \def\colorgray#1{}%
  \else
    \ifGPcolor
      \def\colorrgb#1{\color[rgb]{#1}}%
      \def\colorgray#1{\color[gray]{#1}}%
      \expandafter\def\csname LTw\endcsname{\color{white}}%
      \expandafter\def\csname LTb\endcsname{\color{black}}%
      \expandafter\def\csname LTa\endcsname{\color{black}}%
      \expandafter\def\csname LT0\endcsname{\color[rgb]{1,0,0}}%
      \expandafter\def\csname LT1\endcsname{\color[rgb]{0,1,0}}%
      \expandafter\def\csname LT2\endcsname{\color[rgb]{0,0,1}}%
      \expandafter\def\csname LT3\endcsname{\color[rgb]{1,0,1}}%
      \expandafter\def\csname LT4\endcsname{\color[rgb]{0,1,1}}%
      \expandafter\def\csname LT5\endcsname{\color[rgb]{1,1,0}}%
      \expandafter\def\csname LT6\endcsname{\color[rgb]{0,0,0}}%
      \expandafter\def\csname LT7\endcsname{\color[rgb]{1,0.3,0}}%
      \expandafter\def\csname LT8\endcsname{\color[rgb]{0.5,0.5,0.5}}%
    \else
      \def\colorrgb#1{\color{black}}%
      \def\colorgray#1{\color[gray]{#1}}%
      \expandafter\def\csname LTw\endcsname{\color{white}}%
      \expandafter\def\csname LTb\endcsname{\color{black}}%
      \expandafter\def\csname LTa\endcsname{\color{black}}%
      \expandafter\def\csname LT0\endcsname{\color{black}}%
      \expandafter\def\csname LT1\endcsname{\color{black}}%
      \expandafter\def\csname LT2\endcsname{\color{black}}%
      \expandafter\def\csname LT3\endcsname{\color{black}}%
      \expandafter\def\csname LT4\endcsname{\color{black}}%
      \expandafter\def\csname LT5\endcsname{\color{black}}%
      \expandafter\def\csname LT6\endcsname{\color{black}}%
      \expandafter\def\csname LT7\endcsname{\color{black}}%
      \expandafter\def\csname LT8\endcsname{\color{black}}%
    \fi
  \fi
  \setlength{\unitlength}{0.0500bp}%
  \begin{picture}(7200.00,5040.00)%
    \gplgaddtomacro\gplbacktext{%
      \csname LTb\endcsname%
      \put(770,660){\makebox(0,0)[r]{\strut{} 0}}%
      \put(770,1125){\makebox(0,0)[r]{\strut{} 0.1}}%
      \put(770,1590){\makebox(0,0)[r]{\strut{} 0.2}}%
      \put(770,2055){\makebox(0,0)[r]{\strut{} 0.3}}%
      \put(770,2520){\makebox(0,0)[r]{\strut{} 0.4}}%
      \put(770,2985){\makebox(0,0)[r]{\strut{} 0.5}}%
      \put(770,3450){\makebox(0,0)[r]{\strut{} 0.6}}%
      \put(770,3915){\makebox(0,0)[r]{\strut{} 0.7}}%
      \put(770,4380){\makebox(0,0)[r]{\strut{} 0.8}}%
      \put(902,440){\makebox(0,0){\strut{} 0}}%
      \put(1643,440){\makebox(0,0){\strut{} 0.5}}%
      \put(2383,440){\makebox(0,0){\strut{} 1}}%
      \put(3124,440){\makebox(0,0){\strut{} 1.5}}%
      \put(3864,440){\makebox(0,0){\strut{} 2}}%
      \put(4605,440){\makebox(0,0){\strut{} 2.5}}%
      \put(5345,440){\makebox(0,0){\strut{} 3}}%
      \put(6086,440){\makebox(0,0){\strut{} 3.5}}%
      \put(6826,440){\makebox(0,0){\strut{} 4}}%
      \put(3864,110){\makebox(0,0){\strut{}$\Delta$}}%
      \put(3864,4710){\makebox(0,0){\strut{}Running of ground state weights}}%
    }%
    \gplgaddtomacro\gplfronttext{%
      \csname LTb\endcsname%
      \put(5839,4207){\makebox(0,0)[r]{\strut{}$b$}}%
      \csname LTb\endcsname%
      \put(5839,3987){\makebox(0,0)[r]{\strut{}$\beta$}}%
      \csname LTb\endcsname%
      \put(5839,3767){\makebox(0,0)[r]{\strut{}$\gamma$}}%
      \csname LTb\endcsname%
      \put(5839,3547){\makebox(0,0)[r]{\strut{}$\alpha$}}%
      \csname LTb\endcsname%
      \put(5839,3327){\makebox(0,0)[r]{\strut{}$a$}}%
    }%
    \gplbacktext
    \put(0,0){\includegraphics{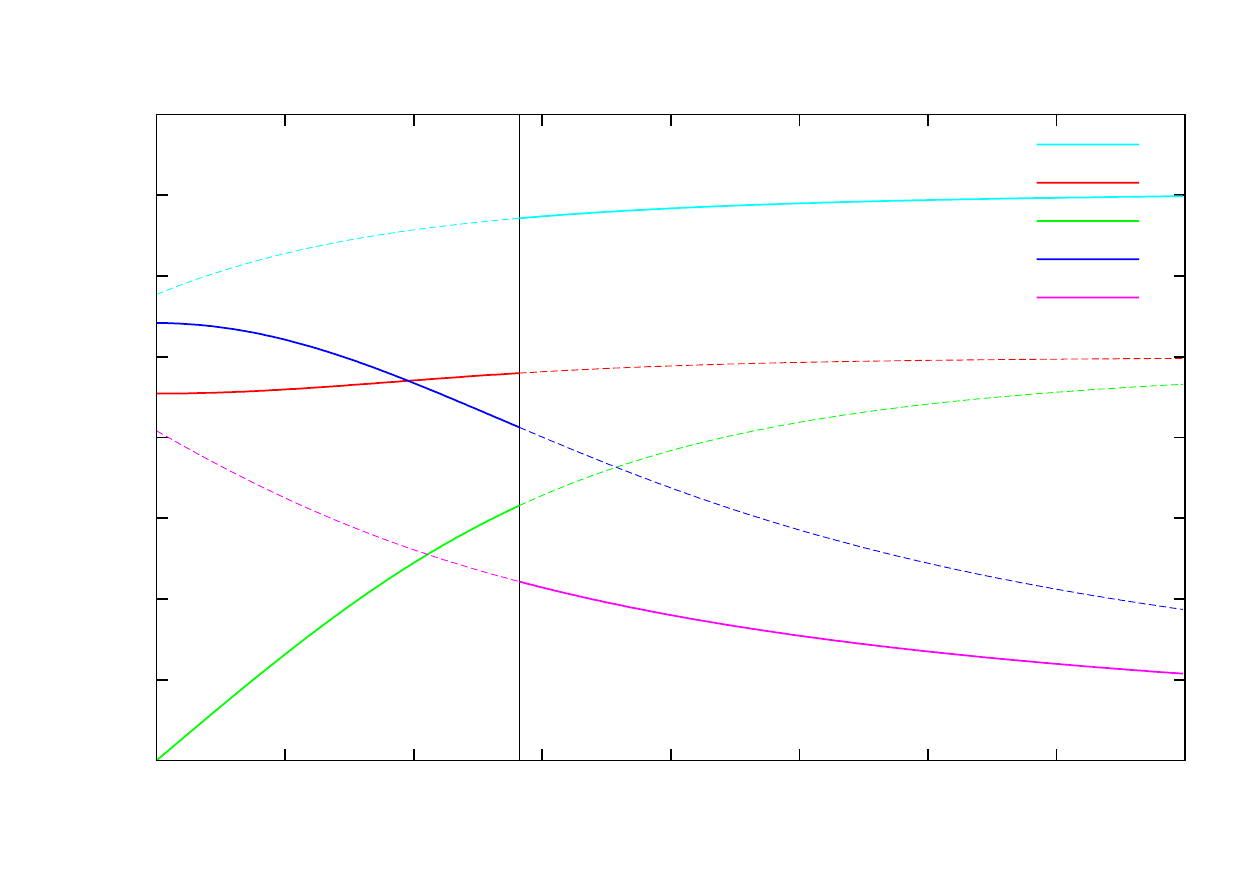}}%
    \gplfronttext
  \end{picture}%
\endgroup

%% file: tex-double-occ-delta-U-B200-3d.tex
\begingroup
  \makeatletter
  \providecommand\color[2][]{%
    \GenericError{(gnuplot) \space\space\space\@spaces}{%
      Package color not loaded in conjunction with
      terminal option `colourtext'%
    }{See the gnuplot documentation for explanation.%
    }{Either use 'blacktext' in gnuplot or load the package
      color.sty in LaTeX.}%
    \renewcommand\color[2][]{}%
  }%
  \providecommand\includegraphics[2][]{%
    \GenericError{(gnuplot) \space\space\space\@spaces}{%
      Package graphicx or graphics not loaded%
    }{See the gnuplot documentation for explanation.%
    }{The gnuplot epslatex terminal needs graphicx.sty or graphics.sty.}%
    \renewcommand\includegraphics[2][]{}%
  }%
  \providecommand\rotatebox[2]{#2}%
  \@ifundefined{ifGPcolor}{%
    \newif\ifGPcolor
    \GPcolortrue
  }{}%
  \@ifundefined{ifGPblacktext}{%
    \newif\ifGPblacktext
    \GPblacktexttrue
  }{}%
  \let\gplgaddtomacro\g@addto@macro
  \gdef\gplbacktext{}%
  \gdef\gplfronttext{}%
  \makeatother
  \ifGPblacktext
    \def\colorrgb#1{}%
    \def\colorgray#1{}%
  \else
    \ifGPcolor
      \def\colorrgb#1{\color[rgb]{#1}}%
      \def\colorgray#1{\color[gray]{#1}}%
      \expandafter\def\csname LTw\endcsname{\color{white}}%
      \expandafter\def\csname LTb\endcsname{\color{black}}%
      \expandafter\def\csname LTa\endcsname{\color{black}}%
      \expandafter\def\csname LT0\endcsname{\color[rgb]{1,0,0}}%
      \expandafter\def\csname LT1\endcsname{\color[rgb]{0,1,0}}%
      \expandafter\def\csname LT2\endcsname{\color[rgb]{0,0,1}}%
      \expandafter\def\csname LT3\endcsname{\color[rgb]{1,0,1}}%
      \expandafter\def\csname LT4\endcsname{\color[rgb]{0,1,1}}%
      \expandafter\def\csname LT5\endcsname{\color[rgb]{1,1,0}}%
      \expandafter\def\csname LT6\endcsname{\color[rgb]{0,0,0}}%
      \expandafter\def\csname LT7\endcsname{\color[rgb]{1,0.3,0}}%
      \expandafter\def\csname LT8\endcsname{\color[rgb]{0.5,0.5,0.5}}%
    \else
      \def\colorrgb#1{\color{black}}%
      \def\colorgray#1{\color[gray]{#1}}%
      \expandafter\def\csname LTw\endcsname{\color{white}}%
      \expandafter\def\csname LTb\endcsname{\color{black}}%
      \expandafter\def\csname LTa\endcsname{\color{black}}%
      \expandafter\def\csname LT0\endcsname{\color{black}}%
      \expandafter\def\csname LT1\endcsname{\color{black}}%
      \expandafter\def\csname LT2\endcsname{\color{black}}%
      \expandafter\def\csname LT3\endcsname{\color{black}}%
      \expandafter\def\csname LT4\endcsname{\color{black}}%
      \expandafter\def\csname LT5\endcsname{\color{black}}%
      \expandafter\def\csname LT6\endcsname{\color{black}}%
      \expandafter\def\csname LT7\endcsname{\color{black}}%
      \expandafter\def\csname LT8\endcsname{\color{black}}%
    \fi
  \fi
  \setlength{\unitlength}{0.0500bp}%
  \begin{picture}(7200.00,5040.00)%
    \gplgaddtomacro\gplbacktext{%
      \csname LTb\endcsname%
      \put(3599,4312){\makebox(0,0){\strut{}Double occupancy}}%
    }%
    \gplgaddtomacro\gplfronttext{%
      \csname LTb\endcsname%
      \put(1170,772){\makebox(0,0){\strut{} 0}}%
      \put(1777,772){\makebox(0,0){\strut{} 0.5}}%
      \put(2385,772){\makebox(0,0){\strut{} 1}}%
      \put(2993,772){\makebox(0,0){\strut{} 1.5}}%
      \put(3600,772){\makebox(0,0){\strut{} 2}}%
      \put(4207,772){\makebox(0,0){\strut{} 2.5}}%
      \put(4815,772){\makebox(0,0){\strut{} 3}}%
      \put(5423,772){\makebox(0,0){\strut{} 3.5}}%
      \put(6030,772){\makebox(0,0){\strut{} 4}}%
      \put(3600,442){\makebox(0,0){\strut{}$ U $}}%
      \put(998,1058){\makebox(0,0)[r]{\strut{} 0}}%
      \put(998,1424){\makebox(0,0)[r]{\strut{} 1}}%
      \put(998,1789){\makebox(0,0)[r]{\strut{} 2}}%
      \put(998,2155){\makebox(0,0)[r]{\strut{} 3}}%
      \put(998,2520){\makebox(0,0)[r]{\strut{} 4}}%
      \put(998,2885){\makebox(0,0)[r]{\strut{} 5}}%
      \put(998,3251){\makebox(0,0)[r]{\strut{} 6}}%
      \put(998,3616){\makebox(0,0)[r]{\strut{} 7}}%
      \put(998,3982){\makebox(0,0)[r]{\strut{} 8}}%
      \put(668,2520){\rotatebox{90}{\makebox(0,0){\strut{}$ \Delta $}}}%
      \put(6527,1057){\makebox(0,0)[l]{\strut{} 0}}%
      \put(6527,1349){\makebox(0,0)[l]{\strut{} 0.05}}%
      \put(6527,1642){\makebox(0,0)[l]{\strut{} 0.1}}%
      \put(6527,1934){\makebox(0,0)[l]{\strut{} 0.15}}%
      \put(6527,2227){\makebox(0,0)[l]{\strut{} 0.2}}%
      \put(6527,2519){\makebox(0,0)[l]{\strut{} 0.25}}%
      \put(6527,2812){\makebox(0,0)[l]{\strut{} 0.3}}%
      \put(6527,3104){\makebox(0,0)[l]{\strut{} 0.35}}%
      \put(6527,3397){\makebox(0,0)[l]{\strut{} 0.4}}%
      \put(6527,3689){\makebox(0,0)[l]{\strut{} 0.45}}%
      \put(6527,3981){\makebox(0,0)[l]{\strut{} 0.5}}%
    }%
    \gplbacktext
    \put(0,0){\includegraphics{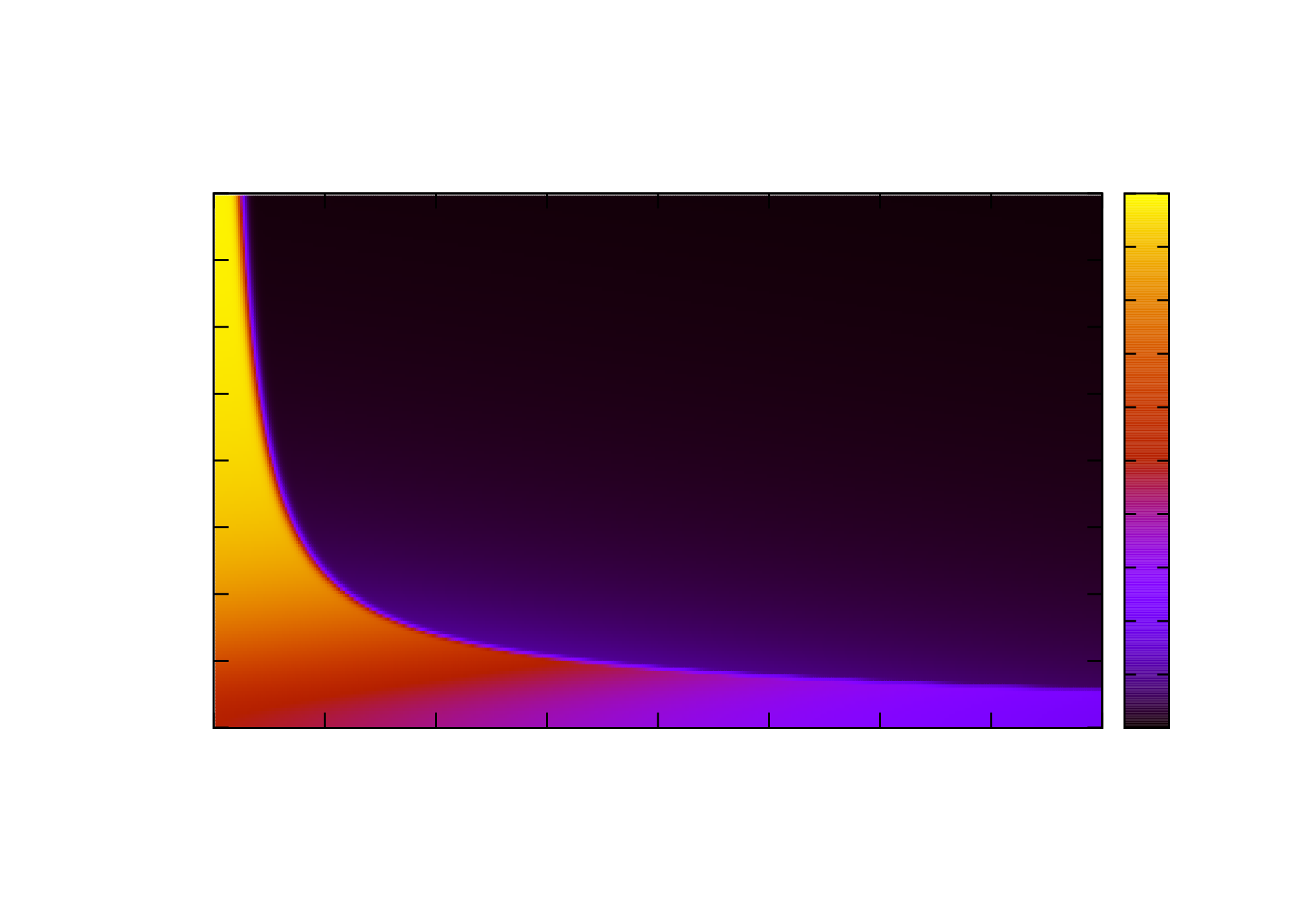}}%
    \gplfronttext
  \end{picture}%
\endgroup

%% file: tex-spectralfunction-B200-3d.tex
\begingroup
  \makeatletter
  \providecommand\color[2][]{%
    \GenericError{(gnuplot) \space\space\space\@spaces}{%
      Package color not loaded in conjunction with
      terminal option `colourtext'%
    }{See the gnuplot documentation for explanation.%
    }{Either use 'blacktext' in gnuplot or load the package
      color.sty in LaTeX.}%
    \renewcommand\color[2][]{}%
  }%
  \providecommand\includegraphics[2][]{%
    \GenericError{(gnuplot) \space\space\space\@spaces}{%
      Package graphicx or graphics not loaded%
    }{See the gnuplot documentation for explanation.%
    }{The gnuplot epslatex terminal needs graphicx.sty or graphics.sty.}%
    \renewcommand\includegraphics[2][]{}%
  }%
  \providecommand\rotatebox[2]{#2}%
  \@ifundefined{ifGPcolor}{%
    \newif\ifGPcolor
    \GPcolortrue
  }{}%
  \@ifundefined{ifGPblacktext}{%
    \newif\ifGPblacktext
    \GPblacktexttrue
  }{}%
  \let\gplgaddtomacro\g@addto@macro
  \gdef\gplbacktext{}%
  \gdef\gplfronttext{}%
  \makeatother
  \ifGPblacktext
    \def\colorrgb#1{}%
    \def\colorgray#1{}%
  \else
    \ifGPcolor
      \def\colorrgb#1{\color[rgb]{#1}}%
      \def\colorgray#1{\color[gray]{#1}}%
      \expandafter\def\csname LTw\endcsname{\color{white}}%
      \expandafter\def\csname LTb\endcsname{\color{black}}%
      \expandafter\def\csname LTa\endcsname{\color{black}}%
      \expandafter\def\csname LT0\endcsname{\color[rgb]{1,0,0}}%
      \expandafter\def\csname LT1\endcsname{\color[rgb]{0,1,0}}%
      \expandafter\def\csname LT2\endcsname{\color[rgb]{0,0,1}}%
      \expandafter\def\csname LT3\endcsname{\color[rgb]{1,0,1}}%
      \expandafter\def\csname LT4\endcsname{\color[rgb]{0,1,1}}%
      \expandafter\def\csname LT5\endcsname{\color[rgb]{1,1,0}}%
      \expandafter\def\csname LT6\endcsname{\color[rgb]{0,0,0}}%
      \expandafter\def\csname LT7\endcsname{\color[rgb]{1,0.3,0}}%
      \expandafter\def\csname LT8\endcsname{\color[rgb]{0.5,0.5,0.5}}%
    \else
      \def\colorrgb#1{\color{black}}%
      \def\colorgray#1{\color[gray]{#1}}%
      \expandafter\def\csname LTw\endcsname{\color{white}}%
      \expandafter\def\csname LTb\endcsname{\color{black}}%
      \expandafter\def\csname LTa\endcsname{\color{black}}%
      \expandafter\def\csname LT0\endcsname{\color{black}}%
      \expandafter\def\csname LT1\endcsname{\color{black}}%
      \expandafter\def\csname LT2\endcsname{\color{black}}%
      \expandafter\def\csname LT3\endcsname{\color{black}}%
      \expandafter\def\csname LT4\endcsname{\color{black}}%
      \expandafter\def\csname LT5\endcsname{\color{black}}%
      \expandafter\def\csname LT6\endcsname{\color{black}}%
      \expandafter\def\csname LT7\endcsname{\color{black}}%
      \expandafter\def\csname LT8\endcsname{\color{black}}%
    \fi
  \fi
  \setlength{\unitlength}{0.0500bp}%
  \begin{picture}(7200.00,7560.00)%
    \gplgaddtomacro\gplbacktext{%
      \csname LTb\endcsname%
      \put(3599,6508){\makebox(0,0){\strut{}Spectral function}}%
    }%
    \gplgaddtomacro\gplfronttext{%
      \csname LTb\endcsname%
      \put(6018,6616){\makebox(0,0)[r]{\strut{}$A_{\up \up}(\omega)$}}%
      \csname LTb\endcsname%
      \put(1170,1096){\makebox(0,0){\strut{} 0}}%
      \put(1980,1096){\makebox(0,0){\strut{} 0.5}}%
      \put(2790,1096){\makebox(0,0){\strut{} 1}}%
      \put(3600,1096){\makebox(0,0){\strut{} 1.5}}%
      \put(4410,1096){\makebox(0,0){\strut{} 2}}%
      \put(5220,1096){\makebox(0,0){\strut{} 2.5}}%
      \put(6030,1096){\makebox(0,0){\strut{} 3}}%
      \put(3600,766){\makebox(0,0){\strut{}$\Delta$}}%
      \put(998,1382){\makebox(0,0)[r]{\strut{}-3}}%
      \put(998,2182){\makebox(0,0)[r]{\strut{}-2}}%
      \put(998,2981){\makebox(0,0)[r]{\strut{}-1}}%
      \put(998,3780){\makebox(0,0)[r]{\strut{} 0}}%
      \put(998,4579){\makebox(0,0)[r]{\strut{} 1}}%
      \put(998,5378){\makebox(0,0)[r]{\strut{} 2}}%
      \put(998,6178){\makebox(0,0)[r]{\strut{} 3}}%
      \put(668,3780){\rotatebox{90}{\makebox(0,0){\strut{}$\omega$}}}%
      \put(6527,1381){\makebox(0,0)[l]{\strut{} 0}}%
      \put(6527,2340){\makebox(0,0)[l]{\strut{} 5}}%
      \put(6527,3299){\makebox(0,0)[l]{\strut{} 10}}%
      \put(6527,4259){\makebox(0,0)[l]{\strut{} 15}}%
      \put(6527,5218){\makebox(0,0)[l]{\strut{} 20}}%
      \put(6527,6178){\makebox(0,0)[l]{\strut{} 25}}%
    }%
    \gplbacktext
    \put(0,0){\includegraphics{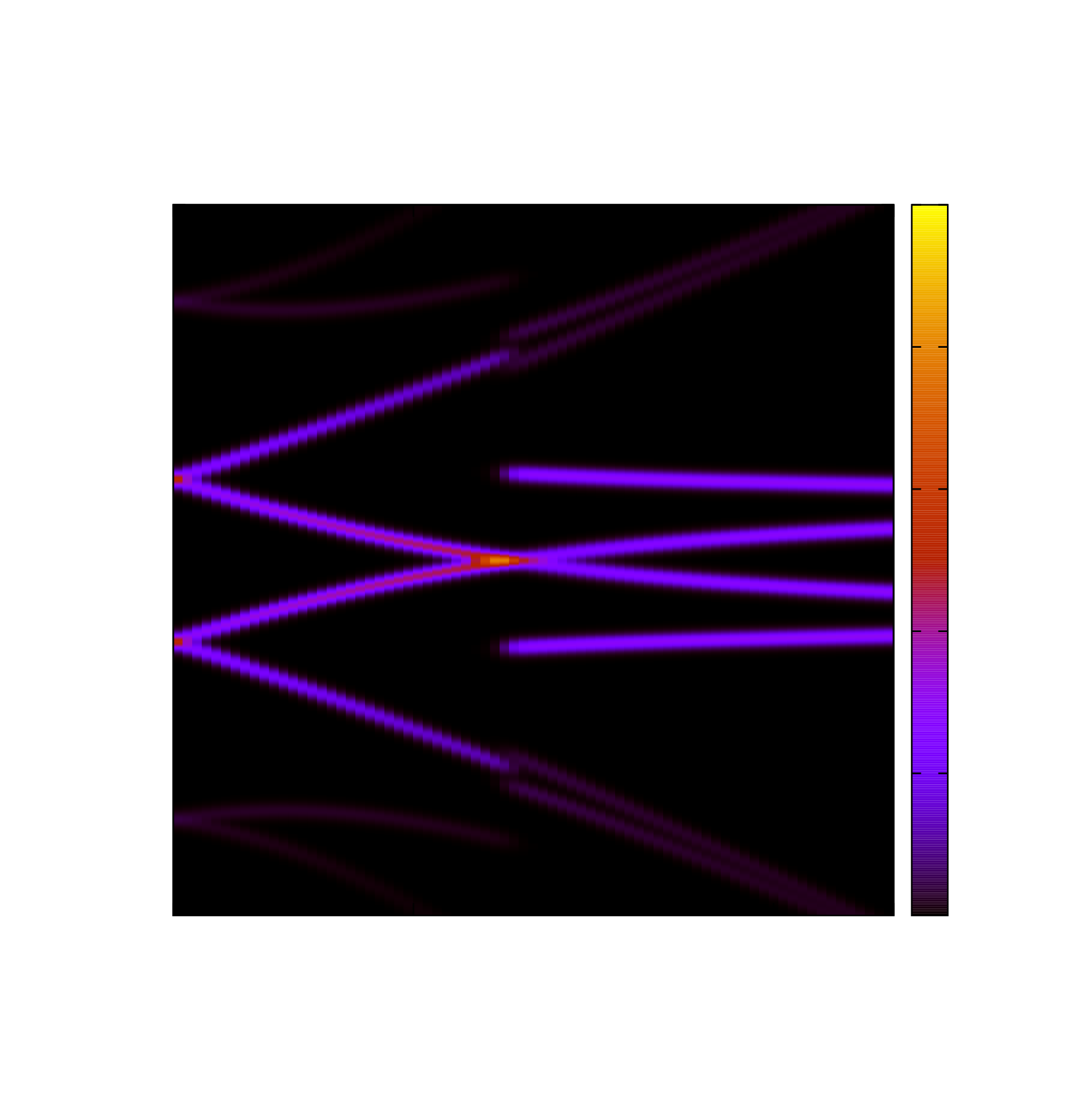}}%
    \gplfronttext
  \end{picture}%
\endgroup

%% file: tex-spinstructure-eff-3d.tex
\begingroup
  \makeatletter
  \providecommand\color[2][]{%
    \GenericError{(gnuplot) \space\space\space\@spaces}{%
      Package color not loaded in conjunction with
      terminal option `colourtext'%
    }{See the gnuplot documentation for explanation.%
    }{Either use 'blacktext' in gnuplot or load the package
      color.sty in LaTeX.}%
    \renewcommand\color[2][]{}%
  }%
  \providecommand\includegraphics[2][]{%
    \GenericError{(gnuplot) \space\space\space\@spaces}{%
      Package graphicx or graphics not loaded%
    }{See the gnuplot documentation for explanation.%
    }{The gnuplot epslatex terminal needs graphicx.sty or graphics.sty.}%
    \renewcommand\includegraphics[2][]{}%
  }%
  \providecommand\rotatebox[2]{#2}%
  \@ifundefined{ifGPcolor}{%
    \newif\ifGPcolor
    \GPcolortrue
  }{}%
  \@ifundefined{ifGPblacktext}{%
    \newif\ifGPblacktext
    \GPblacktexttrue
  }{}%
  \let\gplgaddtomacro\g@addto@macro
  \gdef\gplbacktext{}%
  \gdef\gplfronttext{}%
  \makeatother
  \ifGPblacktext
    \def\colorrgb#1{}%
    \def\colorgray#1{}%
  \else
    \ifGPcolor
      \def\colorrgb#1{\color[rgb]{#1}}%
      \def\colorgray#1{\color[gray]{#1}}%
      \expandafter\def\csname LTw\endcsname{\color{white}}%
      \expandafter\def\csname LTb\endcsname{\color{black}}%
      \expandafter\def\csname LTa\endcsname{\color{black}}%
      \expandafter\def\csname LT0\endcsname{\color[rgb]{1,0,0}}%
      \expandafter\def\csname LT1\endcsname{\color[rgb]{0,1,0}}%
      \expandafter\def\csname LT2\endcsname{\color[rgb]{0,0,1}}%
      \expandafter\def\csname LT3\endcsname{\color[rgb]{1,0,1}}%
      \expandafter\def\csname LT4\endcsname{\color[rgb]{0,1,1}}%
      \expandafter\def\csname LT5\endcsname{\color[rgb]{1,1,0}}%
      \expandafter\def\csname LT6\endcsname{\color[rgb]{0,0,0}}%
      \expandafter\def\csname LT7\endcsname{\color[rgb]{1,0.3,0}}%
      \expandafter\def\csname LT8\endcsname{\color[rgb]{0.5,0.5,0.5}}%
    \else
      \def\colorrgb#1{\color{black}}%
      \def\colorgray#1{\color[gray]{#1}}%
      \expandafter\def\csname LTw\endcsname{\color{white}}%
      \expandafter\def\csname LTb\endcsname{\color{black}}%
      \expandafter\def\csname LTa\endcsname{\color{black}}%
      \expandafter\def\csname LT0\endcsname{\color{black}}%
      \expandafter\def\csname LT1\endcsname{\color{black}}%
      \expandafter\def\csname LT2\endcsname{\color{black}}%
      \expandafter\def\csname LT3\endcsname{\color{black}}%
      \expandafter\def\csname LT4\endcsname{\color{black}}%
      \expandafter\def\csname LT5\endcsname{\color{black}}%
      \expandafter\def\csname LT6\endcsname{\color{black}}%
      \expandafter\def\csname LT7\endcsname{\color{black}}%
      \expandafter\def\csname LT8\endcsname{\color{black}}%
    \fi
  \fi
  \setlength{\unitlength}{0.0500bp}%
  \begin{picture}(7200.00,5040.00)%
    \gplgaddtomacro\gplbacktext{%
      \csname LTb\endcsname%
      \put(3599,4312){\makebox(0,0){\strut{}Dynamical spin structure}}%
    }%
    \gplgaddtomacro\gplfronttext{%
      \csname LTb\endcsname%
      \put(6018,4096){\makebox(0,0)[r]{\strut{}$S(\omega)$}}%
      \csname LTb\endcsname%
      \put(1170,772){\makebox(0,0){\strut{} 0}}%
      \put(1980,772){\makebox(0,0){\strut{} 0.5}}%
      \put(2790,772){\makebox(0,0){\strut{} 1}}%
      \put(3600,772){\makebox(0,0){\strut{} 1.5}}%
      \put(4410,772){\makebox(0,0){\strut{} 2}}%
      \put(5220,772){\makebox(0,0){\strut{} 2.5}}%
      \put(6030,772){\makebox(0,0){\strut{} 3}}%
      \put(3600,442){\makebox(0,0){\strut{}$\Delta$}}%
      \put(998,1058){\makebox(0,0)[r]{\strut{} 0}}%
      \put(998,1546){\makebox(0,0)[r]{\strut{} 0.5}}%
      \put(998,2033){\makebox(0,0)[r]{\strut{} 1}}%
      \put(998,2520){\makebox(0,0)[r]{\strut{} 1.5}}%
      \put(998,3007){\makebox(0,0)[r]{\strut{} 2}}%
      \put(998,3494){\makebox(0,0)[r]{\strut{} 2.5}}%
      \put(998,3982){\makebox(0,0)[r]{\strut{} 3}}%
      \put(404,2520){\rotatebox{90}{\makebox(0,0){\strut{}$\omega$}}}%
      \put(6527,1057){\makebox(0,0)[l]{\strut{} 0}}%
      \put(6527,1642){\makebox(0,0)[l]{\strut{} 5}}%
      \put(6527,2227){\makebox(0,0)[l]{\strut{} 10}}%
      \put(6527,2812){\makebox(0,0)[l]{\strut{} 15}}%
      \put(6527,3397){\makebox(0,0)[l]{\strut{} 20}}%
      \put(6527,3982){\makebox(0,0)[l]{\strut{} 25}}%
    }%
    \gplbacktext
    \put(0,0){\includegraphics{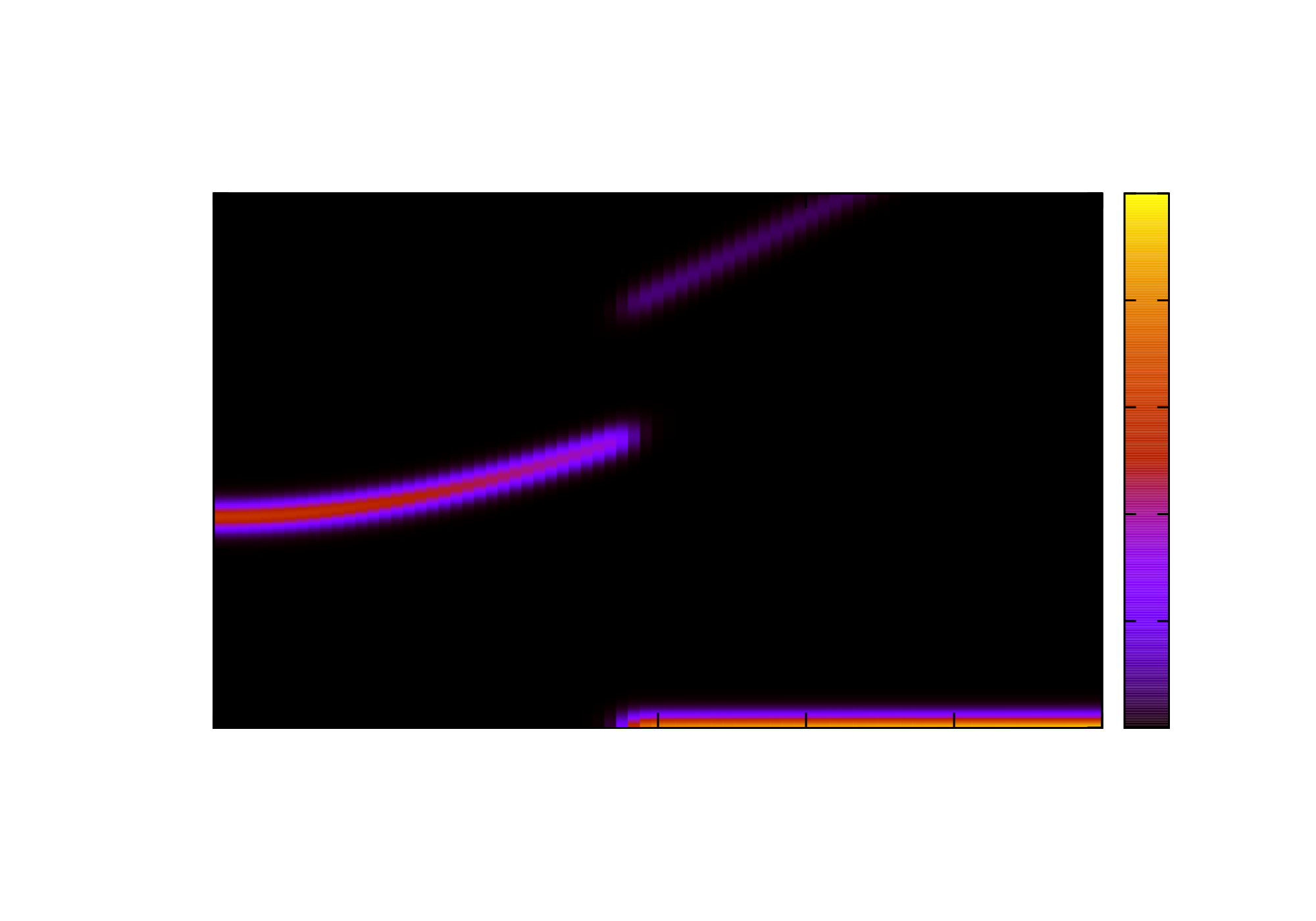}}%
    \gplfronttext
  \end{picture}%
\endgroup

%% file: tex-chargestructure-b200-eff-3d.tex
\begingroup
  \makeatletter
  \providecommand\color[2][]{%
    \GenericError{(gnuplot) \space\space\space\@spaces}{%
      Package color not loaded in conjunction with
      terminal option `colourtext'%
    }{See the gnuplot documentation for explanation.%
    }{Either use 'blacktext' in gnuplot or load the package
      color.sty in LaTeX.}%
    \renewcommand\color[2][]{}%
  }%
  \providecommand\includegraphics[2][]{%
    \GenericError{(gnuplot) \space\space\space\@spaces}{%
      Package graphicx or graphics not loaded%
    }{See the gnuplot documentation for explanation.%
    }{The gnuplot epslatex terminal needs graphicx.sty or graphics.sty.}%
    \renewcommand\includegraphics[2][]{}%
  }%
  \providecommand\rotatebox[2]{#2}%
  \@ifundefined{ifGPcolor}{%
    \newif\ifGPcolor
    \GPcolortrue
  }{}%
  \@ifundefined{ifGPblacktext}{%
    \newif\ifGPblacktext
    \GPblacktexttrue
  }{}%
  \let\gplgaddtomacro\g@addto@macro
  \gdef\gplbacktext{}%
  \gdef\gplfronttext{}%
  \makeatother
  \ifGPblacktext
    \def\colorrgb#1{}%
    \def\colorgray#1{}%
  \else
    \ifGPcolor
      \def\colorrgb#1{\color[rgb]{#1}}%
      \def\colorgray#1{\color[gray]{#1}}%
      \expandafter\def\csname LTw\endcsname{\color{white}}%
      \expandafter\def\csname LTb\endcsname{\color{black}}%
      \expandafter\def\csname LTa\endcsname{\color{black}}%
      \expandafter\def\csname LT0\endcsname{\color[rgb]{1,0,0}}%
      \expandafter\def\csname LT1\endcsname{\color[rgb]{0,1,0}}%
      \expandafter\def\csname LT2\endcsname{\color[rgb]{0,0,1}}%
      \expandafter\def\csname LT3\endcsname{\color[rgb]{1,0,1}}%
      \expandafter\def\csname LT4\endcsname{\color[rgb]{0,1,1}}%
      \expandafter\def\csname LT5\endcsname{\color[rgb]{1,1,0}}%
      \expandafter\def\csname LT6\endcsname{\color[rgb]{0,0,0}}%
      \expandafter\def\csname LT7\endcsname{\color[rgb]{1,0.3,0}}%
      \expandafter\def\csname LT8\endcsname{\color[rgb]{0.5,0.5,0.5}}%
    \else
      \def\colorrgb#1{\color{black}}%
      \def\colorgray#1{\color[gray]{#1}}%
      \expandafter\def\csname LTw\endcsname{\color{white}}%
      \expandafter\def\csname LTb\endcsname{\color{black}}%
      \expandafter\def\csname LTa\endcsname{\color{black}}%
      \expandafter\def\csname LT0\endcsname{\color{black}}%
      \expandafter\def\csname LT1\endcsname{\color{black}}%
      \expandafter\def\csname LT2\endcsname{\color{black}}%
      \expandafter\def\csname LT3\endcsname{\color{black}}%
      \expandafter\def\csname LT4\endcsname{\color{black}}%
      \expandafter\def\csname LT5\endcsname{\color{black}}%
      \expandafter\def\csname LT6\endcsname{\color{black}}%
      \expandafter\def\csname LT7\endcsname{\color{black}}%
      \expandafter\def\csname LT8\endcsname{\color{black}}%
    \fi
  \fi
  \setlength{\unitlength}{0.0500bp}%
  \begin{picture}(7200.00,5040.00)%
    \gplgaddtomacro\gplbacktext{%
      \csname LTb\endcsname%
      \put(3599,4312){\makebox(0,0){\strut{}Dynamical charge structure}}%
    }%
    \gplgaddtomacro\gplfronttext{%
      \csname LTb\endcsname%
      \put(6018,4096){\makebox(0,0)[r]{\strut{}$N(\omega)$}}%
      \csname LTb\endcsname%
      \put(1170,772){\makebox(0,0){\strut{} 0}}%
      \put(1980,772){\makebox(0,0){\strut{} 0.5}}%
      \put(2790,772){\makebox(0,0){\strut{} 1}}%
      \put(3600,772){\makebox(0,0){\strut{} 1.5}}%
      \put(4410,772){\makebox(0,0){\strut{} 2}}%
      \put(5220,772){\makebox(0,0){\strut{} 2.5}}%
      \put(6030,772){\makebox(0,0){\strut{} 3}}%
      \put(3600,442){\makebox(0,0){\strut{}$\Delta$}}%
      \put(998,1058){\makebox(0,0)[r]{\strut{} 0}}%
      \put(998,1546){\makebox(0,0)[r]{\strut{} 0.5}}%
      \put(998,2033){\makebox(0,0)[r]{\strut{} 1}}%
      \put(998,2520){\makebox(0,0)[r]{\strut{} 1.5}}%
      \put(998,3007){\makebox(0,0)[r]{\strut{} 2}}%
      \put(998,3494){\makebox(0,0)[r]{\strut{} 2.5}}%
      \put(998,3982){\makebox(0,0)[r]{\strut{} 3}}%
      \put(404,2520){\rotatebox{90}{\makebox(0,0){\strut{}$\omega$}}}%
      \put(6527,1057){\makebox(0,0)[l]{\strut{} 0}}%
      \put(6527,1349){\makebox(0,0)[l]{\strut{} 2}}%
      \put(6527,1642){\makebox(0,0)[l]{\strut{} 4}}%
      \put(6527,1934){\makebox(0,0)[l]{\strut{} 6}}%
      \put(6527,2227){\makebox(0,0)[l]{\strut{} 8}}%
      \put(6527,2519){\makebox(0,0)[l]{\strut{} 10}}%
      \put(6527,2812){\makebox(0,0)[l]{\strut{} 12}}%
      \put(6527,3104){\makebox(0,0)[l]{\strut{} 14}}%
      \put(6527,3397){\makebox(0,0)[l]{\strut{} 16}}%
      \put(6527,3689){\makebox(0,0)[l]{\strut{} 18}}%
      \put(6527,3982){\makebox(0,0)[l]{\strut{} 20}}%
    }%
    \gplbacktext
    \put(0,0){\includegraphics{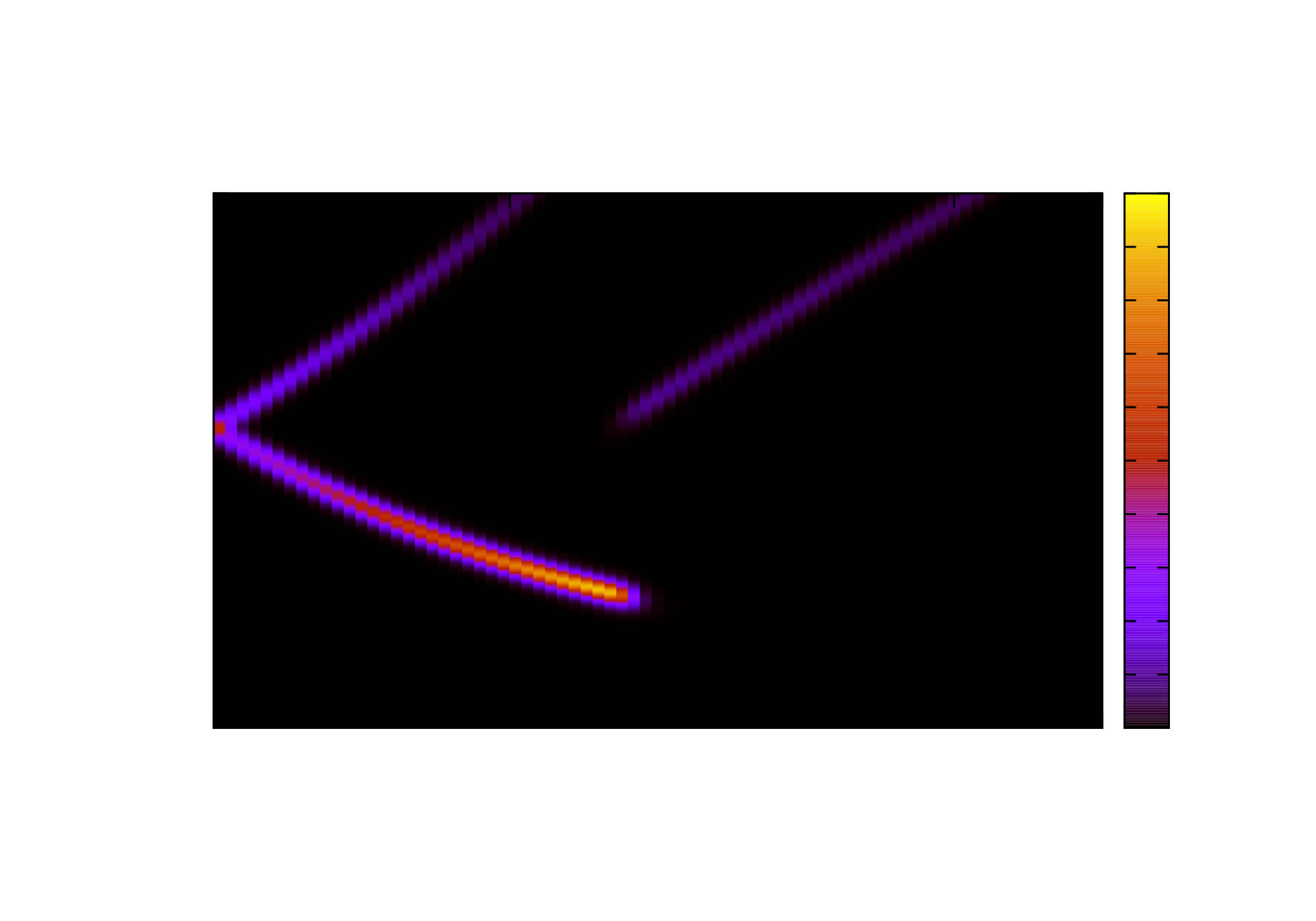}}%
    \gplfronttext
  \end{picture}%
\endgroup

%% file: tex-josephson_pi_shift_1.tex
\begingroup
  \makeatletter
  \providecommand\color[2][]{%
    \GenericError{(gnuplot) \space\space\space\@spaces}{%
      Package color not loaded in conjunction with
      terminal option `colourtext'%
    }{See the gnuplot documentation for explanation.%
    }{Either use 'blacktext' in gnuplot or load the package
      color.sty in LaTeX.}%
    \renewcommand\color[2][]{}%
  }%
  \providecommand\includegraphics[2][]{%
    \GenericError{(gnuplot) \space\space\space\@spaces}{%
      Package graphicx or graphics not loaded%
    }{See the gnuplot documentation for explanation.%
    }{The gnuplot epslatex terminal needs graphicx.sty or graphics.sty.}%
    \renewcommand\includegraphics[2][]{}%
  }%
  \providecommand\rotatebox[2]{#2}%
  \@ifundefined{ifGPcolor}{%
    \newif\ifGPcolor
    \GPcolortrue
  }{}%
  \@ifundefined{ifGPblacktext}{%
    \newif\ifGPblacktext
    \GPblacktexttrue
  }{}%
  \let\gplgaddtomacro\g@addto@macro
  \gdef\gplbacktext{}%
  \gdef\gplfronttext{}%
  \makeatother
  \ifGPblacktext
    \def\colorrgb#1{}%
    \def\colorgray#1{}%
  \else
    \ifGPcolor
      \def\colorrgb#1{\color[rgb]{#1}}%
      \def\colorgray#1{\color[gray]{#1}}%
      \expandafter\def\csname LTw\endcsname{\color{white}}%
      \expandafter\def\csname LTb\endcsname{\color{black}}%
      \expandafter\def\csname LTa\endcsname{\color{black}}%
      \expandafter\def\csname LT0\endcsname{\color[rgb]{1,0,0}}%
      \expandafter\def\csname LT1\endcsname{\color[rgb]{0,1,0}}%
      \expandafter\def\csname LT2\endcsname{\color[rgb]{0,0,1}}%
      \expandafter\def\csname LT3\endcsname{\color[rgb]{1,0,1}}%
      \expandafter\def\csname LT4\endcsname{\color[rgb]{0,1,1}}%
      \expandafter\def\csname LT5\endcsname{\color[rgb]{1,1,0}}%
      \expandafter\def\csname LT6\endcsname{\color[rgb]{0,0,0}}%
      \expandafter\def\csname LT7\endcsname{\color[rgb]{1,0.3,0}}%
      \expandafter\def\csname LT8\endcsname{\color[rgb]{0.5,0.5,0.5}}%
    \else
      \def\colorrgb#1{\color{black}}%
      \def\colorgray#1{\color[gray]{#1}}%
      \expandafter\def\csname LTw\endcsname{\color{white}}%
      \expandafter\def\csname LTb\endcsname{\color{black}}%
      \expandafter\def\csname LTa\endcsname{\color{black}}%
      \expandafter\def\csname LT0\endcsname{\color{black}}%
      \expandafter\def\csname LT1\endcsname{\color{black}}%
      \expandafter\def\csname LT2\endcsname{\color{black}}%
      \expandafter\def\csname LT3\endcsname{\color{black}}%
      \expandafter\def\csname LT4\endcsname{\color{black}}%
      \expandafter\def\csname LT5\endcsname{\color{black}}%
      \expandafter\def\csname LT6\endcsname{\color{black}}%
      \expandafter\def\csname LT7\endcsname{\color{black}}%
      \expandafter\def\csname LT8\endcsname{\color{black}}%
    \fi
  \fi
  \setlength{\unitlength}{0.0500bp}%
  \begin{picture}(7200.00,4536.00)%
    \gplgaddtomacro\gplbacktext{%
      \csname LTb\endcsname%
      \put(1254,829){\makebox(0,0)[r]{\strut{} 0}}%
      \put(1254,1168){\makebox(0,0)[r]{\strut{} 0.001}}%
      \put(1254,1506){\makebox(0,0)[r]{\strut{} 0.002}}%
      \put(1254,1845){\makebox(0,0)[r]{\strut{} 0.003}}%
      \put(1254,2183){\makebox(0,0)[r]{\strut{} 0.004}}%
      \put(1254,2522){\makebox(0,0)[r]{\strut{} 0.005}}%
      \put(1254,2860){\makebox(0,0)[r]{\strut{} 0.006}}%
      \put(1254,3199){\makebox(0,0)[r]{\strut{} 0.007}}%
      \put(1254,3537){\makebox(0,0)[r]{\strut{} 0.008}}%
      \put(1254,3876){\makebox(0,0)[r]{\strut{} 0.009}}%
      \put(1386,440){\makebox(0,0){\strut{} 0}}%
      \put(2163,440){\makebox(0,0){\strut{} 0.5}}%
      \put(2940,440){\makebox(0,0){\strut{} 1}}%
      \put(3717,440){\makebox(0,0){\strut{} 1.5}}%
      \put(4495,440){\makebox(0,0){\strut{} 2}}%
      \put(5272,440){\makebox(0,0){\strut{} 2.5}}%
      \put(6049,440){\makebox(0,0){\strut{} 3}}%
      \put(6826,440){\makebox(0,0){\strut{} 3.5}}%
      \put(220,2268){\rotatebox{90}{\makebox(0,0){\strut{}$I_j$}}}%
      \put(4106,110){\makebox(0,0){\strut{}$\phi$}}%
      \put(4106,4206){\makebox(0,0){\strut{}Josephson current for different values of $\Delta$}}%
      \put(1619,3673){\makebox(0,0)[l]{\strut{}$\beta=50$, $U=1$, $V=0.5$, $\mu=0$, $\epsilon_d=0$}}%
    }%
    \gplgaddtomacro\gplfronttext{%
      \csname LTb\endcsname%
      \put(5971,3427){\makebox(0,0)[r]{\strut{}$\Delta=0.10$}}%
      \csname LTb\endcsname%
      \put(5971,3207){\makebox(0,0)[r]{\strut{}$\Delta=0.09$}}%
      \csname LTb\endcsname%
      \put(5971,2987){\makebox(0,0)[r]{\strut{}$\Delta=0.08$}}%
      \csname LTb\endcsname%
      \put(5971,2767){\makebox(0,0)[r]{\strut{}$\Delta=0.05$}}%
      \csname LTb\endcsname%
      \put(5971,2547){\makebox(0,0)[r]{\strut{}$\Delta=0.03$}}%
      \csname LTb\endcsname%
      \put(5971,2327){\makebox(0,0)[r]{\strut{}$\Delta=0.01$}}%
    }%
    \gplbacktext
    \put(0,0){\includegraphics{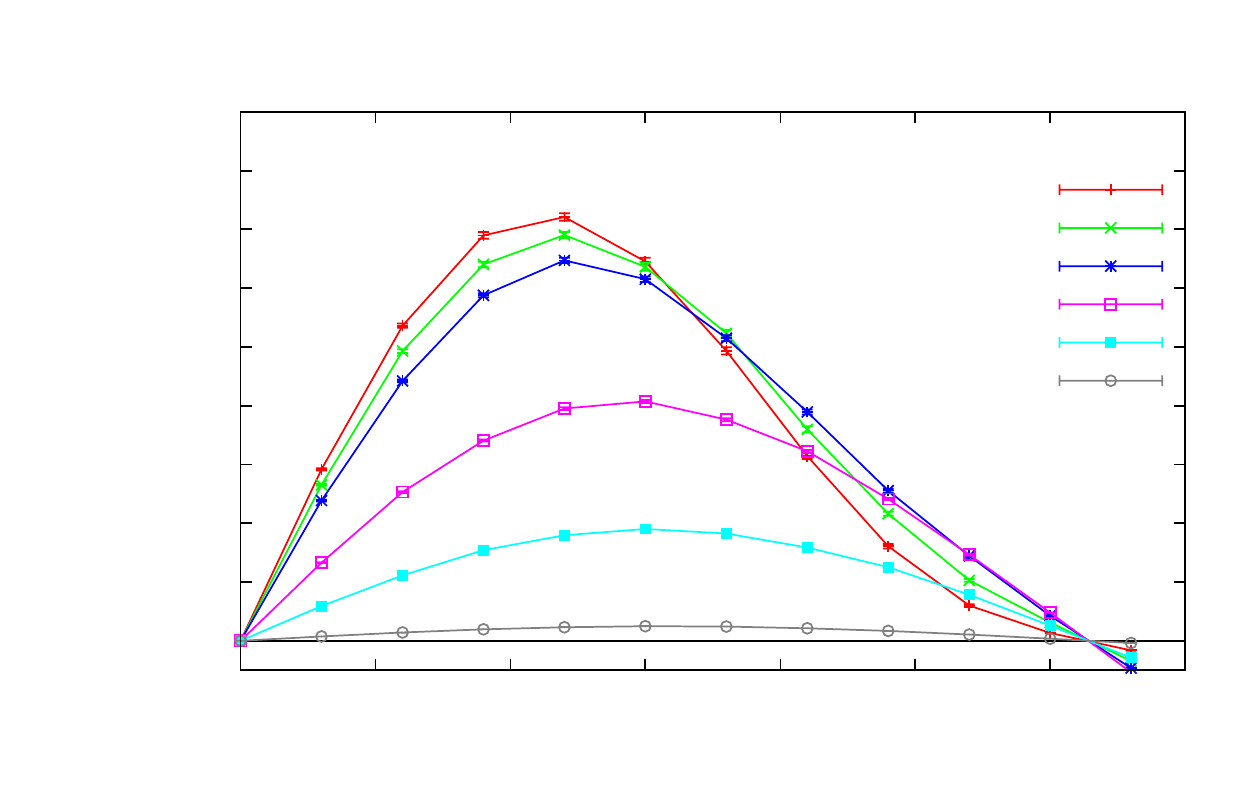}}%
    \gplfronttext
  \end{picture}%
\endgroup

%% file: tex-josephson_pi_shift_2.tex
\begingroup
  \makeatletter
  \providecommand\color[2][]{%
    \GenericError{(gnuplot) \space\space\space\@spaces}{%
      Package color not loaded in conjunction with
      terminal option `colourtext'%
    }{See the gnuplot documentation for explanation.%
    }{Either use 'blacktext' in gnuplot or load the package
      color.sty in LaTeX.}%
    \renewcommand\color[2][]{}%
  }%
  \providecommand\includegraphics[2][]{%
    \GenericError{(gnuplot) \space\space\space\@spaces}{%
      Package graphicx or graphics not loaded%
    }{See the gnuplot documentation for explanation.%
    }{The gnuplot epslatex terminal needs graphicx.sty or graphics.sty.}%
    \renewcommand\includegraphics[2][]{}%
  }%
  \providecommand\rotatebox[2]{#2}%
  \@ifundefined{ifGPcolor}{%
    \newif\ifGPcolor
    \GPcolortrue
  }{}%
  \@ifundefined{ifGPblacktext}{%
    \newif\ifGPblacktext
    \GPblacktexttrue
  }{}%
  \let\gplgaddtomacro\g@addto@macro
  \gdef\gplbacktext{}%
  \gdef\gplfronttext{}%
  \makeatother
  \ifGPblacktext
    \def\colorrgb#1{}%
    \def\colorgray#1{}%
  \else
    \ifGPcolor
      \def\colorrgb#1{\color[rgb]{#1}}%
      \def\colorgray#1{\color[gray]{#1}}%
      \expandafter\def\csname LTw\endcsname{\color{white}}%
      \expandafter\def\csname LTb\endcsname{\color{black}}%
      \expandafter\def\csname LTa\endcsname{\color{black}}%
      \expandafter\def\csname LT0\endcsname{\color[rgb]{1,0,0}}%
      \expandafter\def\csname LT1\endcsname{\color[rgb]{0,1,0}}%
      \expandafter\def\csname LT2\endcsname{\color[rgb]{0,0,1}}%
      \expandafter\def\csname LT3\endcsname{\color[rgb]{1,0,1}}%
      \expandafter\def\csname LT4\endcsname{\color[rgb]{0,1,1}}%
      \expandafter\def\csname LT5\endcsname{\color[rgb]{1,1,0}}%
      \expandafter\def\csname LT6\endcsname{\color[rgb]{0,0,0}}%
      \expandafter\def\csname LT7\endcsname{\color[rgb]{1,0.3,0}}%
      \expandafter\def\csname LT8\endcsname{\color[rgb]{0.5,0.5,0.5}}%
    \else
      \def\colorrgb#1{\color{black}}%
      \def\colorgray#1{\color[gray]{#1}}%
      \expandafter\def\csname LTw\endcsname{\color{white}}%
      \expandafter\def\csname LTb\endcsname{\color{black}}%
      \expandafter\def\csname LTa\endcsname{\color{black}}%
      \expandafter\def\csname LT0\endcsname{\color{black}}%
      \expandafter\def\csname LT1\endcsname{\color{black}}%
      \expandafter\def\csname LT2\endcsname{\color{black}}%
      \expandafter\def\csname LT3\endcsname{\color{black}}%
      \expandafter\def\csname LT4\endcsname{\color{black}}%
      \expandafter\def\csname LT5\endcsname{\color{black}}%
      \expandafter\def\csname LT6\endcsname{\color{black}}%
      \expandafter\def\csname LT7\endcsname{\color{black}}%
      \expandafter\def\csname LT8\endcsname{\color{black}}%
    \fi
  \fi
  \setlength{\unitlength}{0.0500bp}%
  \begin{picture}(7200.00,4536.00)%
    \gplgaddtomacro\gplbacktext{%
      \csname LTb\endcsname%
      \put(1254,660){\makebox(0,0)[r]{\strut{}-0.015}}%
      \put(1254,1119){\makebox(0,0)[r]{\strut{}-0.01}}%
      \put(1254,1579){\makebox(0,0)[r]{\strut{}-0.005}}%
      \put(1254,2038){\makebox(0,0)[r]{\strut{} 0}}%
      \put(1254,2498){\makebox(0,0)[r]{\strut{} 0.005}}%
      \put(1254,2957){\makebox(0,0)[r]{\strut{} 0.01}}%
      \put(1254,3417){\makebox(0,0)[r]{\strut{} 0.015}}%
      \put(1254,3876){\makebox(0,0)[r]{\strut{} 0.02}}%
      \put(1386,440){\makebox(0,0){\strut{} 0}}%
      \put(2163,440){\makebox(0,0){\strut{} 0.5}}%
      \put(2940,440){\makebox(0,0){\strut{} 1}}%
      \put(3717,440){\makebox(0,0){\strut{} 1.5}}%
      \put(4495,440){\makebox(0,0){\strut{} 2}}%
      \put(5272,440){\makebox(0,0){\strut{} 2.5}}%
      \put(6049,440){\makebox(0,0){\strut{} 3}}%
      \put(6826,440){\makebox(0,0){\strut{} 3.5}}%
      \put(220,2268){\rotatebox{90}{\makebox(0,0){\strut{}$I_j$}}}%
      \put(4106,110){\makebox(0,0){\strut{}$\phi$}}%
      \put(4106,4206){\makebox(0,0){\strut{}Josephson current for different values of $\Delta$}}%
      \put(1386,3600){\makebox(0,0)[l]{\strut{}$\beta=50$, $U=1$, $V=0.5$, $\mu=0$, $\epsilon_d=0$}}%
    }%
    \gplgaddtomacro\gplfronttext{%
      \csname LTb\endcsname%
      \put(5971,3766){\makebox(0,0)[r]{\strut{}$\Delta=0.10$}}%
      \csname LTb\endcsname%
      \put(5971,3546){\makebox(0,0)[r]{\strut{}$\Delta=0.15$}}%
      \csname LTb\endcsname%
      \put(5971,3326){\makebox(0,0)[r]{\strut{}$\Delta=0.20$}}%
      \csname LTb\endcsname%
      \put(5971,3106){\makebox(0,0)[r]{\strut{}$\Delta=0.25$}}%
      \csname LTb\endcsname%
      \put(5971,2886){\makebox(0,0)[r]{\strut{}$\Delta=0.30$}}%
      \csname LTb\endcsname%
      \put(5971,2666){\makebox(0,0)[r]{\strut{}$\Delta=0.35$}}%
      \csname LTb\endcsname%
      \put(5971,2446){\makebox(0,0)[r]{\strut{}$\Delta=0.40$}}%
    }%
    \gplbacktext
    \put(0,0){\includegraphics{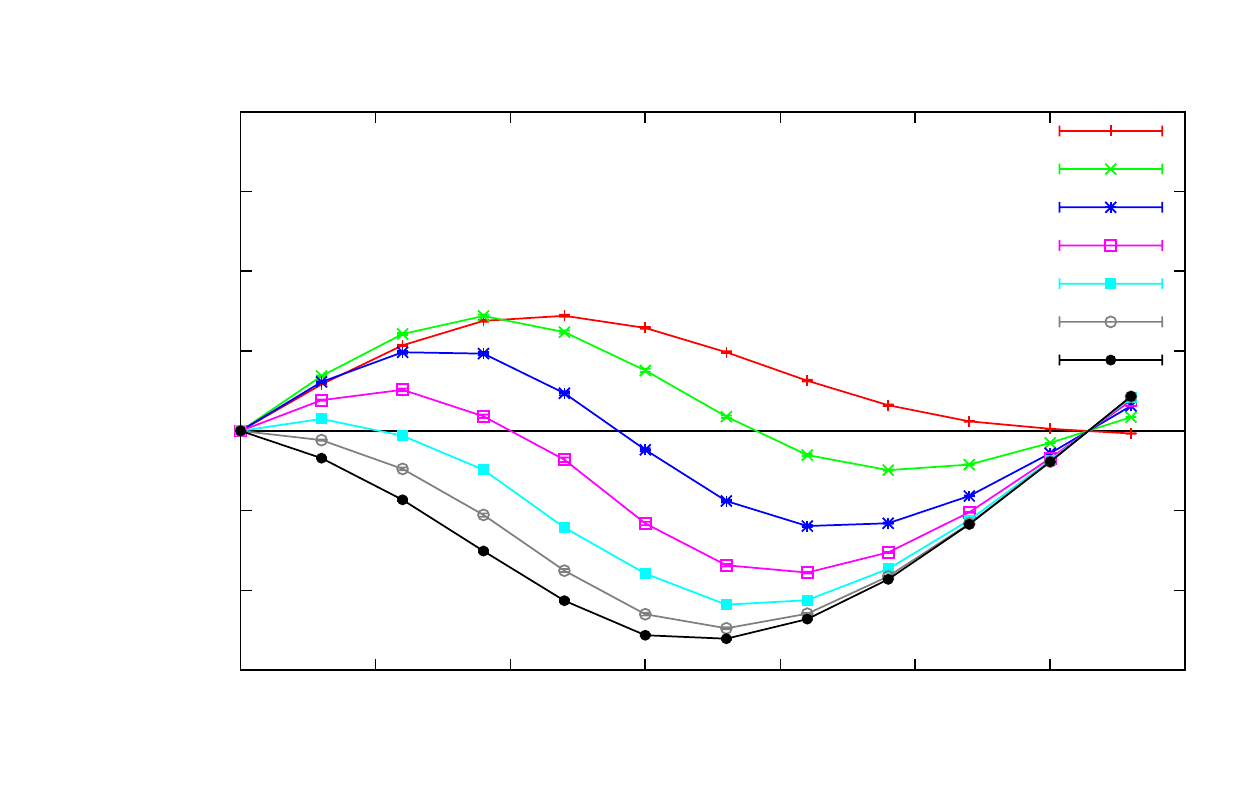}}%
    \gplfronttext
  \end{picture}%
\endgroup

%% file: tex-josephson_pi_shift_3.tex
\begingroup
  \makeatletter
  \providecommand\color[2][]{%
    \GenericError{(gnuplot) \space\space\space\@spaces}{%
      Package color not loaded in conjunction with
      terminal option `colourtext'%
    }{See the gnuplot documentation for explanation.%
    }{Either use 'blacktext' in gnuplot or load the package
      color.sty in LaTeX.}%
    \renewcommand\color[2][]{}%
  }%
  \providecommand\includegraphics[2][]{%
    \GenericError{(gnuplot) \space\space\space\@spaces}{%
      Package graphicx or graphics not loaded%
    }{See the gnuplot documentation for explanation.%
    }{The gnuplot epslatex terminal needs graphicx.sty or graphics.sty.}%
    \renewcommand\includegraphics[2][]{}%
  }%
  \providecommand\rotatebox[2]{#2}%
  \@ifundefined{ifGPcolor}{%
    \newif\ifGPcolor
    \GPcolortrue
  }{}%
  \@ifundefined{ifGPblacktext}{%
    \newif\ifGPblacktext
    \GPblacktexttrue
  }{}%
  \let\gplgaddtomacro\g@addto@macro
  \gdef\gplbacktext{}%
  \gdef\gplfronttext{}%
  \makeatother
  \ifGPblacktext
    \def\colorrgb#1{}%
    \def\colorgray#1{}%
  \else
    \ifGPcolor
      \def\colorrgb#1{\color[rgb]{#1}}%
      \def\colorgray#1{\color[gray]{#1}}%
      \expandafter\def\csname LTw\endcsname{\color{white}}%
      \expandafter\def\csname LTb\endcsname{\color{black}}%
      \expandafter\def\csname LTa\endcsname{\color{black}}%
      \expandafter\def\csname LT0\endcsname{\color[rgb]{1,0,0}}%
      \expandafter\def\csname LT1\endcsname{\color[rgb]{0,1,0}}%
      \expandafter\def\csname LT2\endcsname{\color[rgb]{0,0,1}}%
      \expandafter\def\csname LT3\endcsname{\color[rgb]{1,0,1}}%
      \expandafter\def\csname LT4\endcsname{\color[rgb]{0,1,1}}%
      \expandafter\def\csname LT5\endcsname{\color[rgb]{1,1,0}}%
      \expandafter\def\csname LT6\endcsname{\color[rgb]{0,0,0}}%
      \expandafter\def\csname LT7\endcsname{\color[rgb]{1,0.3,0}}%
      \expandafter\def\csname LT8\endcsname{\color[rgb]{0.5,0.5,0.5}}%
    \else
      \def\colorrgb#1{\color{black}}%
      \def\colorgray#1{\color[gray]{#1}}%
      \expandafter\def\csname LTw\endcsname{\color{white}}%
      \expandafter\def\csname LTb\endcsname{\color{black}}%
      \expandafter\def\csname LTa\endcsname{\color{black}}%
      \expandafter\def\csname LT0\endcsname{\color{black}}%
      \expandafter\def\csname LT1\endcsname{\color{black}}%
      \expandafter\def\csname LT2\endcsname{\color{black}}%
      \expandafter\def\csname LT3\endcsname{\color{black}}%
      \expandafter\def\csname LT4\endcsname{\color{black}}%
      \expandafter\def\csname LT5\endcsname{\color{black}}%
      \expandafter\def\csname LT6\endcsname{\color{black}}%
      \expandafter\def\csname LT7\endcsname{\color{black}}%
      \expandafter\def\csname LT8\endcsname{\color{black}}%
    \fi
  \fi
  \setlength{\unitlength}{0.0500bp}%
  \begin{picture}(7200.00,4536.00)%
    \gplgaddtomacro\gplbacktext{%
      \csname LTb\endcsname%
      \put(1254,660){\makebox(0,0)[r]{\strut{}-0.018}}%
      \put(1254,982){\makebox(0,0)[r]{\strut{}-0.016}}%
      \put(1254,1303){\makebox(0,0)[r]{\strut{}-0.014}}%
      \put(1254,1625){\makebox(0,0)[r]{\strut{}-0.012}}%
      \put(1254,1946){\makebox(0,0)[r]{\strut{}-0.01}}%
      \put(1254,2268){\makebox(0,0)[r]{\strut{}-0.008}}%
      \put(1254,2590){\makebox(0,0)[r]{\strut{}-0.006}}%
      \put(1254,2911){\makebox(0,0)[r]{\strut{}-0.004}}%
      \put(1254,3233){\makebox(0,0)[r]{\strut{}-0.002}}%
      \put(1254,3554){\makebox(0,0)[r]{\strut{} 0}}%
      \put(1254,3876){\makebox(0,0)[r]{\strut{} 0.002}}%
      \put(1386,440){\makebox(0,0){\strut{} 0}}%
      \put(2163,440){\makebox(0,0){\strut{} 0.5}}%
      \put(2940,440){\makebox(0,0){\strut{} 1}}%
      \put(3717,440){\makebox(0,0){\strut{} 1.5}}%
      \put(4495,440){\makebox(0,0){\strut{} 2}}%
      \put(5272,440){\makebox(0,0){\strut{} 2.5}}%
      \put(6049,440){\makebox(0,0){\strut{} 3}}%
      \put(6826,440){\makebox(0,0){\strut{} 3.5}}%
      \put(220,2268){\rotatebox{90}{\makebox(0,0){\strut{}$I_j$}}}%
      \put(4106,110){\makebox(0,0){\strut{}$\phi$}}%
      \put(4106,4206){\makebox(0,0){\strut{}Josephson current for different values of $\Delta$}}%
      \put(1433,821){\makebox(0,0)[l]{\strut{}$\beta=50$, $U=1$, $V=0.5$, $\mu=0$, $\epsilon_d=0$}}%
    }%
    \gplgaddtomacro\gplfronttext{%
      \csname LTb\endcsname%
      \put(6049,2319){\makebox(0,0)[r]{\strut{}$\Delta=2.5$}}%
      \csname LTb\endcsname%
      \put(6049,2099){\makebox(0,0)[r]{\strut{}$\Delta=2.0$}}%
      \csname LTb\endcsname%
      \put(6049,1879){\makebox(0,0)[r]{\strut{}$\Delta=1.5$}}%
      \csname LTb\endcsname%
      \put(6049,1659){\makebox(0,0)[r]{\strut{}$\Delta=1.2$}}%
      \csname LTb\endcsname%
      \put(6049,1439){\makebox(0,0)[r]{\strut{}$\Delta=1.0$}}%
      \csname LTb\endcsname%
      \put(6049,1219){\makebox(0,0)[r]{\strut{}$\Delta=0.8$}}%
      \csname LTb\endcsname%
      \put(6049,999){\makebox(0,0)[r]{\strut{}$\Delta=0.7$}}%
      \csname LTb\endcsname%
      \put(6049,779){\makebox(0,0)[r]{\strut{}$\Delta=0.5$}}%
    }%
    \gplbacktext
    \put(0,0){\includegraphics{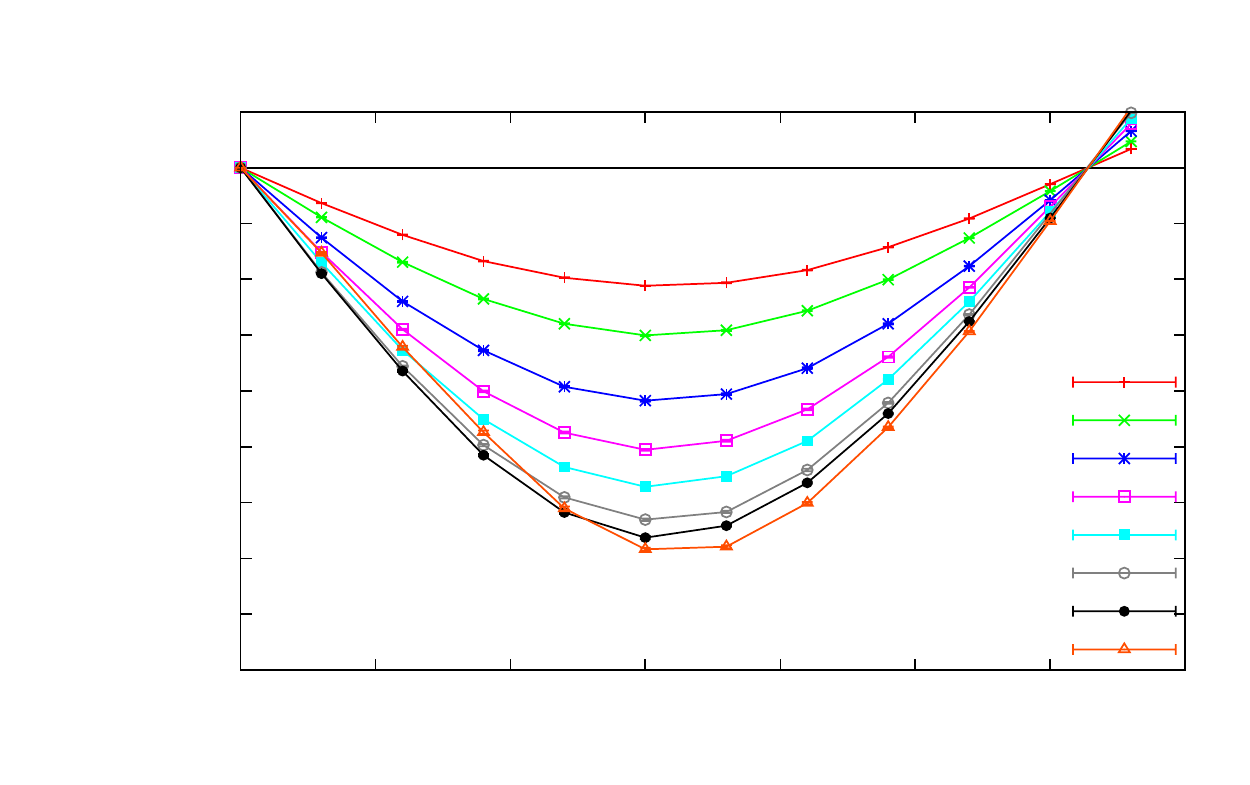}}%
    \gplfronttext
  \end{picture}%
\endgroup

%% file: tex-josephson-curr-temp.tex
\begingroup
  \makeatletter
  \providecommand\color[2][]{%
    \GenericError{(gnuplot) \space\space\space\@spaces}{%
      Package color not loaded in conjunction with
      terminal option `colourtext'%
    }{See the gnuplot documentation for explanation.%
    }{Either use 'blacktext' in gnuplot or load the package
      color.sty in LaTeX.}%
    \renewcommand\color[2][]{}%
  }%
  \providecommand\includegraphics[2][]{%
    \GenericError{(gnuplot) \space\space\space\@spaces}{%
      Package graphicx or graphics not loaded%
    }{See the gnuplot documentation for explanation.%
    }{The gnuplot epslatex terminal needs graphicx.sty or graphics.sty.}%
    \renewcommand\includegraphics[2][]{}%
  }%
  \providecommand\rotatebox[2]{#2}%
  \@ifundefined{ifGPcolor}{%
    \newif\ifGPcolor
    \GPcolortrue
  }{}%
  \@ifundefined{ifGPblacktext}{%
    \newif\ifGPblacktext
    \GPblacktexttrue
  }{}%
  \let\gplgaddtomacro\g@addto@macro
  \gdef\gplbacktext{}%
  \gdef\gplfronttext{}%
  \makeatother
  \ifGPblacktext
    \def\colorrgb#1{}%
    \def\colorgray#1{}%
  \else
    \ifGPcolor
      \def\colorrgb#1{\color[rgb]{#1}}%
      \def\colorgray#1{\color[gray]{#1}}%
      \expandafter\def\csname LTw\endcsname{\color{white}}%
      \expandafter\def\csname LTb\endcsname{\color{black}}%
      \expandafter\def\csname LTa\endcsname{\color{black}}%
      \expandafter\def\csname LT0\endcsname{\color[rgb]{1,0,0}}%
      \expandafter\def\csname LT1\endcsname{\color[rgb]{0,1,0}}%
      \expandafter\def\csname LT2\endcsname{\color[rgb]{0,0,1}}%
      \expandafter\def\csname LT3\endcsname{\color[rgb]{1,0,1}}%
      \expandafter\def\csname LT4\endcsname{\color[rgb]{0,1,1}}%
      \expandafter\def\csname LT5\endcsname{\color[rgb]{1,1,0}}%
      \expandafter\def\csname LT6\endcsname{\color[rgb]{0,0,0}}%
      \expandafter\def\csname LT7\endcsname{\color[rgb]{1,0.3,0}}%
      \expandafter\def\csname LT8\endcsname{\color[rgb]{0.5,0.5,0.5}}%
    \else
      \def\colorrgb#1{\color{black}}%
      \def\colorgray#1{\color[gray]{#1}}%
      \expandafter\def\csname LTw\endcsname{\color{white}}%
      \expandafter\def\csname LTb\endcsname{\color{black}}%
      \expandafter\def\csname LTa\endcsname{\color{black}}%
      \expandafter\def\csname LT0\endcsname{\color{black}}%
      \expandafter\def\csname LT1\endcsname{\color{black}}%
      \expandafter\def\csname LT2\endcsname{\color{black}}%
      \expandafter\def\csname LT3\endcsname{\color{black}}%
      \expandafter\def\csname LT4\endcsname{\color{black}}%
      \expandafter\def\csname LT5\endcsname{\color{black}}%
      \expandafter\def\csname LT6\endcsname{\color{black}}%
      \expandafter\def\csname LT7\endcsname{\color{black}}%
      \expandafter\def\csname LT8\endcsname{\color{black}}%
    \fi
  \fi
  \setlength{\unitlength}{0.0500bp}%
  \begin{picture}(7200.00,4536.00)%
    \gplgaddtomacro\gplbacktext{%
      \csname LTb\endcsname%
      \put(1254,660){\makebox(0,0)[r]{\strut{}-0.006}}%
      \put(1254,1062){\makebox(0,0)[r]{\strut{}-0.004}}%
      \put(1254,1464){\makebox(0,0)[r]{\strut{}-0.002}}%
      \put(1254,1866){\makebox(0,0)[r]{\strut{} 0}}%
      \put(1254,2268){\makebox(0,0)[r]{\strut{} 0.002}}%
      \put(1254,2670){\makebox(0,0)[r]{\strut{} 0.004}}%
      \put(1254,3072){\makebox(0,0)[r]{\strut{} 0.006}}%
      \put(1254,3474){\makebox(0,0)[r]{\strut{} 0.008}}%
      \put(1254,3876){\makebox(0,0)[r]{\strut{} 0.01}}%
      \put(1386,440){\makebox(0,0){\strut{} 0}}%
      \put(2163,440){\makebox(0,0){\strut{} 0.5}}%
      \put(2940,440){\makebox(0,0){\strut{} 1}}%
      \put(3717,440){\makebox(0,0){\strut{} 1.5}}%
      \put(4495,440){\makebox(0,0){\strut{} 2}}%
      \put(5272,440){\makebox(0,0){\strut{} 2.5}}%
      \put(6049,440){\makebox(0,0){\strut{} 3}}%
      \put(6826,440){\makebox(0,0){\strut{} 3.5}}%
      \put(220,2268){\rotatebox{90}{\makebox(0,0){\strut{}$I_j$}}}%
      \put(4106,110){\makebox(0,0){\strut{}$\phi$}}%
      \put(4106,4206){\makebox(0,0){\strut{}Temperature dependence of the Josephson current}}%
      \put(1541,861){\makebox(0,0)[l]{\strut{}$U=1$, $V=0.5$, $\Delta=0.15$, $\mu=0$, $\epsilon_d=0$}}%
    }%
    \gplgaddtomacro\gplfronttext{%
      \csname LTb\endcsname%
      \put(5839,3703){\makebox(0,0)[r]{\strut{}$\beta=25$}}%
      \csname LTb\endcsname%
      \put(5839,3483){\makebox(0,0)[r]{\strut{}$\beta=50$}}%
      \csname LTb\endcsname%
      \put(5839,3263){\makebox(0,0)[r]{\strut{}$\beta=75$}}%
    }%
    \gplbacktext
    \put(0,0){\includegraphics{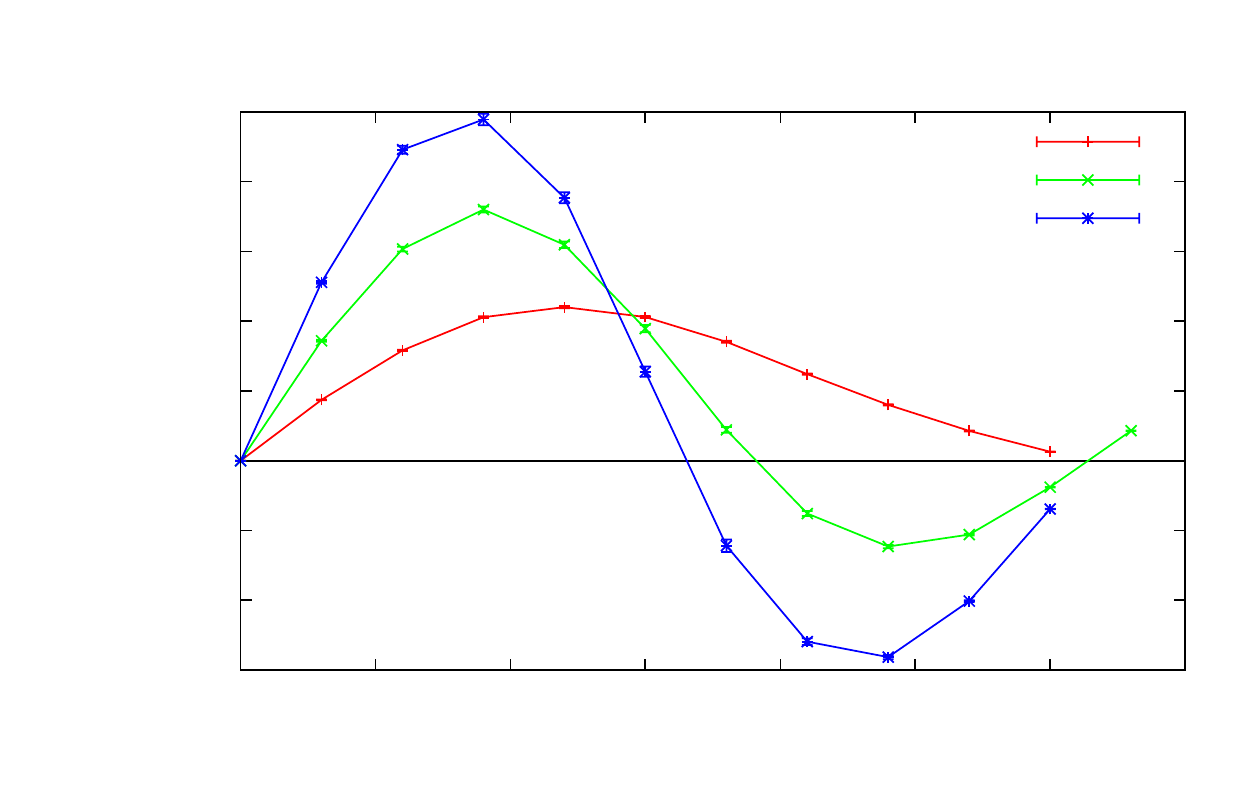}}%
    \gplfronttext
  \end{picture}%
\endgroup

%% file: tex-double-occ-measurements-D1.tex
\begingroup
  \makeatletter
  \providecommand\color[2][]{%
    \GenericError{(gnuplot) \space\space\space\@spaces}{%
      Package color not loaded in conjunction with
      terminal option `colourtext'%
    }{See the gnuplot documentation for explanation.%
    }{Either use 'blacktext' in gnuplot or load the package
      color.sty in LaTeX.}%
    \renewcommand\color[2][]{}%
  }%
  \providecommand\includegraphics[2][]{%
    \GenericError{(gnuplot) \space\space\space\@spaces}{%
      Package graphicx or graphics not loaded%
    }{See the gnuplot documentation for explanation.%
    }{The gnuplot epslatex terminal needs graphicx.sty or graphics.sty.}%
    \renewcommand\includegraphics[2][]{}%
  }%
  \providecommand\rotatebox[2]{#2}%
  \@ifundefined{ifGPcolor}{%
    \newif\ifGPcolor
    \GPcolortrue
  }{}%
  \@ifundefined{ifGPblacktext}{%
    \newif\ifGPblacktext
    \GPblacktexttrue
  }{}%
  \let\gplgaddtomacro\g@addto@macro
  \gdef\gplbacktext{}%
  \gdef\gplfronttext{}%
  \makeatother
  \ifGPblacktext
    \def\colorrgb#1{}%
    \def\colorgray#1{}%
  \else
    \ifGPcolor
      \def\colorrgb#1{\color[rgb]{#1}}%
      \def\colorgray#1{\color[gray]{#1}}%
      \expandafter\def\csname LTw\endcsname{\color{white}}%
      \expandafter\def\csname LTb\endcsname{\color{black}}%
      \expandafter\def\csname LTa\endcsname{\color{black}}%
      \expandafter\def\csname LT0\endcsname{\color[rgb]{1,0,0}}%
      \expandafter\def\csname LT1\endcsname{\color[rgb]{0,1,0}}%
      \expandafter\def\csname LT2\endcsname{\color[rgb]{0,0,1}}%
      \expandafter\def\csname LT3\endcsname{\color[rgb]{1,0,1}}%
      \expandafter\def\csname LT4\endcsname{\color[rgb]{0,1,1}}%
      \expandafter\def\csname LT5\endcsname{\color[rgb]{1,1,0}}%
      \expandafter\def\csname LT6\endcsname{\color[rgb]{0,0,0}}%
      \expandafter\def\csname LT7\endcsname{\color[rgb]{1,0.3,0}}%
      \expandafter\def\csname LT8\endcsname{\color[rgb]{0.5,0.5,0.5}}%
    \else
      \def\colorrgb#1{\color{black}}%
      \def\colorgray#1{\color[gray]{#1}}%
      \expandafter\def\csname LTw\endcsname{\color{white}}%
      \expandafter\def\csname LTb\endcsname{\color{black}}%
      \expandafter\def\csname LTa\endcsname{\color{black}}%
      \expandafter\def\csname LT0\endcsname{\color{black}}%
      \expandafter\def\csname LT1\endcsname{\color{black}}%
      \expandafter\def\csname LT2\endcsname{\color{black}}%
      \expandafter\def\csname LT3\endcsname{\color{black}}%
      \expandafter\def\csname LT4\endcsname{\color{black}}%
      \expandafter\def\csname LT5\endcsname{\color{black}}%
      \expandafter\def\csname LT6\endcsname{\color{black}}%
      \expandafter\def\csname LT7\endcsname{\color{black}}%
      \expandafter\def\csname LT8\endcsname{\color{black}}%
    \fi
  \fi
  \setlength{\unitlength}{0.0500bp}%
  \begin{picture}(7200.00,5040.00)%
    \gplgaddtomacro\gplbacktext{%
      \csname LTb\endcsname%
      \put(1122,660){\makebox(0,0)[r]{\strut{} 0.05}}%
      \put(1122,1191){\makebox(0,0)[r]{\strut{} 0.1}}%
      \put(1122,1723){\makebox(0,0)[r]{\strut{} 0.15}}%
      \put(1122,2254){\makebox(0,0)[r]{\strut{} 0.2}}%
      \put(1122,2786){\makebox(0,0)[r]{\strut{} 0.25}}%
      \put(1122,3317){\makebox(0,0)[r]{\strut{} 0.3}}%
      \put(1122,3849){\makebox(0,0)[r]{\strut{} 0.35}}%
      \put(1122,4380){\makebox(0,0)[r]{\strut{} 0.4}}%
      \put(1254,440){\makebox(0,0){\strut{} 0.1}}%
      \put(1951,440){\makebox(0,0){\strut{} 0.2}}%
      \put(2647,440){\makebox(0,0){\strut{} 0.3}}%
      \put(3344,440){\makebox(0,0){\strut{} 0.4}}%
      \put(4040,440){\makebox(0,0){\strut{} 0.5}}%
      \put(4737,440){\makebox(0,0){\strut{} 0.6}}%
      \put(5433,440){\makebox(0,0){\strut{} 0.7}}%
      \put(6130,440){\makebox(0,0){\strut{} 0.8}}%
      \put(6826,440){\makebox(0,0){\strut{} 0.9}}%
      \put(220,2520){\rotatebox{90}{\makebox(0,0){\strut{}$\langle \hat{n}_\up \hat{n}_\dw \rangle$}}}%
      \put(4040,110){\makebox(0,0){\strut{}$ U $}}%
      \put(4040,4710){\makebox(0,0){\strut{}Double occupancy}}%
      \put(1602,979){\makebox(0,0)[l]{\strut{}$V=0.5$, $\Delta = 1.0$, $\mu=0$, $\epsilon_d=0$}}%
    }%
    \gplgaddtomacro\gplfronttext{%
      \csname LTb\endcsname%
      \put(5839,4207){\makebox(0,0)[r]{\strut{}$\beta = 75$}}%
      \csname LTb\endcsname%
      \put(5839,3987){\makebox(0,0)[r]{\strut{}$\beta = 100$}}%
      \csname LTb\endcsname%
      \put(5839,3767){\makebox(0,0)[r]{\strut{}$\beta = 125$}}%
      \csname LTb\endcsname%
      \put(5839,3547){\makebox(0,0)[r]{\strut{}$\beta = 150$}}%
      \csname LTb\endcsname%
      \put(5839,3327){\makebox(0,0)[r]{\strut{}$\beta = 200$}}%
    }%
    \gplbacktext
    \put(0,0){\includegraphics{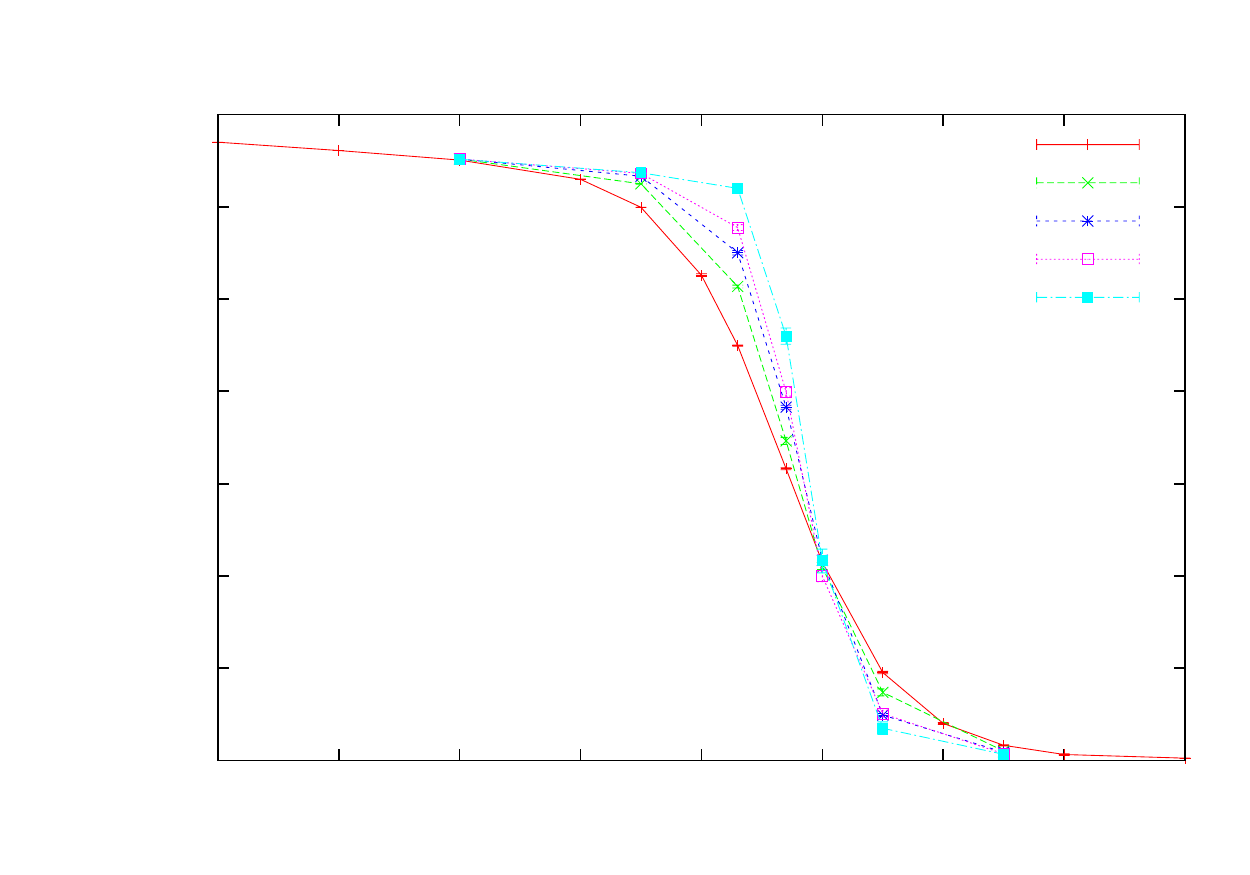}}%
    \gplfronttext
  \end{picture}%
\endgroup

%% file: tex-josephson_double_occ1.tex
\begingroup
  \makeatletter
  \providecommand\color[2][]{%
    \GenericError{(gnuplot) \space\space\space\@spaces}{%
      Package color not loaded in conjunction with
      terminal option `colourtext'%
    }{See the gnuplot documentation for explanation.%
    }{Either use 'blacktext' in gnuplot or load the package
      color.sty in LaTeX.}%
    \renewcommand\color[2][]{}%
  }%
  \providecommand\includegraphics[2][]{%
    \GenericError{(gnuplot) \space\space\space\@spaces}{%
      Package graphicx or graphics not loaded%
    }{See the gnuplot documentation for explanation.%
    }{The gnuplot epslatex terminal needs graphicx.sty or graphics.sty.}%
    \renewcommand\includegraphics[2][]{}%
  }%
  \providecommand\rotatebox[2]{#2}%
  \@ifundefined{ifGPcolor}{%
    \newif\ifGPcolor
    \GPcolortrue
  }{}%
  \@ifundefined{ifGPblacktext}{%
    \newif\ifGPblacktext
    \GPblacktexttrue
  }{}%
  \let\gplgaddtomacro\g@addto@macro
  \gdef\gplbacktext{}%
  \gdef\gplfronttext{}%
  \makeatother
  \ifGPblacktext
    \def\colorrgb#1{}%
    \def\colorgray#1{}%
  \else
    \ifGPcolor
      \def\colorrgb#1{\color[rgb]{#1}}%
      \def\colorgray#1{\color[gray]{#1}}%
      \expandafter\def\csname LTw\endcsname{\color{white}}%
      \expandafter\def\csname LTb\endcsname{\color{black}}%
      \expandafter\def\csname LTa\endcsname{\color{black}}%
      \expandafter\def\csname LT0\endcsname{\color[rgb]{1,0,0}}%
      \expandafter\def\csname LT1\endcsname{\color[rgb]{0,1,0}}%
      \expandafter\def\csname LT2\endcsname{\color[rgb]{0,0,1}}%
      \expandafter\def\csname LT3\endcsname{\color[rgb]{1,0,1}}%
      \expandafter\def\csname LT4\endcsname{\color[rgb]{0,1,1}}%
      \expandafter\def\csname LT5\endcsname{\color[rgb]{1,1,0}}%
      \expandafter\def\csname LT6\endcsname{\color[rgb]{0,0,0}}%
      \expandafter\def\csname LT7\endcsname{\color[rgb]{1,0.3,0}}%
      \expandafter\def\csname LT8\endcsname{\color[rgb]{0.5,0.5,0.5}}%
    \else
      \def\colorrgb#1{\color{black}}%
      \def\colorgray#1{\color[gray]{#1}}%
      \expandafter\def\csname LTw\endcsname{\color{white}}%
      \expandafter\def\csname LTb\endcsname{\color{black}}%
      \expandafter\def\csname LTa\endcsname{\color{black}}%
      \expandafter\def\csname LT0\endcsname{\color{black}}%
      \expandafter\def\csname LT1\endcsname{\color{black}}%
      \expandafter\def\csname LT2\endcsname{\color{black}}%
      \expandafter\def\csname LT3\endcsname{\color{black}}%
      \expandafter\def\csname LT4\endcsname{\color{black}}%
      \expandafter\def\csname LT5\endcsname{\color{black}}%
      \expandafter\def\csname LT6\endcsname{\color{black}}%
      \expandafter\def\csname LT7\endcsname{\color{black}}%
      \expandafter\def\csname LT8\endcsname{\color{black}}%
    \fi
  \fi
  \setlength{\unitlength}{0.0500bp}%
  \begin{picture}(7200.00,4536.00)%
    \gplgaddtomacro\gplbacktext{%
      \csname LTb\endcsname%
      \put(1122,660){\makebox(0,0)[r]{\strut{} 0.07}}%
      \put(1122,1017){\makebox(0,0)[r]{\strut{} 0.08}}%
      \put(1122,1375){\makebox(0,0)[r]{\strut{} 0.09}}%
      \put(1122,1732){\makebox(0,0)[r]{\strut{} 0.1}}%
      \put(1122,2089){\makebox(0,0)[r]{\strut{} 0.11}}%
      \put(1122,2447){\makebox(0,0)[r]{\strut{} 0.12}}%
      \put(1122,2804){\makebox(0,0)[r]{\strut{} 0.13}}%
      \put(1122,3161){\makebox(0,0)[r]{\strut{} 0.14}}%
      \put(1122,3519){\makebox(0,0)[r]{\strut{} 0.15}}%
      \put(1122,3876){\makebox(0,0)[r]{\strut{} 0.16}}%
      \put(1254,440){\makebox(0,0){\strut{} 0}}%
      \put(1847,440){\makebox(0,0){\strut{} 0.5}}%
      \put(2440,440){\makebox(0,0){\strut{} 1}}%
      \put(3032,440){\makebox(0,0){\strut{} 1.5}}%
      \put(3625,440){\makebox(0,0){\strut{} 2}}%
      \put(4218,440){\makebox(0,0){\strut{} 2.5}}%
      \put(4811,440){\makebox(0,0){\strut{} 3}}%
      \put(5403,440){\makebox(0,0){\strut{} 3.5}}%
      \put(5996,440){\makebox(0,0){\strut{} 4}}%
      \put(6589,440){\makebox(0,0){\strut{} 4.5}}%
      \put(220,2268){\rotatebox{90}{\makebox(0,0){\strut{}$\thavg{\hat{n}_{d,\up} \hat{n}_{d,\dw}}$}}}%
      \put(4040,110){\makebox(0,0){\strut{}$\phi$}}%
      \put(4040,4206){\makebox(0,0){\strut{}Double occupancy of the Quantum dot}}%
    }%
    \gplgaddtomacro\gplfronttext{%
      \csname LTb\endcsname%
      \put(6090,3587){\makebox(0,0)[r]{\strut{}$\Delta=0.01$}}%
      \csname LTb\endcsname%
      \put(6090,3367){\makebox(0,0)[r]{\strut{}$\Delta=0.03$}}%
      \csname LTb\endcsname%
      \put(6090,3147){\makebox(0,0)[r]{\strut{}$\Delta=0.05$}}%
      \csname LTb\endcsname%
      \put(6090,2927){\makebox(0,0)[r]{\strut{}$\Delta=0.08$}}%
      \csname LTb\endcsname%
      \put(6090,2707){\makebox(0,0)[r]{\strut{}$\Delta=0.09$}}%
      \csname LTb\endcsname%
      \put(6090,2487){\makebox(0,0)[r]{\strut{}$\Delta=0.10$}}%
      \csname LTb\endcsname%
      \put(6090,2267){\makebox(0,0)[r]{\strut{}$\Delta=0.15$}}%
      \csname LTb\endcsname%
      \put(6090,2047){\makebox(0,0)[r]{\strut{}$\Delta=0.20$}}%
      \csname LTb\endcsname%
      \put(6090,1827){\makebox(0,0)[r]{\strut{}$\Delta=0.25$}}%
      \csname LTb\endcsname%
      \put(6090,1607){\makebox(0,0)[r]{\strut{}$\Delta=0.30$}}%
      \csname LTb\endcsname%
      \put(6090,1387){\makebox(0,0)[r]{\strut{}$\Delta=0.35$}}%
      \csname LTb\endcsname%
      \put(6090,1167){\makebox(0,0)[r]{\strut{}$\Delta=0.40$}}%
      \csname LTb\endcsname%
      \put(6090,947){\makebox(0,0)[r]{\strut{}$\Delta=0.50$}}%
    }%
    \gplbacktext
    \put(0,0){\includegraphics{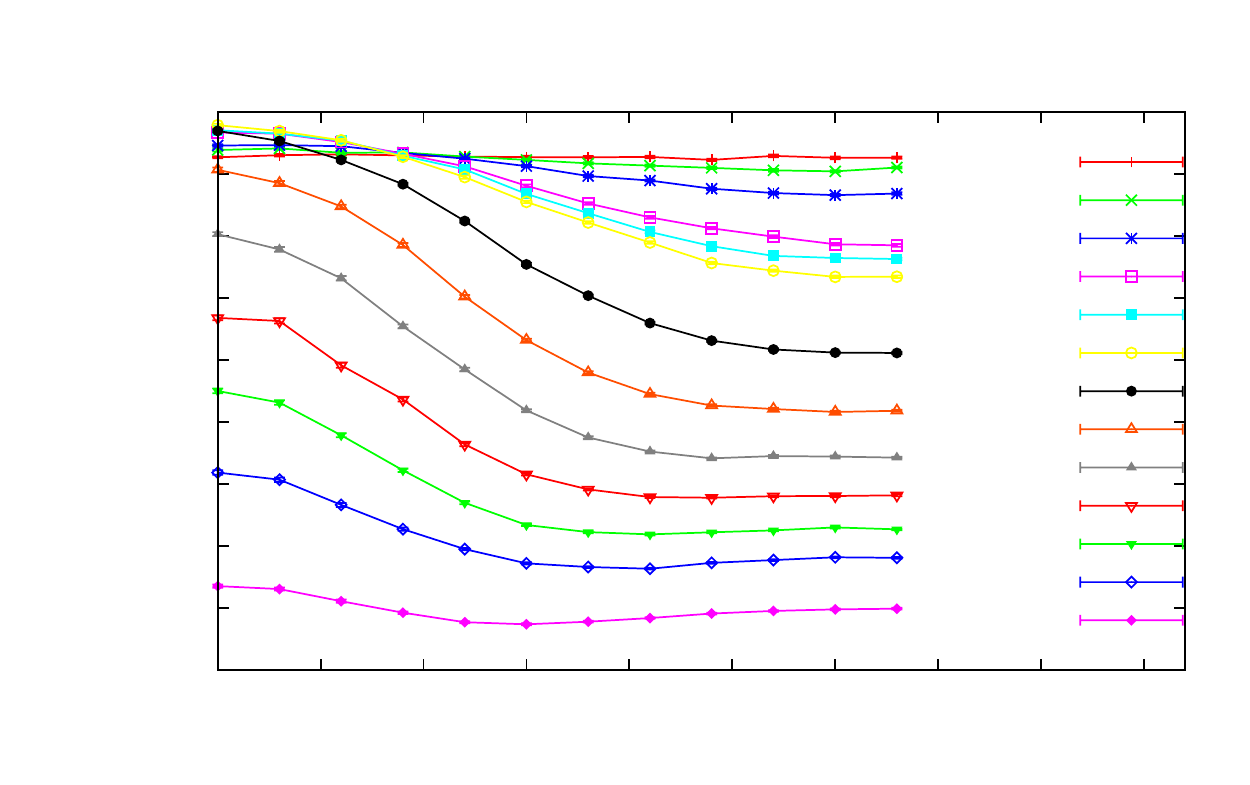}}%
    \gplfronttext
  \end{picture}%
\endgroup

%% file: tex-local_pair_correlation.tex
\begingroup
  \makeatletter
  \providecommand\color[2][]{%
    \GenericError{(gnuplot) \space\space\space\@spaces}{%
      Package color not loaded in conjunction with
      terminal option `colourtext'%
    }{See the gnuplot documentation for explanation.%
    }{Either use 'blacktext' in gnuplot or load the package
      color.sty in LaTeX.}%
    \renewcommand\color[2][]{}%
  }%
  \providecommand\includegraphics[2][]{%
    \GenericError{(gnuplot) \space\space\space\@spaces}{%
      Package graphicx or graphics not loaded%
    }{See the gnuplot documentation for explanation.%
    }{The gnuplot epslatex terminal needs graphicx.sty or graphics.sty.}%
    \renewcommand\includegraphics[2][]{}%
  }%
  \providecommand\rotatebox[2]{#2}%
  \@ifundefined{ifGPcolor}{%
    \newif\ifGPcolor
    \GPcolortrue
  }{}%
  \@ifundefined{ifGPblacktext}{%
    \newif\ifGPblacktext
    \GPblacktexttrue
  }{}%
  \let\gplgaddtomacro\g@addto@macro
  \gdef\gplbacktext{}%
  \gdef\gplfronttext{}%
  \makeatother
  \ifGPblacktext
    \def\colorrgb#1{}%
    \def\colorgray#1{}%
  \else
    \ifGPcolor
      \def\colorrgb#1{\color[rgb]{#1}}%
      \def\colorgray#1{\color[gray]{#1}}%
      \expandafter\def\csname LTw\endcsname{\color{white}}%
      \expandafter\def\csname LTb\endcsname{\color{black}}%
      \expandafter\def\csname LTa\endcsname{\color{black}}%
      \expandafter\def\csname LT0\endcsname{\color[rgb]{1,0,0}}%
      \expandafter\def\csname LT1\endcsname{\color[rgb]{0,1,0}}%
      \expandafter\def\csname LT2\endcsname{\color[rgb]{0,0,1}}%
      \expandafter\def\csname LT3\endcsname{\color[rgb]{1,0,1}}%
      \expandafter\def\csname LT4\endcsname{\color[rgb]{0,1,1}}%
      \expandafter\def\csname LT5\endcsname{\color[rgb]{1,1,0}}%
      \expandafter\def\csname LT6\endcsname{\color[rgb]{0,0,0}}%
      \expandafter\def\csname LT7\endcsname{\color[rgb]{1,0.3,0}}%
      \expandafter\def\csname LT8\endcsname{\color[rgb]{0.5,0.5,0.5}}%
    \else
      \def\colorrgb#1{\color{black}}%
      \def\colorgray#1{\color[gray]{#1}}%
      \expandafter\def\csname LTw\endcsname{\color{white}}%
      \expandafter\def\csname LTb\endcsname{\color{black}}%
      \expandafter\def\csname LTa\endcsname{\color{black}}%
      \expandafter\def\csname LT0\endcsname{\color{black}}%
      \expandafter\def\csname LT1\endcsname{\color{black}}%
      \expandafter\def\csname LT2\endcsname{\color{black}}%
      \expandafter\def\csname LT3\endcsname{\color{black}}%
      \expandafter\def\csname LT4\endcsname{\color{black}}%
      \expandafter\def\csname LT5\endcsname{\color{black}}%
      \expandafter\def\csname LT6\endcsname{\color{black}}%
      \expandafter\def\csname LT7\endcsname{\color{black}}%
      \expandafter\def\csname LT8\endcsname{\color{black}}%
    \fi
  \fi
  \setlength{\unitlength}{0.0500bp}%
  \begin{picture}(7200.00,5040.00)%
    \gplgaddtomacro\gplbacktext{%
      \csname LTb\endcsname%
      \put(1122,660){\makebox(0,0)[r]{\strut{}-0.04}}%
      \put(1122,1280){\makebox(0,0)[r]{\strut{}-0.02}}%
      \put(1122,1900){\makebox(0,0)[r]{\strut{} 0}}%
      \put(1122,2520){\makebox(0,0)[r]{\strut{} 0.02}}%
      \put(1122,3140){\makebox(0,0)[r]{\strut{} 0.04}}%
      \put(1122,3760){\makebox(0,0)[r]{\strut{} 0.06}}%
      \put(1122,4380){\makebox(0,0)[r]{\strut{} 0.08}}%
      \put(3546,440){\makebox(0,0){\strut{} 0.1}}%
      \put(6826,440){\makebox(0,0){\strut{} 1}}%
      \put(220,2520){\rotatebox{90}{\makebox(0,0){\strut{}$\langle \tilde{d}_\up^\dagger \tilde{d}_\dw^\dagger \rangle$}}}%
      \put(4040,110){\makebox(0,0){\strut{}$ \Delta $}}%
      \put(4040,4710){\makebox(0,0){\strut{}Local pairing correlation}}%
      \put(1832,970){\makebox(0,0)[l]{\strut{}$V=0.5$, $ U = 1.0$, $\mu=0$, $\epsilon_d=0$}}%
    }%
    \gplgaddtomacro\gplfronttext{%
      \csname LTb\endcsname%
      \put(5839,4207){\makebox(0,0)[r]{\strut{}$\beta=100$}}%
    }%
    \gplbacktext
    \put(0,0){\includegraphics{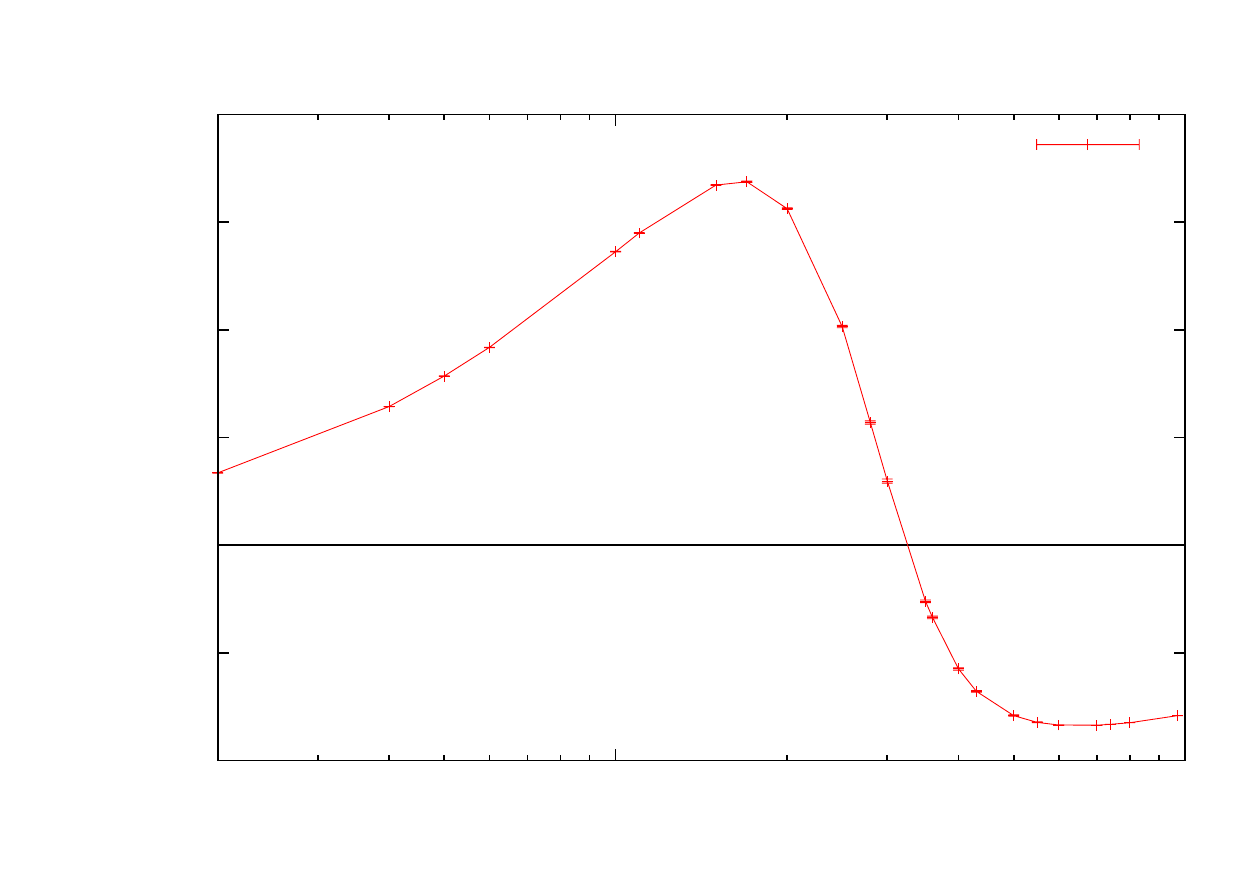}}%
    \gplfronttext
  \end{picture}%
\endgroup

%% file: tex-impurity-Aom-of-Delta-3d.tex
\begingroup
  \makeatletter
  \providecommand\color[2][]{%
    \GenericError{(gnuplot) \space\space\space\@spaces}{%
      Package color not loaded in conjunction with
      terminal option `colourtext'%
    }{See the gnuplot documentation for explanation.%
    }{Either use 'blacktext' in gnuplot or load the package
      color.sty in LaTeX.}%
    \renewcommand\color[2][]{}%
  }%
  \providecommand\includegraphics[2][]{%
    \GenericError{(gnuplot) \space\space\space\@spaces}{%
      Package graphicx or graphics not loaded%
    }{See the gnuplot documentation for explanation.%
    }{The gnuplot epslatex terminal needs graphicx.sty or graphics.sty.}%
    \renewcommand\includegraphics[2][]{}%
  }%
  \providecommand\rotatebox[2]{#2}%
  \@ifundefined{ifGPcolor}{%
    \newif\ifGPcolor
    \GPcolortrue
  }{}%
  \@ifundefined{ifGPblacktext}{%
    \newif\ifGPblacktext
    \GPblacktexttrue
  }{}%
  \let\gplgaddtomacro\g@addto@macro
  \gdef\gplbacktext{}%
  \gdef\gplfronttext{}%
  \makeatother
  \ifGPblacktext
    \def\colorrgb#1{}%
    \def\colorgray#1{}%
  \else
    \ifGPcolor
      \def\colorrgb#1{\color[rgb]{#1}}%
      \def\colorgray#1{\color[gray]{#1}}%
      \expandafter\def\csname LTw\endcsname{\color{white}}%
      \expandafter\def\csname LTb\endcsname{\color{black}}%
      \expandafter\def\csname LTa\endcsname{\color{black}}%
      \expandafter\def\csname LT0\endcsname{\color[rgb]{1,0,0}}%
      \expandafter\def\csname LT1\endcsname{\color[rgb]{0,1,0}}%
      \expandafter\def\csname LT2\endcsname{\color[rgb]{0,0,1}}%
      \expandafter\def\csname LT3\endcsname{\color[rgb]{1,0,1}}%
      \expandafter\def\csname LT4\endcsname{\color[rgb]{0,1,1}}%
      \expandafter\def\csname LT5\endcsname{\color[rgb]{1,1,0}}%
      \expandafter\def\csname LT6\endcsname{\color[rgb]{0,0,0}}%
      \expandafter\def\csname LT7\endcsname{\color[rgb]{1,0.3,0}}%
      \expandafter\def\csname LT8\endcsname{\color[rgb]{0.5,0.5,0.5}}%
    \else
      \def\colorrgb#1{\color{black}}%
      \def\colorgray#1{\color[gray]{#1}}%
      \expandafter\def\csname LTw\endcsname{\color{white}}%
      \expandafter\def\csname LTb\endcsname{\color{black}}%
      \expandafter\def\csname LTa\endcsname{\color{black}}%
      \expandafter\def\csname LT0\endcsname{\color{black}}%
      \expandafter\def\csname LT1\endcsname{\color{black}}%
      \expandafter\def\csname LT2\endcsname{\color{black}}%
      \expandafter\def\csname LT3\endcsname{\color{black}}%
      \expandafter\def\csname LT4\endcsname{\color{black}}%
      \expandafter\def\csname LT5\endcsname{\color{black}}%
      \expandafter\def\csname LT6\endcsname{\color{black}}%
      \expandafter\def\csname LT7\endcsname{\color{black}}%
      \expandafter\def\csname LT8\endcsname{\color{black}}%
    \fi
  \fi
  \setlength{\unitlength}{0.0500bp}%
  \begin{picture}(7200.00,7560.00)%
    \gplgaddtomacro\gplbacktext{%
      \csname LTb\endcsname%
      \put(3599,6508){\makebox(0,0){\strut{}Spectral function}}%
    }%
    \gplgaddtomacro\gplfronttext{%
      \csname LTb\endcsname%
      \put(6018,6616){\makebox(0,0)[r]{\strut{}$A(\omega)$}}%
      \color{white}%
      \color{white}%
      \color{white}%
      \color{white}%
      \color{black}%
      \color{black}%
      \color{black}%
      \color{black}%
      \color{black}%
      \color{black}%
      \color{black}%
      \color{black}%
      \color{black}%
      \color{black}%
      \color{black}%
      \color{black}%
      \color{black}%
      \color{black}%
      \color{black}%
      \color{black}%
      \color{black}%
      \color{black}%
      \color{black}%
      \color{black}%
      \color{black}%
      \color{black}%
      \color{black}%
      \color{black}%
      \color{black}%
      \color{black}%
      \color{black}%
      \color{black}%
      \color{black}%
      \color{black}%
      \color{black}%
      \color{black}%
      \color{black}%
      \color{black}%
      \color{black}%
      \color{black}%
      \color{black}%
      \color{black}%
      \color{black}%
      \color{black}%
      \color{black}%
      \color{black}%
      \color{black}%
      \color{black}%
      \color{black}%
      \color{black}%
      \color{black}%
      \color{black}%
      \color{black}%
      \color{black}%
      \color{white}%
      \color{white}%
      \color{black}%
      \color{black}%
      \color{black}%
      \color{black}%
      \color{black}%
      \color{black}%
      \color{black}%
      \color{black}%
      \color{black}%
      \color{black}%
      \color{black}%
      \color{black}%
      \color{black}%
      \color{black}%
      \color{black}%
      \color{black}%
      \color{black}%
      \color{black}%
      \color{black}%
      \color{black}%
      \color{black}%
      \color{black}%
      \color{black}%
      \color{black}%
      \color{black}%
      \color{black}%
      \color{black}%
      \color{black}%
      \color{white}%
      \color{black}%
      \color{black}%
      \color{black}%
      \color{black}%
      \color{black}%
      \color{black}%
      \color{black}%
      \color{black}%
      \color{black}%
      \color{black}%
      \color{black}%
      \color{black}%
      \color{black}%
      \color{black}%
      \color{black}%
      \color{black}%
      \color{black}%
      \color{black}%
      \color{black}%
      \color{black}%
      \color{black}%
      \color{black}%
      \color{black}%
      \color{black}%
      \color{black}%
      \color{black}%
      \color{black}%
      \color{black}%
      \color{black}%
      \color{black}%
      \color{black}%
      \color{black}%
      \color{black}%
      \color{black}%
      \color{black}%
      \color{black}%
      \color{black}%
      \color{black}%
      \color{black}%
      \color{black}%
      \color{black}%
      \color{black}%
      \color{black}%
      \color{black}%
      \color{black}%
      \color{black}%
      \color{black}%
      \color{black}%
      \color{black}%
      \color{black}%
      \color{black}%
      \color{black}%
      \color{black}%
      \color{black}%
      \color{black}%
      \color{black}%
      \color{black}%
      \color{black}%
      \color{black}%
      \color{black}%
      \color{black}%
      \color{black}%
      \color{black}%
      \color{black}%
      \color{black}%
      \color{black}%
      \color{black}%
      \color{black}%
      \color{black}%
      \color{black}%
      \color{black}%
      \color{black}%
      \color{black}%
      \color{black}%
      \color{black}%
      \color{black}%
      \color{black}%
      \color{black}%
      \color{black}%
      \color{black}%
      \color{black}%
      \color{black}%
      \color{black}%
      \color{black}%
      \color{black}%
      \color{black}%
      \color{black}%
      \color{black}%
      \color{black}%
      \color{black}%
      \color{black}%
      \color{black}%
      \color{black}%
      \color{black}%
      \color{black}%
      \color{black}%
      \color{black}%
      \color{black}%
      \color{black}%
      \color{black}%
      \color{black}%
      \color{black}%
      \color{black}%
      \color{black}%
      \color{black}%
      \color{black}%
      \color{black}%
      \color{black}%
      \color{black}%
      \color{black}%
      \color{black}%
      \color{black}%
      \color{black}%
      \color{black}%
      \color{black}%
      \color{black}%
      \color{black}%
      \color{black}%
      \color{black}%
      \color{black}%
      \color{black}%
      \color{black}%
      \color{black}%
      \color{black}%
      \color{black}%
      \color{black}%
      \color{black}%
      \color{black}%
      \color{black}%
      \color{black}%
      \color{black}%
      \color{black}%
      \color{black}%
      \color{black}%
      \color{black}%
      \color{black}%
      \color{black}%
      \color{black}%
      \color{black}%
      \color{black}%
      \color{black}%
      \color{black}%
      \color{black}%
      \color{black}%
      \color{black}%
      \color{black}%
      \color{black}%
      \color{black}%
      \color{black}%
      \color{black}%
      \color{black}%
      \color{black}%
      \color{black}%
      \color{black}%
      \color{black}%
      \color{black}%
      \color{black}%
      \color{black}%
      \color{black}%
      \color{black}%
      \color{black}%
      \color{black}%
      \color{black}%
      \color{black}%
      \color{black}%
      \color{black}%
      \color{black}%
      \color{black}%
      \color{black}%
      \color{black}%
      \color{black}%
      \color{black}%
      \color{black}%
      \color{black}%
      \color{black}%
      \color{black}%
      \color{black}%
      \color{black}%
      \color{black}%
      \color{black}%
      \color{black}%
      \color{black}%
      \color{black}%
      \color{black}%
      \color{black}%
      \color{black}%
      \color{black}%
      \color{black}%
      \color{black}%
      \color{black}%
      \color{black}%
      \color{black}%
      \color{black}%
      \color{black}%
      \color{black}%
      \color{black}%
      \color{black}%
      \color{black}%
      \color{black}%
      \color{black}%
      \color{black}%
      \color{black}%
      \color{black}%
      \color{black}%
      \color{black}%
      \color{black}%
      \color{black}%
      \color{black}%
      \color{black}%
      \color{black}%
      \color{black}%
      \color{black}%
      \color{black}%
      \color{black}%
      \color{black}%
      \color{black}%
      \color{black}%
      \color{black}%
      \color{black}%
      \color{black}%
      \color{black}%
      \color{black}%
      \color{black}%
      \color{black}%
      \color{black}%
      \color{black}%
      \color{black}%
      \color{black}%
      \color{black}%
      \color{black}%
      \color{black}%
      \color{black}%
      \color{black}%
      \color{black}%
      \color{black}%
      \color{black}%
      \color{black}%
      \color{black}%
      \color{black}%
      \color{black}%
      \color{black}%
      \color{black}%
      \color{black}%
      \color{black}%
      \color{black}%
      \color{black}%
      \color{black}%
      \color{black}%
      \color{black}%
      \color{black}%
      \color{black}%
      \color{black}%
      \color{black}%
      \color{black}%
      \color{black}%
      \color{black}%
      \color{black}%
      \color{black}%
      \color{black}%
      \color{black}%
      \color{black}%
      \color{black}%
      \color{black}%
      \color{black}%
      \color{black}%
      \color{black}%
      \color{black}%
      \color{black}%
      \color{black}%
      \color{black}%
      \color{black}%
      \color{black}%
      \color{black}%
      \color{black}%
      \color{black}%
      \color{black}%
      \color{black}%
      \color{black}%
      \color{black}%
      \color{black}%
      \color{black}%
      \color{black}%
      \color{black}%
      \color{black}%
      \color{black}%
      \color{black}%
      \color{black}%
      \color{black}%
      \color{black}%
      \color{black}%
      \color{black}%
      \color{black}%
      \color{black}%
      \color{black}%
      \color{black}%
      \color{black}%
      \color{black}%
      \color{black}%
      \color{black}%
      \color{black}%
      \color{black}%
      \color{black}%
      \color{black}%
      \color{black}%
      \color{black}%
      \color{black}%
      \color{black}%
      \color{black}%
      \color{black}%
      \color{black}%
      \color{black}%
      \color{black}%
      \color{black}%
      \color{black}%
      \color{black}%
      \color{black}%
      \color{black}%
      \color{black}%
      \color{black}%
      \color{black}%
      \color{black}%
      \color{black}%
      \color{black}%
      \color{black}%
      \color{black}%
      \color{black}%
      \color{black}%
      \color{black}%
      \color{black}%
      \color{black}%
      \color{black}%
      \color{black}%
      \color{black}%
      \color{black}%
      \color{black}%
      \color{black}%
      \color{black}%
      \color{black}%
      \color{black}%
      \color{black}%
      \color{black}%
      \color{black}%
      \color{black}%
      \color{black}%
      \color{black}%
      \color{black}%
      \color{black}%
      \color{black}%
      \color{black}%
      \color{black}%
      \color{black}%
      \color{black}%
      \color{black}%
      \color{black}%
      \color{black}%
      \color{black}%
      \color{black}%
      \color{black}%
      \color{black}%
      \color{black}%
      \color{black}%
      \color{black}%
      \color{black}%
      \color{black}%
      \color{black}%
      \color{black}%
      \color{black}%
      \color{black}%
      \color{black}%
      \color{black}%
      \color{black}%
      \color{black}%
      \color{black}%
      \color{black}%
      \color{black}%
      \color{black}%
      \color{black}%
      \color{black}%
      \color{black}%
      \color{black}%
      \color{black}%
      \color{black}%
      \color{black}%
      \color{black}%
      \color{black}%
      \color{black}%
      \color{black}%
      \color{black}%
      \color{black}%
      \color{black}%
      \color{black}%
      \color{black}%
      \color{black}%
      \color{black}%
      \color{black}%
      \color{black}%
      \color{black}%
      \color{black}%
      \color{black}%
      \color{black}%
      \color{black}%
      \color{black}%
      \color{black}%
      \color{black}%
      \color{black}%
      \color{black}%
      \color{black}%
      \color{black}%
      \color{black}%
      \color{black}%
      \color{black}%
      \color{black}%
      \color{black}%
      \color{black}%
      \color{black}%
      \color{black}%
      \color{black}%
      \color{black}%
      \color{black}%
      \color{black}%
      \color{black}%
      \color{black}%
      \color{black}%
      \color{black}%
      \color{black}%
      \color{black}%
      \color{black}%
      \color{black}%
      \color{black}%
      \color{black}%
      \color{black}%
      \color{black}%
      \color{black}%
      \color{black}%
      \color{black}%
      \color{black}%
      \color{black}%
      \color{black}%
      \color{black}%
      \color{black}%
      \color{black}%
      \color{black}%
      \color{black}%
      \color{black}%
      \color{black}%
      \color{black}%
      \color{black}%
      \color{black}%
      \color{black}%
      \color{black}%
      \color{black}%
      \color{black}%
      \color{black}%
      \color{black}%
      \color{black}%
      \color{black}%
      \color{black}%
      \color{black}%
      \color{white}%
      \color{white}%
      \color{black}%
      \color{black}%
      \color{black}%
      \color{black}%
      \color{black}%
      \color{black}%
      \color{black}%
      \color{black}%
      \color{black}%
      \color{black}%
      \color{black}%
      \color{black}%
      \color{black}%
      \color{black}%
      \color{black}%
      \color{black}%
      \color{black}%
      \color{black}%
      \color{black}%
      \color{black}%
      \color{black}%
      \color{black}%
      \color{black}%
      \color{black}%
      \color{black}%
      \color{black}%
      \color{black}%
      \color{black}%
      \color{black}%
      \color{black}%
      \color{black}%
      \color{black}%
      \color{black}%
      \color{black}%
      \color{black}%
      \color{black}%
      \color{black}%
      \color{black}%
      \color{black}%
      \color{black}%
      \color{black}%
      \color{black}%
      \color{black}%
      \color{black}%
      \color{black}%
      \color{black}%
      \color{black}%
      \color{black}%
      \color{black}%
      \color{black}%
      \color{white}%
      \color{white}%
      \color{black}%
      \color{black}%
      \color{black}%
      \color{black}%
      \color{black}%
      \color{black}%
      \color{black}%
      \color{black}%
      \color{black}%
      \color{black}%
      \color{black}%
      \color{black}%
      \color{black}%
      \color{black}%
      \color{black}%
      \color{black}%
      \color{black}%
      \color{black}%
      \color{black}%
      \color{black}%
      \color{black}%
      \color{black}%
      \color{black}%
      \color{black}%
      \color{black}%
      \color{black}%
      \color{black}%
      \color{black}%
      \color{white}%
      \color{white}%
      \color{white}%
      \color{black}%
      \color{black}%
      \color{black}%
      \color{black}%
      \color{black}%
      \color{black}%
      \color{black}%
      \color{black}%
      \color{black}%
      \color{black}%
      \color{black}%
      \color{black}%
      \color{black}%
      \color{black}%
      \color{black}%
      \color{black}%
      \color{black}%
      \color{black}%
      \color{black}%
      \color{black}%
      \color{black}%
      \color{black}%
      \color{black}%
      \color{black}%
      \color{black}%
      \color{black}%
      \color{black}%
      \color{black}%
      \color{black}%
      \color{black}%
      \color{black}%
      \color{black}%
      \color{black}%
      \color{black}%
      \color{black}%
      \color{black}%
      \color{black}%
      \color{black}%
      \color{black}%
      \color{black}%
      \color{black}%
      \color{black}%
      \color{black}%
      \color{black}%
      \color{black}%
      \color{black}%
      \color{black}%
      \color{black}%
      \color{black}%
      \color{black}%
      \color{black}%
      \color{black}%
      \color{black}%
      \color{black}%
      \color{black}%
      \color{black}%
      \color{black}%
      \color{black}%
      \color{black}%
      \color{black}%
      \color{black}%
      \color{black}%
      \color{black}%
      \color{black}%
      \color{black}%
      \color{black}%
      \color{black}%
      \color{black}%
      \color{black}%
      \color{black}%
      \color{black}%
      \color{black}%
      \color{black}%
      \color{black}%
      \color{black}%
      \color{black}%
      \color{black}%
      \color{black}%
      \color{black}%
      \color{black}%
      \color{black}%
      \color{black}%
      \color{black}%
      \color{black}%
      \color{black}%
      \color{black}%
      \color{black}%
      \color{black}%
      \color{black}%
      \color{black}%
      \color{black}%
      \color{black}%
      \color{black}%
      \color{black}%
      \color{black}%
      \color{black}%
      \color{black}%
      \color{black}%
      \color{black}%
      \color{black}%
      \color{black}%
      \color{black}%
      \color{black}%
      \color{black}%
      \color{black}%
      \color{black}%
      \color{black}%
      \color{black}%
      \color{black}%
      \color{black}%
      \color{black}%
      \color{black}%
      \color{black}%
      \color{black}%
      \color{black}%
      \color{black}%
      \color{black}%
      \color{black}%
      \color{black}%
      \color{black}%
      \color{black}%
      \color{black}%
      \color{black}%
      \color{black}%
      \color{black}%
      \color{black}%
      \color{black}%
      \color{black}%
      \color{black}%
      \color{black}%
      \color{black}%
      \color{black}%
      \color{black}%
      \color{black}%
      \color{black}%
      \color{black}%
      \color{black}%
      \color{black}%
      \color{black}%
      \color{black}%
      \color{black}%
      \color{black}%
      \color{black}%
      \color{black}%
      \color{black}%
      \color{black}%
      \color{black}%
      \color{black}%
      \color{black}%
      \color{black}%
      \color{black}%
      \color{black}%
      \color{black}%
      \color{black}%
      \color{black}%
      \color{black}%
      \color{black}%
      \color{black}%
      \color{black}%
      \color{black}%
      \color{black}%
      \color{black}%
      \color{black}%
      \color{black}%
      \color{black}%
      \color{black}%
      \color{black}%
      \color{black}%
      \color{black}%
      \color{black}%
      \color{black}%
      \color{black}%
      \color{black}%
      \color{black}%
      \color{black}%
      \color{black}%
      \color{black}%
      \color{black}%
      \color{black}%
      \color{black}%
      \color{black}%
      \color{black}%
      \color{black}%
      \color{black}%
      \color{black}%
      \color{black}%
      \color{black}%
      \color{black}%
      \color{black}%
      \color{black}%
      \color{black}%
      \color{black}%
      \color{black}%
      \color{black}%
      \color{black}%
      \color{black}%
      \color{black}%
      \color{black}%
      \color{black}%
      \color{black}%
      \color{black}%
      \color{black}%
      \color{black}%
      \color{black}%
      \color{black}%
      \color{black}%
      \color{black}%
      \color{black}%
      \color{black}%
      \color{black}%
      \color{black}%
      \color{black}%
      \color{black}%
      \color{black}%
      \color{black}%
      \color{black}%
      \color{black}%
      \color{black}%
      \color{black}%
      \color{black}%
      \color{black}%
      \color{black}%
      \color{black}%
      \color{black}%
      \color{black}%
      \color{black}%
      \color{black}%
      \color{black}%
      \color{black}%
      \color{black}%
      \color{black}%
      \color{black}%
      \color{black}%
      \color{black}%
      \color{black}%
      \color{black}%
      \color{black}%
      \color{black}%
      \color{black}%
      \color{black}%
      \color{black}%
      \color{black}%
      \color{black}%
      \color{black}%
      \color{black}%
      \color{black}%
      \color{black}%
      \color{black}%
      \color{black}%
      \color{black}%
      \color{black}%
      \color{black}%
      \color{black}%
      \color{black}%
      \color{black}%
      \color{black}%
      \color{black}%
      \color{black}%
      \color{black}%
      \color{black}%
      \color{black}%
      \color{black}%
      \color{black}%
      \color{black}%
      \color{black}%
      \color{black}%
      \color{black}%
      \color{black}%
      \color{black}%
      \color{black}%
      \color{black}%
      \color{black}%
      \color{black}%
      \color{black}%
      \color{black}%
      \color{black}%
      \color{black}%
      \color{black}%
      \color{black}%
      \color{black}%
      \color{black}%
      \color{black}%
      \color{black}%
      \color{black}%
      \color{black}%
      \color{black}%
      \color{black}%
      \color{black}%
      \color{black}%
      \color{black}%
      \color{black}%
      \color{black}%
      \color{black}%
      \color{black}%
      \color{black}%
      \color{black}%
      \color{black}%
      \color{black}%
      \color{black}%
      \color{black}%
      \color{black}%
      \color{black}%
      \color{black}%
      \color{black}%
      \color{black}%
      \color{black}%
      \color{black}%
      \color{black}%
      \color{black}%
      \color{black}%
      \color{black}%
      \color{black}%
      \color{black}%
      \color{black}%
      \color{black}%
      \color{black}%
      \color{black}%
      \color{black}%
      \color{black}%
      \color{black}%
      \color{black}%
      \color{black}%
      \color{black}%
      \color{black}%
      \color{black}%
      \color{black}%
      \color{black}%
      \color{black}%
      \color{black}%
      \color{black}%
      \color{black}%
      \color{black}%
      \color{black}%
      \color{black}%
      \color{black}%
      \color{black}%
      \color{black}%
      \color{black}%
      \color{black}%
      \color{black}%
      \color{black}%
      \color{black}%
      \color{black}%
      \color{black}%
      \color{black}%
      \color{black}%
      \color{black}%
      \color{black}%
      \color{black}%
      \color{black}%
      \color{black}%
      \color{black}%
      \color{black}%
      \color{black}%
      \color{black}%
      \color{black}%
      \color{black}%
      \color{black}%
      \color{black}%
      \color{black}%
      \color{black}%
      \color{black}%
      \color{black}%
      \color{black}%
      \color{black}%
      \color{black}%
      \color{black}%
      \color{black}%
      \color{black}%
      \color{black}%
      \color{black}%
      \color{black}%
      \color{black}%
      \color{black}%
      \color{black}%
      \color{black}%
      \color{black}%
      \color{black}%
      \color{black}%
      \color{black}%
      \color{black}%
      \color{black}%
      \color{black}%
      \color{black}%
      \color{black}%
      \color{black}%
      \color{black}%
      \color{black}%
      \color{black}%
      \color{black}%
      \color{black}%
      \color{black}%
      \color{black}%
      \color{black}%
      \color{black}%
      \color{black}%
      \color{black}%
      \color{black}%
      \color{black}%
      \color{black}%
      \color{black}%
      \color{black}%
      \color{black}%
      \color{black}%
      \color{black}%
      \color{black}%
      \color{black}%
      \color{black}%
      \color{black}%
      \color{black}%
      \color{black}%
      \color{black}%
      \color{black}%
      \color{black}%
      \color{black}%
      \color{black}%
      \color{black}%
      \color{black}%
      \color{black}%
      \color{black}%
      \color{black}%
      \color{black}%
      \color{black}%
      \color{black}%
      \color{black}%
      \color{black}%
      \color{black}%
      \color{black}%
      \color{black}%
      \color{black}%
      \color{black}%
      \color{black}%
      \color{black}%
      \color{black}%
      \color{black}%
      \color{black}%
      \color{black}%
      \color{black}%
      \color{black}%
      \color{black}%
      \color{black}%
      \color{black}%
      \color{black}%
      \color{black}%
      \color{black}%
      \color{black}%
      \color{black}%
      \color{black}%
      \color{black}%
      \color{black}%
      \color{black}%
      \color{black}%
      \color{black}%
      \color{black}%
      \color{black}%
      \color{black}%
      \color{black}%
      \color{black}%
      \csname LTb\endcsname%
      \put(1170,1096){\makebox(0,0){\strut{} 0}}%
      \put(1777,1096){\makebox(0,0){\strut{} 0.1}}%
      \put(2385,1096){\makebox(0,0){\strut{} 0.2}}%
      \put(2993,1096){\makebox(0,0){\strut{} 0.3}}%
      \put(3600,1096){\makebox(0,0){\strut{} 0.4}}%
      \put(4207,1096){\makebox(0,0){\strut{} 0.5}}%
      \put(4815,1096){\makebox(0,0){\strut{} 0.6}}%
      \put(5423,1096){\makebox(0,0){\strut{} 0.7}}%
      \put(6030,1096){\makebox(0,0){\strut{} 0.8}}%
      \put(3600,766){\makebox(0,0){\strut{}$\Delta$}}%
      \put(998,1382){\makebox(0,0)[r]{\strut{}-1}}%
      \put(998,2581){\makebox(0,0)[r]{\strut{}-0.5}}%
      \put(998,3780){\makebox(0,0)[r]{\strut{} 0}}%
      \put(998,4979){\makebox(0,0)[r]{\strut{} 0.5}}%
      \put(998,6178){\makebox(0,0)[r]{\strut{} 1}}%
      \put(404,3780){\rotatebox{90}{\makebox(0,0){\strut{}$\omega$}}}%
      \put(6527,1380){\makebox(0,0)[l]{\strut{} 0.1}}%
      \put(6527,3779){\makebox(0,0)[l]{\strut{} 1}}%
      \put(6527,6178){\makebox(0,0)[l]{\strut{} 10}}%
    }%
    \gplbacktext
    \put(0,0){\includegraphics{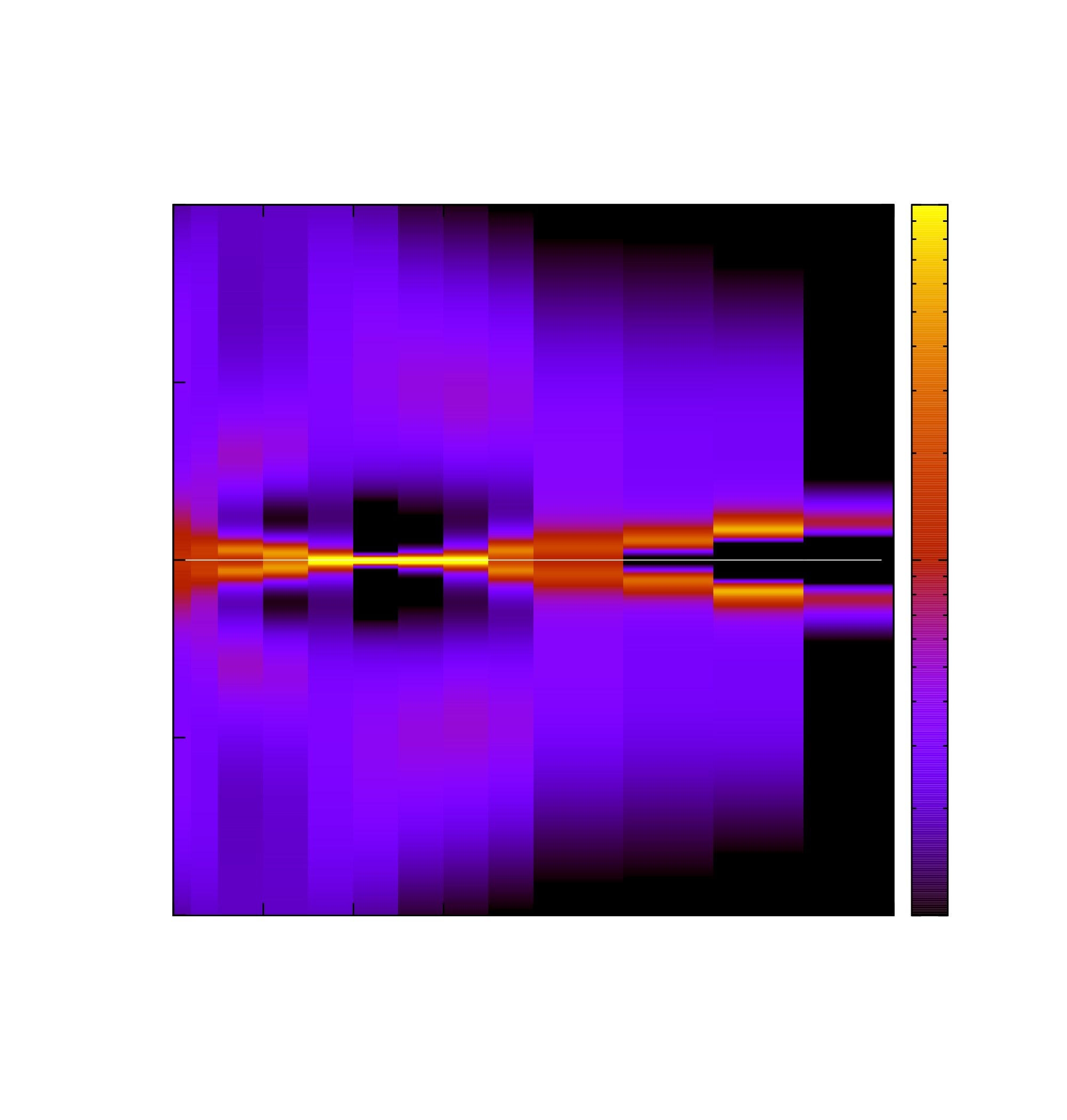}}%
    \gplfronttext
  \end{picture}%
\endgroup

%% file: tex-impurity-Som-of-Delta.tex
\begingroup
  \makeatletter
  \providecommand\color[2][]{%
    \GenericError{(gnuplot) \space\space\space\@spaces}{%
      Package color not loaded in conjunction with
      terminal option `colourtext'%
    }{See the gnuplot documentation for explanation.%
    }{Either use 'blacktext' in gnuplot or load the package
      color.sty in LaTeX.}%
    \renewcommand\color[2][]{}%
  }%
  \providecommand\includegraphics[2][]{%
    \GenericError{(gnuplot) \space\space\space\@spaces}{%
      Package graphicx or graphics not loaded%
    }{See the gnuplot documentation for explanation.%
    }{The gnuplot epslatex terminal needs graphicx.sty or graphics.sty.}%
    \renewcommand\includegraphics[2][]{}%
  }%
  \providecommand\rotatebox[2]{#2}%
  \@ifundefined{ifGPcolor}{%
    \newif\ifGPcolor
    \GPcolortrue
  }{}%
  \@ifundefined{ifGPblacktext}{%
    \newif\ifGPblacktext
    \GPblacktexttrue
  }{}%
  \let\gplgaddtomacro\g@addto@macro
  \gdef\gplbacktext{}%
  \gdef\gplfronttext{}%
  \makeatother
  \ifGPblacktext
    \def\colorrgb#1{}%
    \def\colorgray#1{}%
  \else
    \ifGPcolor
      \def\colorrgb#1{\color[rgb]{#1}}%
      \def\colorgray#1{\color[gray]{#1}}%
      \expandafter\def\csname LTw\endcsname{\color{white}}%
      \expandafter\def\csname LTb\endcsname{\color{black}}%
      \expandafter\def\csname LTa\endcsname{\color{black}}%
      \expandafter\def\csname LT0\endcsname{\color[rgb]{1,0,0}}%
      \expandafter\def\csname LT1\endcsname{\color[rgb]{0,1,0}}%
      \expandafter\def\csname LT2\endcsname{\color[rgb]{0,0,1}}%
      \expandafter\def\csname LT3\endcsname{\color[rgb]{1,0,1}}%
      \expandafter\def\csname LT4\endcsname{\color[rgb]{0,1,1}}%
      \expandafter\def\csname LT5\endcsname{\color[rgb]{1,1,0}}%
      \expandafter\def\csname LT6\endcsname{\color[rgb]{0,0,0}}%
      \expandafter\def\csname LT7\endcsname{\color[rgb]{1,0.3,0}}%
      \expandafter\def\csname LT8\endcsname{\color[rgb]{0.5,0.5,0.5}}%
    \else
      \def\colorrgb#1{\color{black}}%
      \def\colorgray#1{\color[gray]{#1}}%
      \expandafter\def\csname LTw\endcsname{\color{white}}%
      \expandafter\def\csname LTb\endcsname{\color{black}}%
      \expandafter\def\csname LTa\endcsname{\color{black}}%
      \expandafter\def\csname LT0\endcsname{\color{black}}%
      \expandafter\def\csname LT1\endcsname{\color{black}}%
      \expandafter\def\csname LT2\endcsname{\color{black}}%
      \expandafter\def\csname LT3\endcsname{\color{black}}%
      \expandafter\def\csname LT4\endcsname{\color{black}}%
      \expandafter\def\csname LT5\endcsname{\color{black}}%
      \expandafter\def\csname LT6\endcsname{\color{black}}%
      \expandafter\def\csname LT7\endcsname{\color{black}}%
      \expandafter\def\csname LT8\endcsname{\color{black}}%
    \fi
  \fi
  \setlength{\unitlength}{0.0500bp}%
  \begin{picture}(7200.00,8064.00)%
    \gplgaddtomacro\gplbacktext{%
      \csname LTb\endcsname%
      \put(529,22){\makebox(0,0)[r]{\strut{} 0}}%
      \put(529,212){\makebox(0,0)[r]{\strut{} 1}}%
      \put(529,401){\makebox(0,0)[r]{\strut{} 2}}%
      \put(529,591){\makebox(0,0)[r]{\strut{} 3}}%
      \put(529,781){\makebox(0,0)[r]{\strut{} 4}}%
      \put(529,970){\makebox(0,0)[r]{\strut{} 5}}%
      \put(661,-198){\makebox(0,0){\strut{} 0}}%
      \put(1585,-198){\makebox(0,0){\strut{} 0.1}}%
      \put(2509,-198){\makebox(0,0){\strut{} 0.2}}%
      \put(3433,-198){\makebox(0,0){\strut{} 0.3}}%
      \put(4356,-198){\makebox(0,0){\strut{} 0.4}}%
      \put(5280,-198){\makebox(0,0){\strut{} 0.5}}%
      \put(6204,-198){\makebox(0,0){\strut{} 0.6}}%
      \put(7128,-198){\makebox(0,0){\strut{} 0.7}}%
      \put(23,641){\rotatebox{90}{\makebox(0,0){\strut{}$S(\omega)$}}}%
      \put(3894,-528){\makebox(0,0){\strut{}$\omega$}}%
      \put(1308,117){\makebox(0,0)[l]{\strut{}$T_K$}}%
    }%
    \gplgaddtomacro\gplfronttext{%
      \csname LTb\endcsname%
      \put(6141,1087){\makebox(0,0)[r]{\strut{}$\Delta = 0$}}%
    }%
    \gplgaddtomacro\gplbacktext{%
      \csname LTb\endcsname%
      \put(529,1260){\makebox(0,0)[r]{\strut{} 0}}%
      \put(529,1543){\makebox(0,0)[r]{\strut{} 1}}%
      \put(529,1826){\makebox(0,0)[r]{\strut{} 2}}%
      \put(529,2109){\makebox(0,0)[r]{\strut{} 3}}%
    }%
    \gplgaddtomacro\gplfronttext{%
      \csname LTb\endcsname%
      \put(6141,2347){\makebox(0,0)[r]{\strut{}$\Delta= 0.02$}}%
    }%
    \gplgaddtomacro\gplbacktext{%
      \csname LTb\endcsname%
      \put(529,2520){\makebox(0,0)[r]{\strut{} 0}}%
      \put(529,2787){\makebox(0,0)[r]{\strut{} 1}}%
      \put(529,3053){\makebox(0,0)[r]{\strut{} 2}}%
      \put(529,3320){\makebox(0,0)[r]{\strut{} 3}}%
    }%
    \gplgaddtomacro\gplfronttext{%
      \csname LTb\endcsname%
      \put(6141,3607){\makebox(0,0)[r]{\strut{}$\Delta= 0.05$}}%
    }%
    \gplgaddtomacro\gplbacktext{%
      \csname LTb\endcsname%
      \put(529,3780){\makebox(0,0)[r]{\strut{} 0}}%
      \put(529,4035){\makebox(0,0)[r]{\strut{} 3}}%
      \put(529,4290){\makebox(0,0)[r]{\strut{} 6}}%
      \put(529,4544){\makebox(0,0)[r]{\strut{} 9}}%
    }%
    \gplgaddtomacro\gplfronttext{%
      \csname LTb\endcsname%
      \put(6141,4867){\makebox(0,0)[r]{\strut{}$\Delta= 0.1$}}%
    }%
    \gplgaddtomacro\gplbacktext{%
      \csname LTb\endcsname%
      \put(529,5040){\makebox(0,0)[r]{\strut{} 0}}%
      \put(529,5284){\makebox(0,0)[r]{\strut{} 20}}%
      \put(529,5528){\makebox(0,0)[r]{\strut{} 40}}%
      \put(529,5772){\makebox(0,0)[r]{\strut{} 60}}%
    }%
    \gplgaddtomacro\gplfronttext{%
      \csname LTb\endcsname%
      \put(6141,6127){\makebox(0,0)[r]{\strut{}$\Delta= 0.2 $}}%
    }%
    \gplgaddtomacro\gplbacktext{%
      \csname LTb\endcsname%
      \put(529,6300){\makebox(0,0)[r]{\strut{} 0}}%
      \put(529,6489){\makebox(0,0)[r]{\strut{} 40}}%
      \put(529,6678){\makebox(0,0)[r]{\strut{} 80}}%
      \put(529,6867){\makebox(0,0)[r]{\strut{} 120}}%
      \put(529,7056){\makebox(0,0)[r]{\strut{} 160}}%
      \put(529,7245){\makebox(0,0)[r]{\strut{} 200}}%
      \put(3894,7890){\makebox(0,0){\strut{}Dynamical spin structure factor}}%
    }%
    \gplgaddtomacro\gplfronttext{%
      \csname LTb\endcsname%
      \put(6141,7387){\makebox(0,0)[r]{\strut{}$\Delta= 0.3 $}}%
    }%
    \gplbacktext
    \put(0,0){\includegraphics{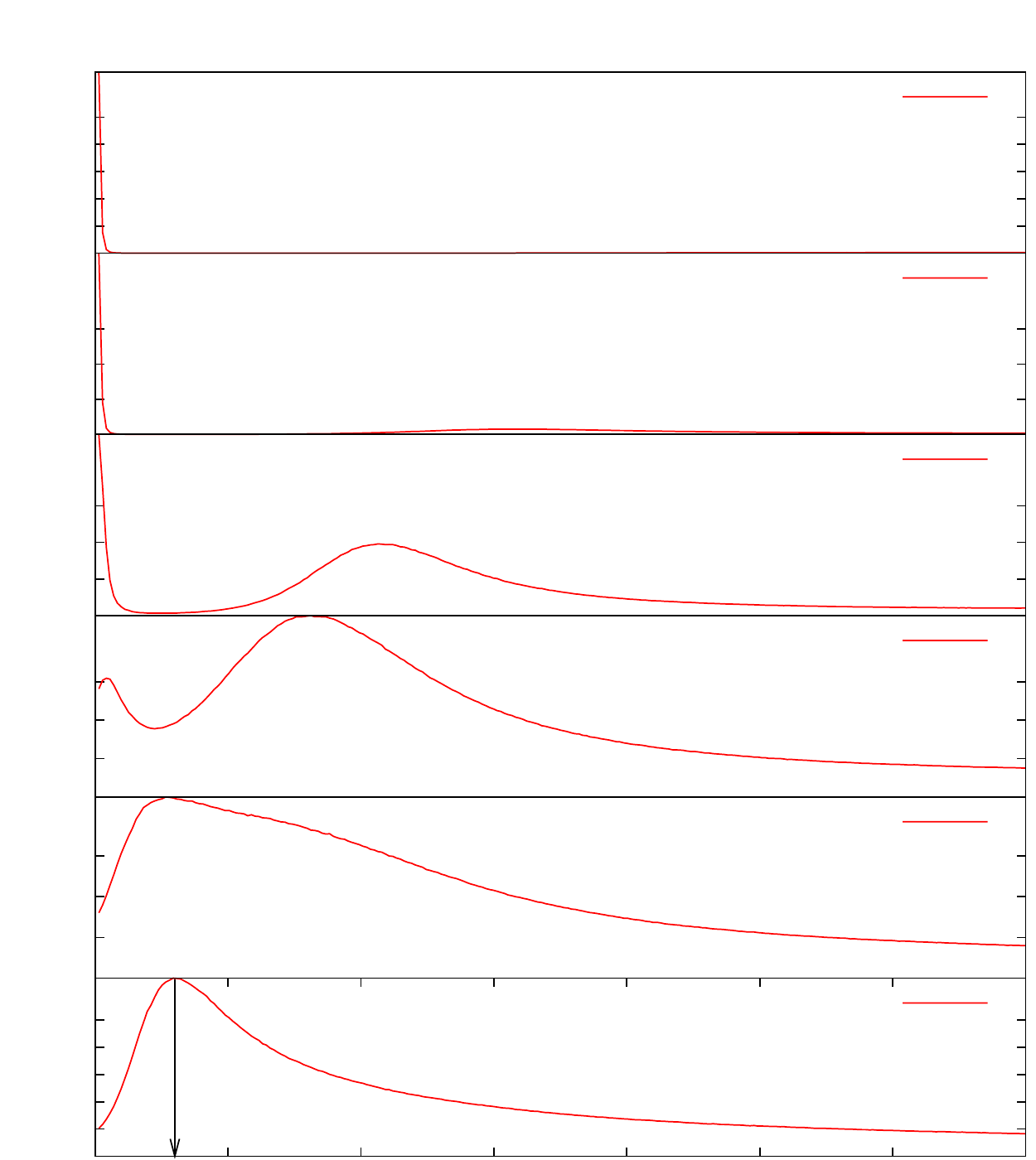}}%
    \gplfronttext
  \end{picture}%
\endgroup

%% file: tex-chargegap.tex
\begingroup
  \makeatletter
  \providecommand\color[2][]{%
    \GenericError{(gnuplot) \space\space\space\@spaces}{%
      Package color not loaded in conjunction with
      terminal option `colourtext'%
    }{See the gnuplot documentation for explanation.%
    }{Either use 'blacktext' in gnuplot or load the package
      color.sty in LaTeX.}%
    \renewcommand\color[2][]{}%
  }%
  \providecommand\includegraphics[2][]{%
    \GenericError{(gnuplot) \space\space\space\@spaces}{%
      Package graphicx or graphics not loaded%
    }{See the gnuplot documentation for explanation.%
    }{The gnuplot epslatex terminal needs graphicx.sty or graphics.sty.}%
    \renewcommand\includegraphics[2][]{}%
  }%
  \providecommand\rotatebox[2]{#2}%
  \@ifundefined{ifGPcolor}{%
    \newif\ifGPcolor
    \GPcolortrue
  }{}%
  \@ifundefined{ifGPblacktext}{%
    \newif\ifGPblacktext
    \GPblacktexttrue
  }{}%
  \let\gplgaddtomacro\g@addto@macro
  \gdef\gplbacktext{}%
  \gdef\gplfronttext{}%
  \makeatother
  \ifGPblacktext
    \def\colorrgb#1{}%
    \def\colorgray#1{}%
  \else
    \ifGPcolor
      \def\colorrgb#1{\color[rgb]{#1}}%
      \def\colorgray#1{\color[gray]{#1}}%
      \expandafter\def\csname LTw\endcsname{\color{white}}%
      \expandafter\def\csname LTb\endcsname{\color{black}}%
      \expandafter\def\csname LTa\endcsname{\color{black}}%
      \expandafter\def\csname LT0\endcsname{\color[rgb]{1,0,0}}%
      \expandafter\def\csname LT1\endcsname{\color[rgb]{0,1,0}}%
      \expandafter\def\csname LT2\endcsname{\color[rgb]{0,0,1}}%
      \expandafter\def\csname LT3\endcsname{\color[rgb]{1,0,1}}%
      \expandafter\def\csname LT4\endcsname{\color[rgb]{0,1,1}}%
      \expandafter\def\csname LT5\endcsname{\color[rgb]{1,1,0}}%
      \expandafter\def\csname LT6\endcsname{\color[rgb]{0,0,0}}%
      \expandafter\def\csname LT7\endcsname{\color[rgb]{1,0.3,0}}%
      \expandafter\def\csname LT8\endcsname{\color[rgb]{0.5,0.5,0.5}}%
    \else
      \def\colorrgb#1{\color{black}}%
      \def\colorgray#1{\color[gray]{#1}}%
      \expandafter\def\csname LTw\endcsname{\color{white}}%
      \expandafter\def\csname LTb\endcsname{\color{black}}%
      \expandafter\def\csname LTa\endcsname{\color{black}}%
      \expandafter\def\csname LT0\endcsname{\color{black}}%
      \expandafter\def\csname LT1\endcsname{\color{black}}%
      \expandafter\def\csname LT2\endcsname{\color{black}}%
      \expandafter\def\csname LT3\endcsname{\color{black}}%
      \expandafter\def\csname LT4\endcsname{\color{black}}%
      \expandafter\def\csname LT5\endcsname{\color{black}}%
      \expandafter\def\csname LT6\endcsname{\color{black}}%
      \expandafter\def\csname LT7\endcsname{\color{black}}%
      \expandafter\def\csname LT8\endcsname{\color{black}}%
    \fi
  \fi
  \setlength{\unitlength}{0.0500bp}%
  \begin{picture}(7200.00,5040.00)%
    \gplgaddtomacro\gplbacktext{%
      \csname LTb\endcsname%
      \put(1122,660){\makebox(0,0)[r]{\strut{} 0.05}}%
      \put(1122,1191){\makebox(0,0)[r]{\strut{} 0.1}}%
      \put(1122,1723){\makebox(0,0)[r]{\strut{} 0.15}}%
      \put(1122,2254){\makebox(0,0)[r]{\strut{} 0.2}}%
      \put(1122,2786){\makebox(0,0)[r]{\strut{} 0.25}}%
      \put(1122,3317){\makebox(0,0)[r]{\strut{} 0.3}}%
      \put(1122,3849){\makebox(0,0)[r]{\strut{} 0.35}}%
      \put(1122,4380){\makebox(0,0)[r]{\strut{} 0.4}}%
      \put(1254,440){\makebox(0,0){\strut{} 0}}%
      \put(1951,440){\makebox(0,0){\strut{} 0.05}}%
      \put(2647,440){\makebox(0,0){\strut{} 0.1}}%
      \put(3344,440){\makebox(0,0){\strut{} 0.15}}%
      \put(4040,440){\makebox(0,0){\strut{} 0.2}}%
      \put(4737,440){\makebox(0,0){\strut{} 0.25}}%
      \put(5433,440){\makebox(0,0){\strut{} 0.3}}%
      \put(6130,440){\makebox(0,0){\strut{} 0.35}}%
      \put(6826,440){\makebox(0,0){\strut{} 0.4}}%
      \put(220,2520){\rotatebox{90}{\makebox(0,0){\strut{}$\Delta_c$}}}%
      \put(4040,110){\makebox(0,0){\strut{}$\Delta$}}%
      \put(4040,4710){\makebox(0,0){\strut{}Chargegap as a function of $\Delta$}}%
    }%
    \gplgaddtomacro\gplfronttext{%
      \csname LTb\endcsname%
      \put(5839,1273){\makebox(0,0)[r]{\strut{}$\Delta_c$ from $C_c(\tau)$}}%
      \csname LTb\endcsname%
      \put(5839,1053){\makebox(0,0)[r]{\strut{}$\Delta_c$ from $N(\omega)$}}%
      \csname LTb\endcsname%
      \put(5839,833){\makebox(0,0)[r]{\strut{}$\Delta_c=0.8038 \Delta + 0.0714$}}%
    }%
    \gplbacktext
    \put(0,0){\includegraphics{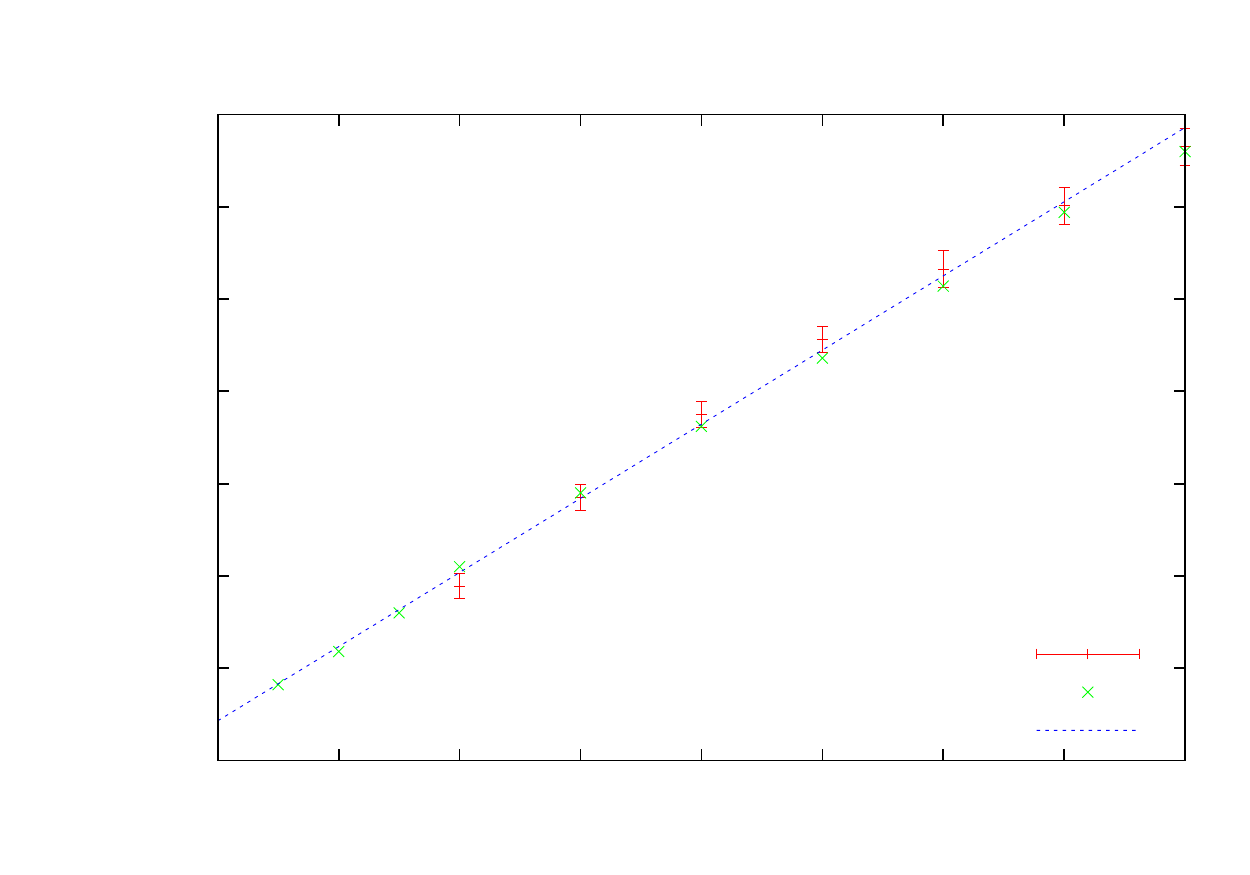}}%
    \gplfronttext
  \end{picture}%
\endgroup

%% file: dmft-results.tex
\label{sec:DMFT}

\subsection{Periodic Anderson Model with BCS conduction band}

In the previous sections, we have studied the first order phase transition in the impurity model
(\ref{eq:qd_supercond_hamiltonian}). As the dynamical mean field theory (DMFT) provides a link
between impurity models and lattice models, we can ask the question if the singlet to doublet phase
transition observed in the impurity model is also realized in a corresponding lattice model.

An appropriate lattice model will of course include a $U(1)$ symmetry breaking term like the impurity model
(\ref{eq:qd_supercond_hamiltonian}) does, and in fact in the framework of the DMFT, a periodic
Anderson model extended by the BCS mean field Hamiltonian (BCS-PAM) for the conduction band electrons corresponds to the impurity
model presented in the previous sections
\footnote{ Strictly speaking, the reference model in the DMFT only has one superconducting bath,
while we introduced a left and a right bath in the Hamiltonian (\ref{eq:qd_supercond_hamiltonian}).
However, in the CTQMC, the reference model is entirely encoded in the bare Green's function, which
can be understood as an action representation in the path integral formalism. The explicit number of
the superconducting baths is therefore unimportant.
}. The Hamiltonian of the BCS-PAM is given by:

\begin{equation}
\label{eq:BCS_PAM}
H = H_c + H_f + H_V
\end{equation}
with
\begin{equation}
H_c = \sum_{k,\sigma} \xi(k) \tilde{c}_{k,\sigma}^\dagger \tilde{c}_{k,\sigma}
- \Delta \sum_k \left( \tilde{c}_{k,\up}^\dagger \tilde{c}_{-k\dw}^\dagger + \text{h.c.} \right)
\label{eq:BCS_PAM_c}
\end{equation}

\begin{equation}
	H_f = \sum_{k,\sigma} \xi_f \tilde{f}_{k,\sigma}^\dagger \tilde{f}_{k,\sigma} + U \sum_{i_f} \left( \tilde{n}_{i_f,\up} - \frac{1}{2} \right)	 \left( \tilde{n}_{i_f,\dw} - \frac{1}{2} \right)
	\label{eq:BCS_PAM_f}
\end{equation}

\begin{equation}
H_V = - V \sum_{k,\sigma} \left( \tilde{c}_{k,\sigma}^\dagger \tilde{f}_{k,\sigma} + \text{h.c.} \right)
	\label{eq:BCS_PAM_V}
\end{equation}
We have considered a square lattice with hopping matrix element $t$ between the conduction electrons such that: 
\begin{equation}
	\label{eq:BCS_PAM_disp}
	\xi(k) = -2t \left( \cos( k a_x) + \cos(k a_y)  \right).
\end{equation}

Note, that the impurity model (\ref{eq:qd_supercond_hamiltonian}) has a large range of applications
in the DMFT
ranging from the attractive Hubbard model with $U(1)$ symmetry broken solutions studied in references 
\cite{0295-5075-85-2-27001,bauer-hewson-dmft-nrg} to the BCS-PAM, which
is considered here.

The treatment of this model within DMFT involves the same steps as for the impurity model (\ref{eq:qd_supercond_hamiltonian}), introducing a particle-hole transformation for the spin down operators. The Hamiltonian can then be cast in the form $H=H_0+H_U$ with
\begin{equation}
	H_0 = \sum_k \vec{c}_k^\dagger \mat{E}(k) \vec{c}_k - V \sum_{k} \left( \vec{c}_k^\dagger \matgr{\sigma}_z \vec{f}_k + \text{h.c.} \right) + \sum_k \vec{f}_k^\dagger \matgr{\epsilon}_f \vec{f}_k 
	\label{eq:BCS_PAM_canon}
\end{equation}
and $H_U= - U \sum_{i_f} \left( n_{i_f,\up} - \frac{1}{2} \right) \left( n_{i_f,\dw} - \frac{1}{2}
\right) $. Here, we have used the same Nambu-spinor notation as in Sec. \ref{sec:impurity-model}
with the exception, that $d$ operators have been renamed $f$ to be consistent with the literature
\cite{hewson,RevModPhys.68.13}.

\subsection{DMFT with superconducting medium}

The standard DMFT can be easily adapted to a superconducting bath using the
Nambu formalism \cite{RevModPhys.68.13}. We obtain the self consistency equation for a finite lattice
with $N$ sites expressed by a $2\times2$ matrix equation:
\begin{equation}
	\label{eq:dmft-selfconsistency}
	\mat{G^{ff}}(i\omega_n) = \frac{1}{N} \sum_\vec{k}
	\left[\mat{G_{kk}^{0,ff}}^{-1}(i\omega_n)-\mat{\Sigma^{ff}}(i\omega_n)\right]^{-1}.
\end{equation}
Here, $\mat{G^{ff}}(i\omega_n) = -\int \limits_0^\beta \D \tau \, \E^{-i\omega_n \tau}
\thavg{T \vec{f}(\tau) \vec{f}^\dagger}$ is the full Matsubara Green's function of the reference model,
$\mat{G_{kk}^{0,ff}}(i\omega_n) $ is the Matsubara  $f$-Green function of the bare lattice model
and $\mat{\Sigma^{ff}}$ is the self energy.
Equation (\ref{eq:dmft-selfconsistency}) can be solved by iteration starting usually at a self
energy $\mat{\Sigma^{ff}} \equiv 0$. From $\mat{G^{ff}}(i\omega_n)$, the bare Green's function $\mat{\mathcal
G^{ff}_0}(i\omega_n)$ of the reference model, can be calculated using Dyson's equation
$\mat{\mathcal G_0^{ff}}^{-1} =  \mat{G^{ff}}^{-1} + \mat{\Sigma^{ff}} $. The reference model, which is now
described by $\mat{\mathcal G^{ff}_0}$ and the interaction part of the Hamiltonian can subsequently be
solved using the CTQMC method yielding $\mat{G^{ff}}(i\omega_n)$ for the next DMFT iteration.

\subsection{Hysteresis}

In the DMFT, we can calculate the double occupancy $\thavg{\tilde{f}_{\up,i}^\dagger
\tilde{f}_{\up,i} \tilde{f}_{\dw,i}^\dagger \tilde{f}_{\dw,i}}$ of the $f$-sites, which is together
with the assumption of a homogeneous system proportional to $\frac{\partial \Omega }{\partial U}$.
Therefore, we expect a jump in the double occupancy to appear at a critical value of $U$, if we have
a first order phase transition as in the impurity problem.

\begin{figure}
\resizebox{\columnwidth}{!}{ \input{tex-DMFT-double-occupancy-hysteresis-B150.tex} }
	\caption{\label{fig:dmft-double-occ-hyst} Double occupancy of the $f$ sites in the
	BCS-PAM. In the proximity of the critical value of $U$, we observe two different solutions
	of the DMFT self consistency cycle. The upper (red) branch is generated, if we start the
	DMFT algorithm with a self energy $\Sigma \equiv 0$, while we obtain the solution shown by
	the lower (blue) branch if we take the self energy of the data point at $U=0.44$ as the
	starting point of the DMFT iterations.  }
\end{figure}
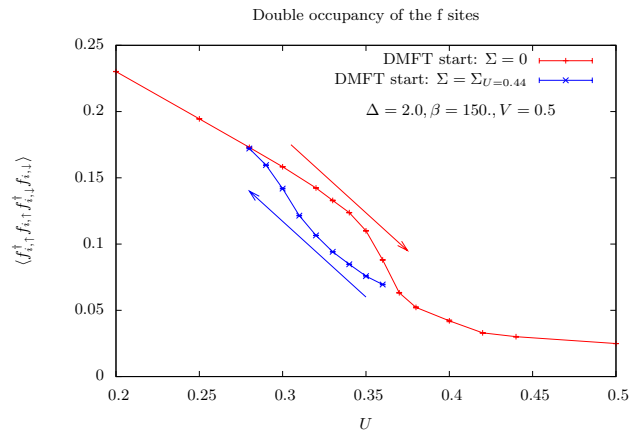

Figure \ref{fig:dmft-double-occ-hyst} shows our result for the double occupancy of the $f$ sites as
a function of $U$. Depending on the  initial choice of the self energy in the DMFT cycle, we obtain two
different solutions. If we start with the local Green's function of the bare lattice model, which
corresponds to a self energy $\Sigma \equiv 0 $, we obtain the upper branch of the hysteresis.
The lower branch is obtained by taking the self energy of the solution in the strong coupling phase
at $U=0.44$ as  starting point for the DMFT cycle.
The coexistence of two solutions is a strong hint that a first order phase transition occurs.

It should be noted that beginning at a value of $U\approx0.34$, the upper branch of the hysteresis
becomes unstable, i.e. the inherent fluctuations of the Monte Carlo results suffice to drop from the
upper branch of the hysteresis to the lower branch after a certain number of iterations. Increasing
the number of Monte Carlo measurements delays the drop to the lower branch to a higher number of
iterations. This behavior can be understood in the following way: In the coexistence region, the
grand potential $\Omega$ of the upper and lower branch of the hysteresis cross at a certain value of
$U$. For small values of $U$, $\Omega$ is minimal on the upper branch, while the lower branch is
metastable, for larger values of $U$, however, the stable solution is the lower branch.

In the strong coupling phase and on the lower branch of the hysteresis, the Monte Carlo results
suddenly develop a finite magnetization corresponding to a frozen spin. This is due to divergent
autocorrelation times in the Monte Carlo simulation and is linked to the physical formation of a
local moment.

\subsection{Local dynamical spin structure factor}
\label{subsec:DMFT-spin}

To further classify the weak and strong coupling phases, we calculate the local dynamical spin
structure factor $S(\omega) = \frac{1}{N} \sum_\vec{q} S(\vec{q},\omega)$. The Lehmann
representation for $S(\omega)$ is given by Eq. (\ref{eq:spinstructure-lehmann_2}), where in this
case $S_+=S_+^{f,i}$. 

\begin{figure}[h]
	\begin{center}
\resizebox{\columnwidth}{!}{ \input{tex-DMFT-spinstructure-U034.tex} }
	\end{center}
	\caption{Dynamical spin structure factor for the upper and the lower branch of the
	hysteresis in Fig. \ref{fig:dmft-double-occ-hyst}. Clearly, the upper branch of the
	hysteresis corresponds to a singlet solution, while the lower branch shows a local moment.
	}
	\label{fig:dmft-spinstructure-hysteresis}
\end{figure}
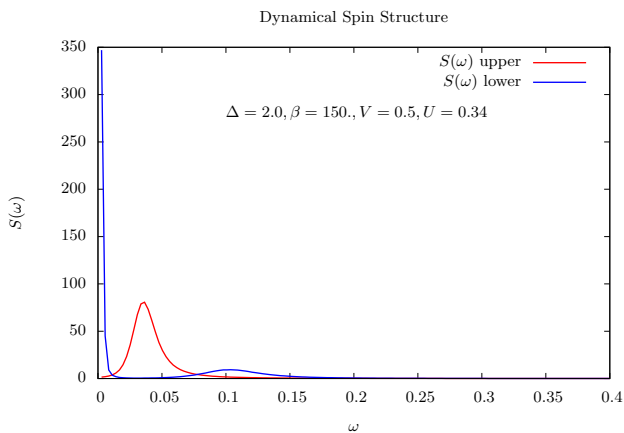

As in the impurity case, $S(\omega)$ is a measure for the energy needed to flip the spin on an
$f$-site. Figure \ref{fig:dmft-spinstructure-hysteresis} shows the result for the local dynamical
spin structure factor on both branches of the hysteresis. The solution corresponding to the upper
branch of the hysteresis is linked to the weak coupling regime and shows a characteristic energy
scale required for flipping a spin. 

The lower branch of the hysteresis represents the strong coupling phase and shows a clear local
moment peak in the dynamical spin structure factor at $\omega=0$.

This behavior reflects exactly the single impurity physics discussed in the previous section where
we observed the Kondo effect in the weak coupling phase and the formation of a local moment in the
strong coupling phase.

\subsection{f-Density of states}

\begin{figure}[h]
	\begin{center}
		\resizebox{\columnwidth}{!}{ \input{tex-DMFT-ff-DOS-B100-3d.tex} }
	\end{center}
\vspace{-1cm}
	\caption{Density of states for the f-electrons as a function of $U$ for the parameters
	$V=0.5$, $\Delta=2$, $\mu=\epsilon_f=0$ and $\beta=100$.}
	\label{fig:dmft-dos}
\end{figure}
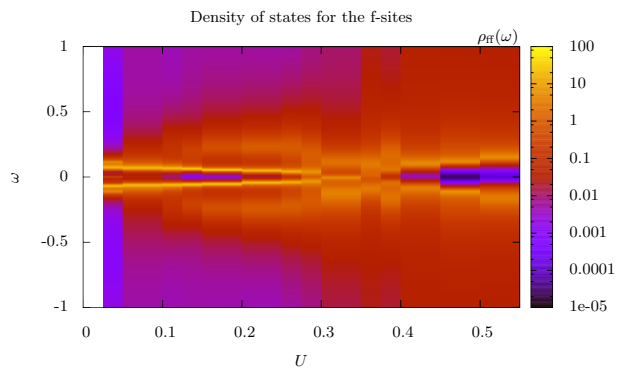

In order to investigate the behavior of the $f$-bands at the phase boundary and to be able to
compare with the single impurity model, we calculate the density of states for the $f$-sites 
$\rho_{\text{ff}}$ directly from the local Green's function $G(\tau)$ using the stochastic analytic 
continuation method for different values of $U$. From Fig. \ref{fig:dmft-dos}, one can recognize the signature 
of the  impurity physics (see Sec. \ref{subsec:impurity-spectral}), namely the crossing of Andreev bound states in the
 vicinity of the first order  transition at $U\approx0.35$.   
Note, that we have only shown the level crossing for the impurity model if $\Delta$ is changed, but
for varying $U$, the crossing of the Andreev bound states in the impurity model
(\ref{eq:qd_supercond_hamiltonian}) has been observed by Bauer et al. \cite{0953-8984-19-48-486211}.
Clearly in the lattice model, one  expects the Andreev bound states to acquire a dispersion relation which shows up 
as a finite width in  $\rho_{\text{ff}}$.

\subsection{Dispersion relation of Andreev bound states}

We have seen in the previous subsections, that the local physics of the single impurity model can 
be carried over to the lattice case within the DMFT approximation.    
Here,  we  concentrate on  unique features of the lattice model (\ref{eq:BCS_PAM}), namely the
dispersion relation of the f-bands as obtained by analyzing the  single particle spectral function.  

Using the  local  self-energy of the DMFT,  $\mat{\Sigma^{ff}}(i\omega_n)$,  this quantity is extracted from the Green 
functions 
\begin{equation}
  \mat{G^{ff}_{\vec{k}\vec{k}}}(i\omega_n) =
  \left[\mat{G^{0,ff}_{\vec{k}\vec{k}}}(i\omega_n)^{-1} - \mat{\Sigma^{ff}}(i\omega_n)
	\right]^{-1}.
	\label{eq:dmft_g_ff_kk}
\end{equation}
and
\begin{equation}
  \mat{G_{\vec{k}\vec{k}}^{{cc}}}(i\omega_n) =
  \mat{G_{\vec{k}\vec{k}}^{{0,cc}}}(i\omega_n) - \mat{G_{\vec{k}\vec{k}}^{{0,cf}}}(i\omega_n) 
  \mat{G_{\vec{k}\vec{k}}^{{ff}}}(i\omega_n)
  \mat{G_{\vec{k}\vec{k}}^{{0,fc}}}(i\omega_n) .
	\label{eq:dmft_g_cc_kk}
\end{equation}
where  $ \mat{G_{\vec{k}\vec{k}}^{{0,cc}}}(i\omega_n) $, $ \mat{G_{\vec{k}\vec{k}}^{{0,ff}}}(i\omega_n) $,
$ \mat{G_{\vec{k}\vec{k}}^{{0,cf}}}(i\omega_n) $,    $ \mat{G_{\vec{k} \vec{k}}^{{0,fc}}} (i\omega_n) $
 denote  the  noninteracting Green functions for the corresponding  orbitals in the unit cell. 

Using the stochastic analytic continuation, these Green's functions can be rotated to real frequencies,
yielding in principle the spectral function $\mat{A}(\vec{k},\omega)$. For each
$\vec{k}$-point and  real frequency this quantity is a  $4\times4$ matrix since we have a $2\times2$ Nambu spectral
function for each combination of $f$ and $c$ orbitals.  Our analysis of the spectral function is based on
the  basis independent quantity  $A(\vec{k},\omega)= \tr \mat{A}(\vec{k},\omega)$.

\begin{figure}[h]
	\begin{center}
\resizebox{\columnwidth}{!}{ \input{tex-DMFT-spectrum-Trace-U0125-3d.tex} }
	\end{center}
\vspace{-1cm}
	\caption{Trace of the spectral function $A(\vec{k},\omega)$ at $\beta=100$ in the singlet regime. The
	parameters of the simulation were given by $U=0.125$, $V=0.5$, $\Delta=2$ and
	$\mu=\epsilon_f=0$.  }
	\label{fig:dmft-spectral-singlet}
\end{figure}
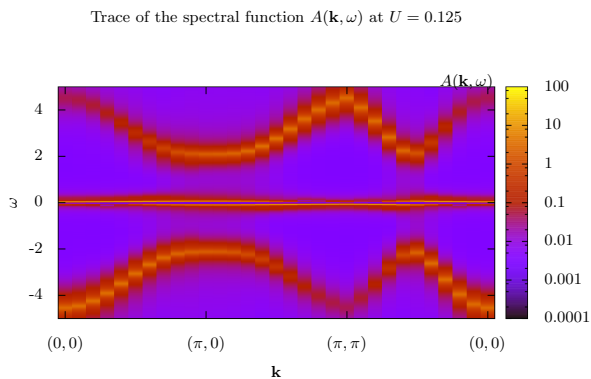

Fig. \ref{fig:dmft-spectral-singlet}  plots this quantity in the singlet phase. 
The overall structure of the spectral function is similar to the structure observed for
the bare BCS-PAM  characterized by the four bands: 
\begin{equation}
	E_{\pm,\pm} (\vec{k})  =  
	\pm \sqrt{ V^2 +  E^{2}(\vec{k})/2   \pm  E(\vec{k}) \sqrt{   V^2 + E^2(\vec{k}) /4  }   } 
\end{equation}
where $ E(\vec{k}) = \sqrt{ \epsilon^2(\vec{k}) + \Delta^2} $.   The bands  with dominant c-character,  
$ E^{c}_{\pm}(\vec{k}) \equiv  E^{\pm,+}(\vec{k}) $,  
at high frequencies  are well separated from the bands of dominant $f$-character at  low frequencies, 
$ E^{f}_{\pm}(\vec{k})  =  E_{\pm,-}(\vec{k}) $.  For the considered bare parameters, $V$ is the smallest scale and sets
the  magnitude of the dispersion relation of the $f$-band. In particular expanding in $V$ gives: 
\begin{equation}
	E^{f}_{\pm}(\vec{k})  =  \pm  \frac{V^2}{E(\vec{k})}    +  {\cal O} 
\left(  \frac{V^4}{E(\vec{k})^3}  \right)
\label{E_f_BCS_PAM.Eq}
\end{equation}
Starting from the point of view of the impurity model,  which as seen above accounts very well for overall form of the
k-integrated $f$-spectral function, $ E^{f}_{\pm}(\vec{k}) $ may be perceived as  the dispersion relation of  the 
Andreev bound states.

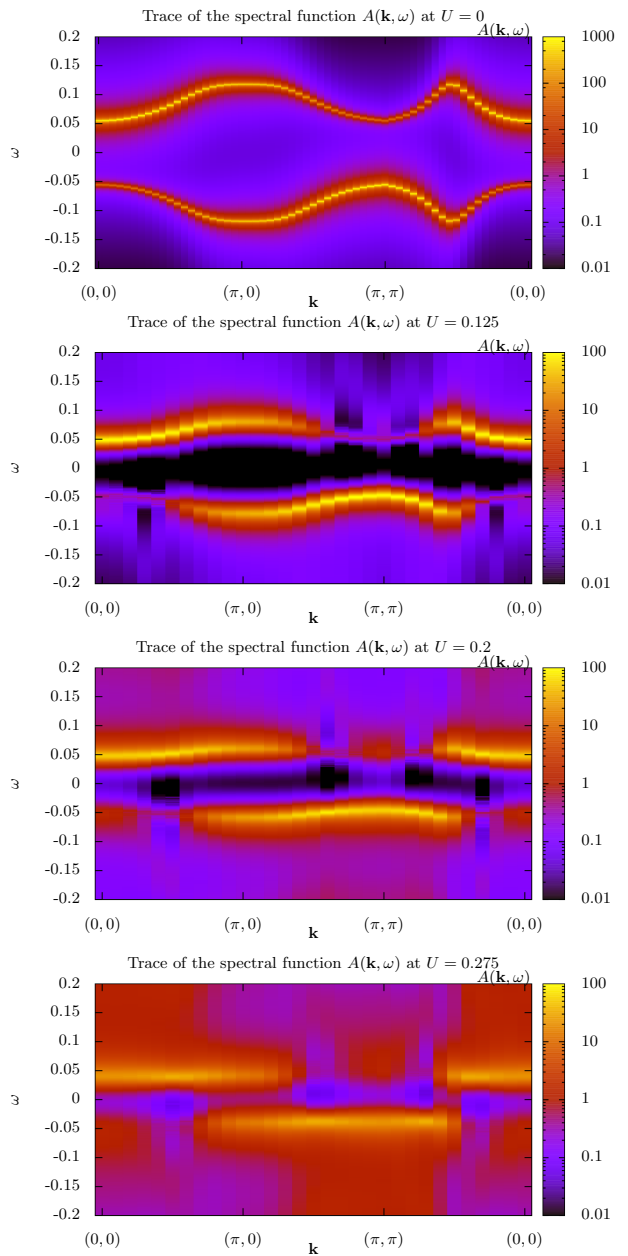
\begin{figure}[h]
\begin{minipage}[h]{\columnwidth}
	\resizebox{\columnwidth}{!}{	\input{tex-DMFT-free-spectrum-Trace-zoomed-3d.tex} }
\end{minipage}
\begin{minipage}[h]{\columnwidth}
\vspace{-1.8cm}
	\resizebox{\columnwidth}{!}{     \input{tex-DMFT-spectrum-Trace-U0125-zoomed-3d.tex} }
\end{minipage}
\begin{minipage}[h]{\columnwidth}
\vspace{-1.8cm}
	\resizebox{\columnwidth}{!}{     \input{tex-DMFT-spectrum-ff-U02-3d.tex}  }
\end{minipage}
\begin{minipage}[h]{\columnwidth}
\vspace{-1.8cm}
	\resizebox{\columnwidth}{!}{      \input{tex-DMFT-spectrum-Trace-U0275-zoomed-3d.tex}  }
\end{minipage}
\vspace{-1cm}
	\caption{Trace of the spectral function $A(\vec{k},\omega)$ at $\beta=100$ in the singlet 
	regime for increasing interaction $U$. The width of the $f$-bands clearly decreases and the
	dispersion becomes weaker.
	The parameters of the simulations were given by $V=0.5$, $\Delta=2$ and
	$\mu=\epsilon_f=0$.  }
	\label{fig:dmft-spectral-singlet-zoom}
\end{figure}

The singlet phase is continuously connected to the $U=0$  point. Starting from this limit, we can account for 
the Hubbard $U$ within a slave boson approximation \cite{Kotliar86} which  will renormalize  the  hybridization 
matrix element to lower values. Owing to Eq. \ref{E_f_BCS_PAM.Eq}  this suppresses the dispersion relation of the $f$-electrons. This  
aspect is clearly  observed in Fig. \ref{fig:dmft-spectral-singlet-zoom}.

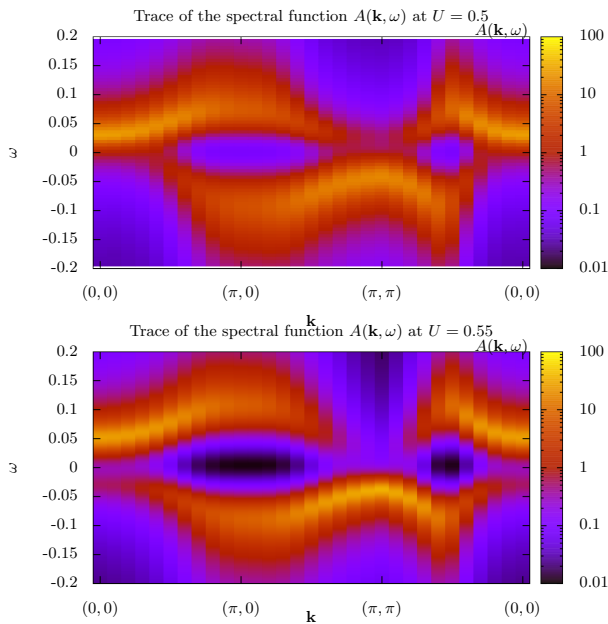
\begin{figure}[h]
	\begin{center}
	  \begin{minipage}[h]{\columnwidth}
\resizebox{\columnwidth}{!}{ \input{tex-DMFT-spectrum-Trace-U05-zoomed-3d.tex} }
\end{minipage}
\begin{minipage}[h]{\columnwidth}
  \vspace{-1.8cm}
\resizebox{\columnwidth}{!}{ \input{tex-DMFT-spectrum-Trace-U055-zoomed-3d.tex} }
\end{minipage}
	\end{center}
\vspace{-0.7cm}
	\caption{Trace of the spectral function $A(\vec{k},\omega)$ at $\beta=100$ in the doublet
	regime for different values of $U$. Here, we only show the $f$-bands.
	The parameters of the simulation were given by $V=0.5$, $\Delta=2$ and
	$\mu=\epsilon_f=0$.  }
	\label{fig:dmft-spectral-doublet-zoom}
\end{figure}

\begin{figure}[h]
	\begin{center}
\resizebox{\columnwidth}{!}{ \input{tex-CPA-spec-3d.tex} }
	\end{center}
\vspace{-0.7cm}
	\caption{Trace of the spectral function $A(\vec{k},\omega)$ as obtained from using Eq.
	\ref{Eq:S_eff_mz} for the impurity action.
	The z-component of the local moment is  sampled from the box distribution $m_z \in  [-M_z, M_z] $.
	The parameters used for this plot were given by $V=0.5$, $U=0.5$, $\Delta=2$ and
	$M_z=0.0375$.   Here, the calculations are carried out on the real time axis such that  no  analytical 
	continuation is required.
	\label{fig:CPA} }
\end{figure}
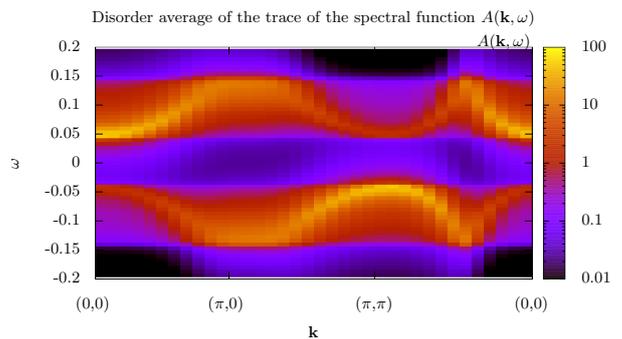

In the doublet phase, $U > U_c$, the paramagnetic  slave-boson mean-field approach  fails.    In this state,
the $f$-spin is frozen and in the DMFT cycle we  have imposed spin symmetric baths thereby  inhibiting  magnetic 
ordering. 
The QMC data of Fig. \ref{fig:dmft-spectral-doublet-zoom}  points to a very incoherent  $f$-spectral function.
It is therefore tempting  to model this state in terms of spin disorder: the spin of the
$f$-electrons on each site 
is static and  points in a random direction.  To  provide some  support for this picture 
we  stay in the dynamical mean field framework but 
consider a  mean-field 
decomposition of the Hubbard term in the action of the  impurity problem:
\begin{equation}
	U  \left( \tilde{n}_{f,\up} - \frac{1}{2} \right)	 \left( \tilde{n}_{\dw} - \frac{1}{2} \right) \rightarrow
    -\frac{Um_z}{2}\left( \tilde{n}_{f,\up} -  \tilde{n}_{f,\dw} \right)
\end{equation}
 This mean field approximation,  accounts for the local moment  formation with $z$-component of spin $m_z$. The  corresponding 
mean-field action of the  impurity  model now reads:
\begin{equation}
S_{MF} =  \int \limits_{0}^{\beta} {\rm d} \tau  \int \limits_{0}^{\beta} {\rm d} \tau' \tilde{\vec{f}}^{\dagger}(\tau)  
{\matgr{\cal G} }^{-1}(\tau - \tau') \tilde{\vec{f}}(\tau')  
	-\frac{U m_z}{2} \int \limits_{0}^{\beta} \D \tau \tilde{\vec{f}}^{\dagger}(\tau) \tilde{\vec{f}}(\tau)
\label{Eq:S_eff_mz}
\end{equation}
where  $ \tilde{\vec{f}}^{\dagger}  = \left(\tilde{f}^{\dagger}_{\uparrow} ,  \tilde{f}_{\downarrow} \right) $   and 
$ \matgr{\cal G }(\tau - \tau')$ corresponds to the bath Green function.  To account for disorder,
the $z$-component of the 
f-spin is sampled from the box distribution  $ m_z \in [-M_z, M_z] $. 
Averaging over disorder at each iteration in the DMFT cycle yields the spectral function shown in Fig. \ref{fig:CPA}.
As apparent, the disorder average generates a finite lifetime.   

%% file: tex-DMFT-double-occupancy-hysteresis-B150.tex
\begingroup
  \makeatletter
  \providecommand\color[2][]{%
    \GenericError{(gnuplot) \space\space\space\@spaces}{%
      Package color not loaded in conjunction with
      terminal option `colourtext'%
    }{See the gnuplot documentation for explanation.%
    }{Either use 'blacktext' in gnuplot or load the package
      color.sty in LaTeX.}%
    \renewcommand\color[2][]{}%
  }%
  \providecommand\includegraphics[2][]{%
    \GenericError{(gnuplot) \space\space\space\@spaces}{%
      Package graphicx or graphics not loaded%
    }{See the gnuplot documentation for explanation.%
    }{The gnuplot epslatex terminal needs graphicx.sty or graphics.sty.}%
    \renewcommand\includegraphics[2][]{}%
  }%
  \providecommand\rotatebox[2]{#2}%
  \@ifundefined{ifGPcolor}{%
    \newif\ifGPcolor
    \GPcolortrue
  }{}%
  \@ifundefined{ifGPblacktext}{%
    \newif\ifGPblacktext
    \GPblacktexttrue
  }{}%
  \let\gplgaddtomacro\g@addto@macro
  \gdef\gplbacktext{}%
  \gdef\gplfronttext{}%
  \makeatother
  \ifGPblacktext
    \def\colorrgb#1{}%
    \def\colorgray#1{}%
  \else
    \ifGPcolor
      \def\colorrgb#1{\color[rgb]{#1}}%
      \def\colorgray#1{\color[gray]{#1}}%
      \expandafter\def\csname LTw\endcsname{\color{white}}%
      \expandafter\def\csname LTb\endcsname{\color{black}}%
      \expandafter\def\csname LTa\endcsname{\color{black}}%
      \expandafter\def\csname LT0\endcsname{\color[rgb]{1,0,0}}%
      \expandafter\def\csname LT1\endcsname{\color[rgb]{0,1,0}}%
      \expandafter\def\csname LT2\endcsname{\color[rgb]{0,0,1}}%
      \expandafter\def\csname LT3\endcsname{\color[rgb]{1,0,1}}%
      \expandafter\def\csname LT4\endcsname{\color[rgb]{0,1,1}}%
      \expandafter\def\csname LT5\endcsname{\color[rgb]{1,1,0}}%
      \expandafter\def\csname LT6\endcsname{\color[rgb]{0,0,0}}%
      \expandafter\def\csname LT7\endcsname{\color[rgb]{1,0.3,0}}%
      \expandafter\def\csname LT8\endcsname{\color[rgb]{0.5,0.5,0.5}}%
    \else
      \def\colorrgb#1{\color{black}}%
      \def\colorgray#1{\color[gray]{#1}}%
      \expandafter\def\csname LTw\endcsname{\color{white}}%
      \expandafter\def\csname LTb\endcsname{\color{black}}%
      \expandafter\def\csname LTa\endcsname{\color{black}}%
      \expandafter\def\csname LT0\endcsname{\color{black}}%
      \expandafter\def\csname LT1\endcsname{\color{black}}%
      \expandafter\def\csname LT2\endcsname{\color{black}}%
      \expandafter\def\csname LT3\endcsname{\color{black}}%
      \expandafter\def\csname LT4\endcsname{\color{black}}%
      \expandafter\def\csname LT5\endcsname{\color{black}}%
      \expandafter\def\csname LT6\endcsname{\color{black}}%
      \expandafter\def\csname LT7\endcsname{\color{black}}%
      \expandafter\def\csname LT8\endcsname{\color{black}}%
    \fi
  \fi
  \setlength{\unitlength}{0.0500bp}%
  \begin{picture}(7200.00,5040.00)%
    \gplgaddtomacro\gplbacktext{%
      \csname LTb\endcsname%
      \put(1122,660){\makebox(0,0)[r]{\strut{} 0}}%
      \put(1122,1404){\makebox(0,0)[r]{\strut{} 0.05}}%
      \put(1122,2148){\makebox(0,0)[r]{\strut{} 0.1}}%
      \put(1122,2892){\makebox(0,0)[r]{\strut{} 0.15}}%
      \put(1122,3636){\makebox(0,0)[r]{\strut{} 0.2}}%
      \put(1122,4380){\makebox(0,0)[r]{\strut{} 0.25}}%
      \put(1254,440){\makebox(0,0){\strut{} 0.2}}%
      \put(2183,440){\makebox(0,0){\strut{} 0.25}}%
      \put(3111,440){\makebox(0,0){\strut{} 0.3}}%
      \put(4040,440){\makebox(0,0){\strut{} 0.35}}%
      \put(4969,440){\makebox(0,0){\strut{} 0.4}}%
      \put(5897,440){\makebox(0,0){\strut{} 0.45}}%
      \put(6826,440){\makebox(0,0){\strut{} 0.5}}%
      \put(220,2520){\rotatebox{90}{\makebox(0,0){\strut{}$\langle f^\dagger_{i,\up} f_{i,\up} f^\dagger_{i,\dw} f_{i,\dw} \rangle$}}}%
      \put(4040,110){\makebox(0,0){\strut{}$U$}}%
      \put(4040,4710){\makebox(0,0){\strut{}Double occupancy of the f sites}}%
      \put(4040,3636){\makebox(0,0)[l]{\strut{}$\Delta=2.0, \beta=150., V=0.5$}}%
    }%
    \gplgaddtomacro\gplfronttext{%
      \csname LTb\endcsname%
      \put(5839,4207){\makebox(0,0)[r]{\strut{}DMFT start: $\Sigma=0$}}%
      \csname LTb\endcsname%
      \put(5839,3987){\makebox(0,0)[r]{\strut{}DMFT start: $\Sigma=\Sigma_{U=0.44}$}}%
    }%
    \gplbacktext
    \put(0,0){\includegraphics{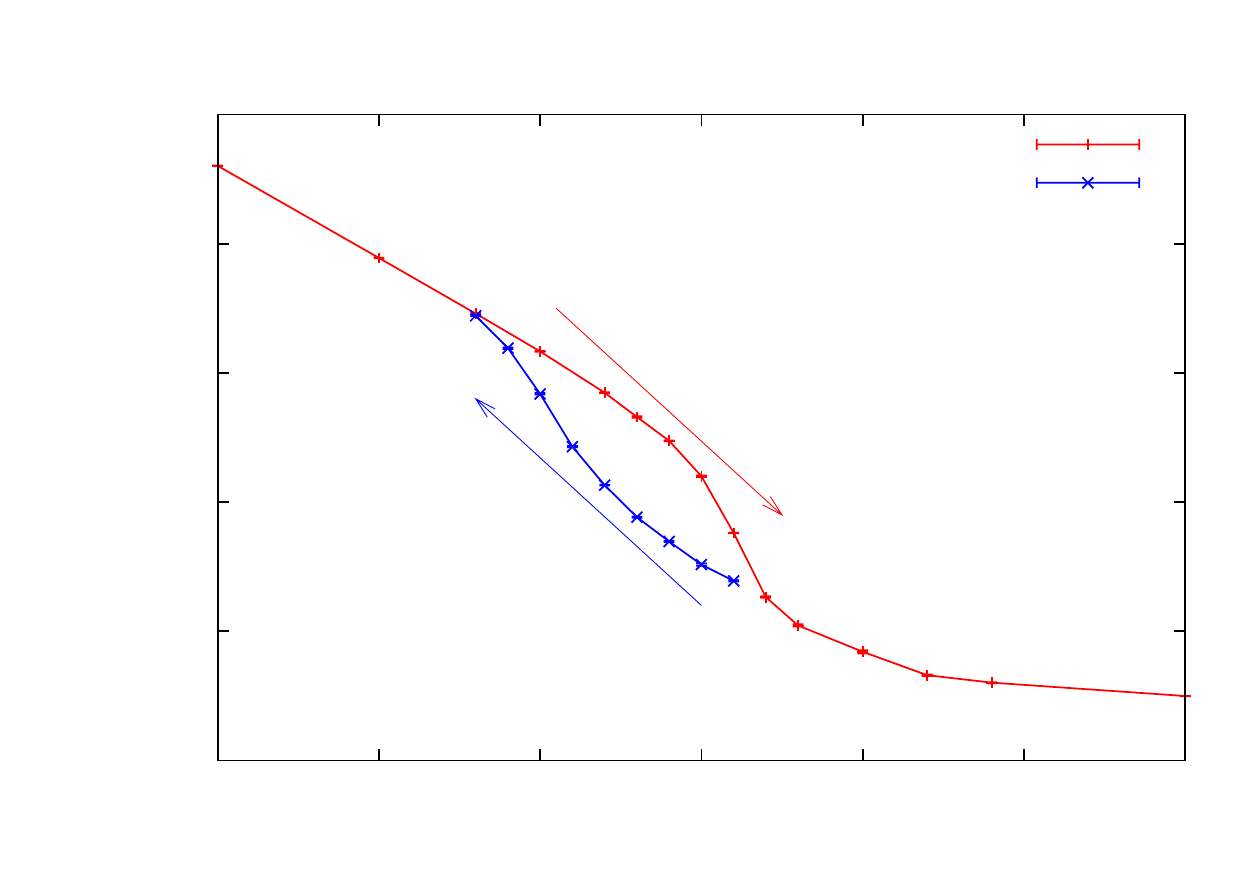}}%
    \gplfronttext
  \end{picture}%
\endgroup

%% file: tex-DMFT-spinstructure-U034.tex
\begingroup
  \makeatletter
  \providecommand\color[2][]{%
    \GenericError{(gnuplot) \space\space\space\@spaces}{%
      Package color not loaded in conjunction with
      terminal option `colourtext'%
    }{See the gnuplot documentation for explanation.%
    }{Either use 'blacktext' in gnuplot or load the package
      color.sty in LaTeX.}%
    \renewcommand\color[2][]{}%
  }%
  \providecommand\includegraphics[2][]{%
    \GenericError{(gnuplot) \space\space\space\@spaces}{%
      Package graphicx or graphics not loaded%
    }{See the gnuplot documentation for explanation.%
    }{The gnuplot epslatex terminal needs graphicx.sty or graphics.sty.}%
    \renewcommand\includegraphics[2][]{}%
  }%
  \providecommand\rotatebox[2]{#2}%
  \@ifundefined{ifGPcolor}{%
    \newif\ifGPcolor
    \GPcolortrue
  }{}%
  \@ifundefined{ifGPblacktext}{%
    \newif\ifGPblacktext
    \GPblacktexttrue
  }{}%
  \let\gplgaddtomacro\g@addto@macro
  \gdef\gplbacktext{}%
  \gdef\gplfronttext{}%
  \makeatother
  \ifGPblacktext
    \def\colorrgb#1{}%
    \def\colorgray#1{}%
  \else
    \ifGPcolor
      \def\colorrgb#1{\color[rgb]{#1}}%
      \def\colorgray#1{\color[gray]{#1}}%
      \expandafter\def\csname LTw\endcsname{\color{white}}%
      \expandafter\def\csname LTb\endcsname{\color{black}}%
      \expandafter\def\csname LTa\endcsname{\color{black}}%
      \expandafter\def\csname LT0\endcsname{\color[rgb]{1,0,0}}%
      \expandafter\def\csname LT1\endcsname{\color[rgb]{0,1,0}}%
      \expandafter\def\csname LT2\endcsname{\color[rgb]{0,0,1}}%
      \expandafter\def\csname LT3\endcsname{\color[rgb]{1,0,1}}%
      \expandafter\def\csname LT4\endcsname{\color[rgb]{0,1,1}}%
      \expandafter\def\csname LT5\endcsname{\color[rgb]{1,1,0}}%
      \expandafter\def\csname LT6\endcsname{\color[rgb]{0,0,0}}%
      \expandafter\def\csname LT7\endcsname{\color[rgb]{1,0.3,0}}%
      \expandafter\def\csname LT8\endcsname{\color[rgb]{0.5,0.5,0.5}}%
    \else
      \def\colorrgb#1{\color{black}}%
      \def\colorgray#1{\color[gray]{#1}}%
      \expandafter\def\csname LTw\endcsname{\color{white}}%
      \expandafter\def\csname LTb\endcsname{\color{black}}%
      \expandafter\def\csname LTa\endcsname{\color{black}}%
      \expandafter\def\csname LT0\endcsname{\color{black}}%
      \expandafter\def\csname LT1\endcsname{\color{black}}%
      \expandafter\def\csname LT2\endcsname{\color{black}}%
      \expandafter\def\csname LT3\endcsname{\color{black}}%
      \expandafter\def\csname LT4\endcsname{\color{black}}%
      \expandafter\def\csname LT5\endcsname{\color{black}}%
      \expandafter\def\csname LT6\endcsname{\color{black}}%
      \expandafter\def\csname LT7\endcsname{\color{black}}%
      \expandafter\def\csname LT8\endcsname{\color{black}}%
    \fi
  \fi
  \setlength{\unitlength}{0.0500bp}%
  \begin{picture}(7200.00,5040.00)%
    \gplgaddtomacro\gplbacktext{%
      \csname LTb\endcsname%
      \put(990,660){\makebox(0,0)[r]{\strut{} 0}}%
      \put(990,1191){\makebox(0,0)[r]{\strut{} 50}}%
      \put(990,1723){\makebox(0,0)[r]{\strut{} 100}}%
      \put(990,2254){\makebox(0,0)[r]{\strut{} 150}}%
      \put(990,2786){\makebox(0,0)[r]{\strut{} 200}}%
      \put(990,3317){\makebox(0,0)[r]{\strut{} 250}}%
      \put(990,3849){\makebox(0,0)[r]{\strut{} 300}}%
      \put(990,4380){\makebox(0,0)[r]{\strut{} 350}}%
      \put(1122,440){\makebox(0,0){\strut{} 0}}%
      \put(1835,440){\makebox(0,0){\strut{} 0.05}}%
      \put(2548,440){\makebox(0,0){\strut{} 0.1}}%
      \put(3261,440){\makebox(0,0){\strut{} 0.15}}%
      \put(3974,440){\makebox(0,0){\strut{} 0.2}}%
      \put(4687,440){\makebox(0,0){\strut{} 0.25}}%
      \put(5400,440){\makebox(0,0){\strut{} 0.3}}%
      \put(6113,440){\makebox(0,0){\strut{} 0.35}}%
      \put(6826,440){\makebox(0,0){\strut{} 0.4}}%
      \put(220,2520){\rotatebox{90}{\makebox(0,0){\strut{}$S(\omega)$}}}%
      \put(3974,110){\makebox(0,0){\strut{}$\omega$}}%
      \put(3974,4710){\makebox(0,0){\strut{}Dynamical Spin Structure}}%
      \put(2548,3636){\makebox(0,0)[l]{\strut{}$\Delta=2.0, \beta=150., V=0.5, U=0.34$}}%
    }%
    \gplgaddtomacro\gplfronttext{%
      \csname LTb\endcsname%
      \put(5839,4207){\makebox(0,0)[r]{\strut{}$S(\omega)$ upper}}%
      \csname LTb\endcsname%
      \put(5839,3987){\makebox(0,0)[r]{\strut{}$S(\omega)$ lower}}%
    }%
    \gplbacktext
    \put(0,0){\includegraphics{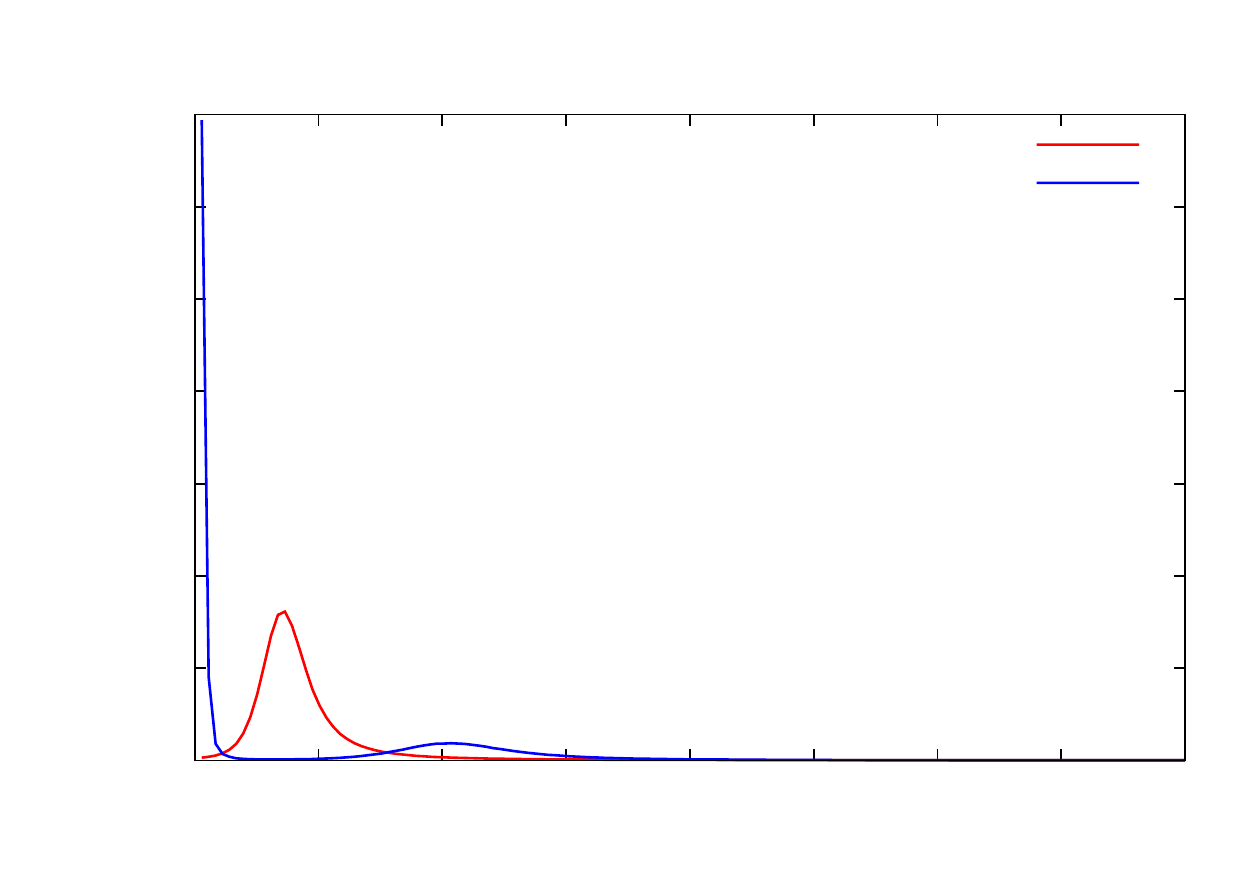}}%
    \gplfronttext
  \end{picture}%
\endgroup

%% file: tex-DMFT-ff-DOS-B100-3d.tex
\begingroup
  \makeatletter
  \providecommand\color[2][]{%
    \GenericError{(gnuplot) \space\space\space\@spaces}{%
      Package color not loaded in conjunction with
      terminal option `colourtext'%
    }{See the gnuplot documentation for explanation.%
    }{Either use 'blacktext' in gnuplot or load the package
      color.sty in LaTeX.}%
    \renewcommand\color[2][]{}%
  }%
  \providecommand\includegraphics[2][]{%
    \GenericError{(gnuplot) \space\space\space\@spaces}{%
      Package graphicx or graphics not loaded%
    }{See the gnuplot documentation for explanation.%
    }{The gnuplot epslatex terminal needs graphicx.sty or graphics.sty.}%
    \renewcommand\includegraphics[2][]{}%
  }%
  \providecommand\rotatebox[2]{#2}%
  \@ifundefined{ifGPcolor}{%
    \newif\ifGPcolor
    \GPcolortrue
  }{}%
  \@ifundefined{ifGPblacktext}{%
    \newif\ifGPblacktext
    \GPblacktexttrue
  }{}%
  \let\gplgaddtomacro\g@addto@macro
  \gdef\gplbacktext{}%
  \gdef\gplfronttext{}%
  \makeatother
  \ifGPblacktext
    \def\colorrgb#1{}%
    \def\colorgray#1{}%
  \else
    \ifGPcolor
      \def\colorrgb#1{\color[rgb]{#1}}%
      \def\colorgray#1{\color[gray]{#1}}%
      \expandafter\def\csname LTw\endcsname{\color{white}}%
      \expandafter\def\csname LTb\endcsname{\color{black}}%
      \expandafter\def\csname LTa\endcsname{\color{black}}%
      \expandafter\def\csname LT0\endcsname{\color[rgb]{1,0,0}}%
      \expandafter\def\csname LT1\endcsname{\color[rgb]{0,1,0}}%
      \expandafter\def\csname LT2\endcsname{\color[rgb]{0,0,1}}%
      \expandafter\def\csname LT3\endcsname{\color[rgb]{1,0,1}}%
      \expandafter\def\csname LT4\endcsname{\color[rgb]{0,1,1}}%
      \expandafter\def\csname LT5\endcsname{\color[rgb]{1,1,0}}%
      \expandafter\def\csname LT6\endcsname{\color[rgb]{0,0,0}}%
      \expandafter\def\csname LT7\endcsname{\color[rgb]{1,0.3,0}}%
      \expandafter\def\csname LT8\endcsname{\color[rgb]{0.5,0.5,0.5}}%
    \else
      \def\colorrgb#1{\color{black}}%
      \def\colorgray#1{\color[gray]{#1}}%
      \expandafter\def\csname LTw\endcsname{\color{white}}%
      \expandafter\def\csname LTb\endcsname{\color{black}}%
      \expandafter\def\csname LTa\endcsname{\color{black}}%
      \expandafter\def\csname LT0\endcsname{\color{black}}%
      \expandafter\def\csname LT1\endcsname{\color{black}}%
      \expandafter\def\csname LT2\endcsname{\color{black}}%
      \expandafter\def\csname LT3\endcsname{\color{black}}%
      \expandafter\def\csname LT4\endcsname{\color{black}}%
      \expandafter\def\csname LT5\endcsname{\color{black}}%
      \expandafter\def\csname LT6\endcsname{\color{black}}%
      \expandafter\def\csname LT7\endcsname{\color{black}}%
      \expandafter\def\csname LT8\endcsname{\color{black}}%
    \fi
  \fi
  \setlength{\unitlength}{0.0500bp}%
  \begin{picture}(7200.00,5040.00)%
    \gplgaddtomacro\gplbacktext{%
      \csname LTb\endcsname%
      \put(3599,4312){\makebox(0,0){\strut{}Density of states for the f-sites}}%
    }%
    \gplgaddtomacro\gplfronttext{%
      \csname LTb\endcsname%
      \put(6018,4096){\makebox(0,0)[r]{\strut{}$\rho_{\text{ff}}(\omega)$}}%
      \csname LTb\endcsname%
      \put(1170,772){\makebox(0,0){\strut{} 0}}%
      \put(2054,772){\makebox(0,0){\strut{} 0.1}}%
      \put(2938,772){\makebox(0,0){\strut{} 0.2}}%
      \put(3820,772){\makebox(0,0){\strut{} 0.3}}%
      \put(4704,772){\makebox(0,0){\strut{} 0.4}}%
      \put(5588,772){\makebox(0,0){\strut{} 0.5}}%
      \put(3600,442){\makebox(0,0){\strut{}$U$}}%
      \put(998,1058){\makebox(0,0)[r]{\strut{}-1}}%
      \put(998,1789){\makebox(0,0)[r]{\strut{}-0.5}}%
      \put(998,2520){\makebox(0,0)[r]{\strut{} 0}}%
      \put(998,3251){\makebox(0,0)[r]{\strut{} 0.5}}%
      \put(998,3982){\makebox(0,0)[r]{\strut{} 1}}%
      \put(404,2520){\rotatebox{90}{\makebox(0,0){\strut{}$\omega$}}}%
      \put(6527,1057){\makebox(0,0)[l]{\strut{} 1e-05}}%
      \put(6527,1474){\makebox(0,0)[l]{\strut{} 0.0001}}%
      \put(6527,1892){\makebox(0,0)[l]{\strut{} 0.001}}%
      \put(6527,2310){\makebox(0,0)[l]{\strut{} 0.01}}%
      \put(6527,2728){\makebox(0,0)[l]{\strut{} 0.1}}%
      \put(6527,3146){\makebox(0,0)[l]{\strut{} 1}}%
      \put(6527,3564){\makebox(0,0)[l]{\strut{} 10}}%
      \put(6527,3982){\makebox(0,0)[l]{\strut{} 100}}%
    }%
    \gplbacktext
    \put(0,0){\includegraphics{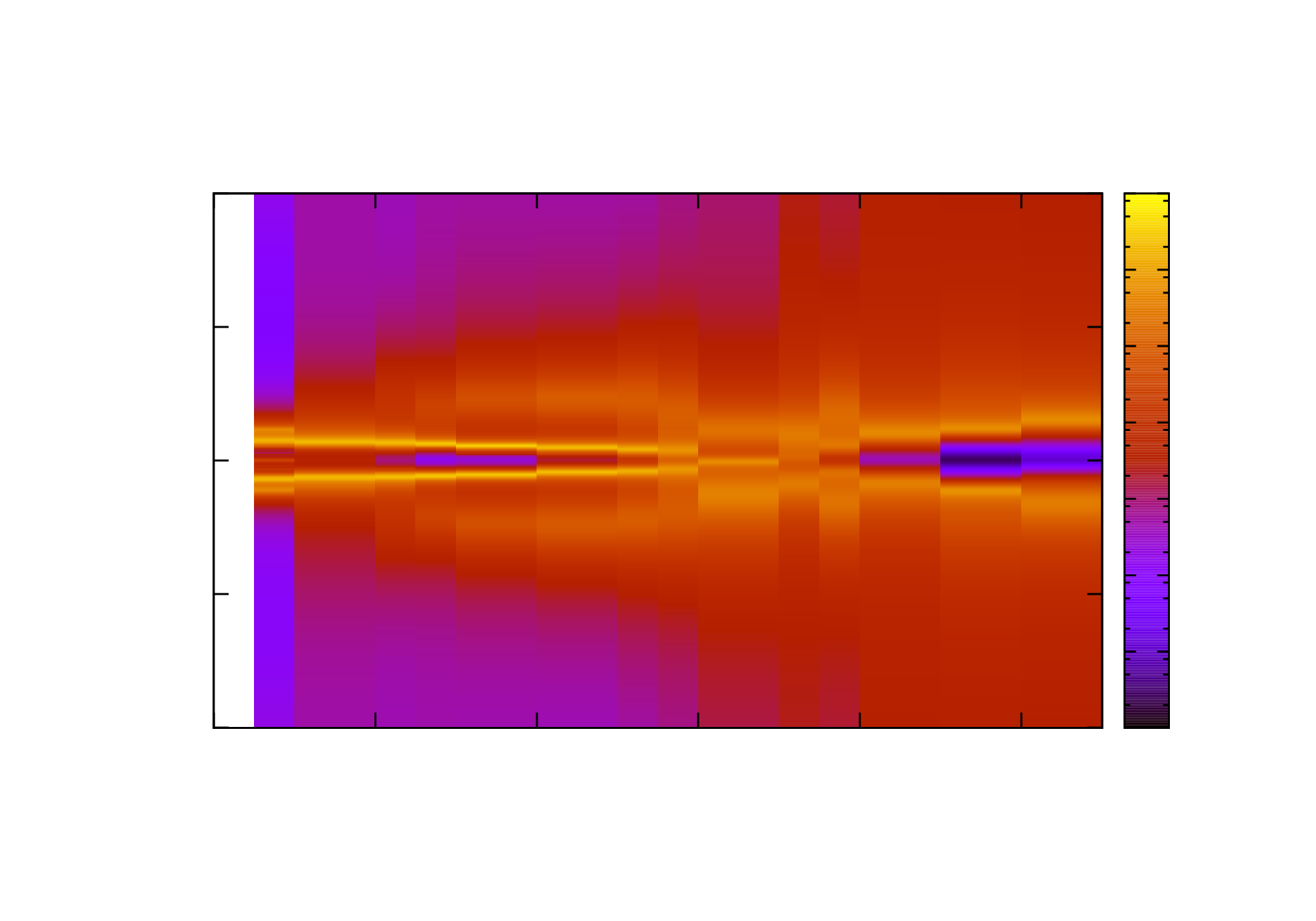}}%
    \gplfronttext
  \end{picture}%
\endgroup

%% file: tex-DMFT-spectrum-Trace-U0125-3d.tex
\begingroup
  \makeatletter
  \providecommand\color[2][]{%
    \GenericError{(gnuplot) \space\space\space\@spaces}{%
      Package color not loaded in conjunction with
      terminal option `colourtext'%
    }{See the gnuplot documentation for explanation.%
    }{Either use 'blacktext' in gnuplot or load the package
      color.sty in LaTeX.}%
    \renewcommand\color[2][]{}%
  }%
  \providecommand\includegraphics[2][]{%
    \GenericError{(gnuplot) \space\space\space\@spaces}{%
      Package graphicx or graphics not loaded%
    }{See the gnuplot documentation for explanation.%
    }{The gnuplot epslatex terminal needs graphicx.sty or graphics.sty.}%
    \renewcommand\includegraphics[2][]{}%
  }%
  \providecommand\rotatebox[2]{#2}%
  \@ifundefined{ifGPcolor}{%
    \newif\ifGPcolor
    \GPcolortrue
  }{}%
  \@ifundefined{ifGPblacktext}{%
    \newif\ifGPblacktext
    \GPblacktextfalse
  }{}%
  \let\gplgaddtomacro\g@addto@macro
  \gdef\gplbacktext{}%
  \gdef\gplfronttext{}%
  \makeatother
  \ifGPblacktext
    \def\colorrgb#1{}%
    \def\colorgray#1{}%
  \else
    \ifGPcolor
      \def\colorrgb#1{\color[rgb]{#1}}%
      \def\colorgray#1{\color[gray]{#1}}%
      \expandafter\def\csname LTw\endcsname{\color{white}}%
      \expandafter\def\csname LTb\endcsname{\color{black}}%
      \expandafter\def\csname LTa\endcsname{\color{black}}%
      \expandafter\def\csname LT0\endcsname{\color[rgb]{1,0,0}}%
      \expandafter\def\csname LT1\endcsname{\color[rgb]{0,1,0}}%
      \expandafter\def\csname LT2\endcsname{\color[rgb]{0,0,1}}%
      \expandafter\def\csname LT3\endcsname{\color[rgb]{1,0,1}}%
      \expandafter\def\csname LT4\endcsname{\color[rgb]{0,1,1}}%
      \expandafter\def\csname LT5\endcsname{\color[rgb]{1,1,0}}%
      \expandafter\def\csname LT6\endcsname{\color[rgb]{0,0,0}}%
      \expandafter\def\csname LT7\endcsname{\color[rgb]{1,0.3,0}}%
      \expandafter\def\csname LT8\endcsname{\color[rgb]{0.5,0.5,0.5}}%
    \else
      \def\colorrgb#1{\color{black}}%
      \def\colorgray#1{\color[gray]{#1}}%
      \expandafter\def\csname LTw\endcsname{\color{white}}%
      \expandafter\def\csname LTb\endcsname{\color{black}}%
      \expandafter\def\csname LTa\endcsname{\color{black}}%
      \expandafter\def\csname LT0\endcsname{\color{black}}%
      \expandafter\def\csname LT1\endcsname{\color{black}}%
      \expandafter\def\csname LT2\endcsname{\color{black}}%
      \expandafter\def\csname LT3\endcsname{\color{black}}%
      \expandafter\def\csname LT4\endcsname{\color{black}}%
      \expandafter\def\csname LT5\endcsname{\color{black}}%
      \expandafter\def\csname LT6\endcsname{\color{black}}%
      \expandafter\def\csname LT7\endcsname{\color{black}}%
      \expandafter\def\csname LT8\endcsname{\color{black}}%
    \fi
  \fi
  \setlength{\unitlength}{0.0500bp}%
  \begin{picture}(7200.00,5040.00)%
    \gplgaddtomacro\gplbacktext{%
      \csname LTb\endcsname%
      \put(3599,4368){\makebox(0,0){\strut{}Trace of the spectral function $A(\vec{k},\omega)$ at $U=0.125$}}%
    }%
    \gplgaddtomacro\gplfronttext{%
      \csname LTb\endcsname%
      \put(6018,3656){\makebox(0,0)[r]{\strut{}$A(\vec{k},\omega)$}}%
      \csname LTb\endcsname%
      \put(1248,716){\makebox(0,0){\strut{}$(0,0)$}}%
      \put(2816,716){\makebox(0,0){\strut{}$(\pi,0)$}}%
      \put(4384,716){\makebox(0,0){\strut{}$(\pi,\pi)$}}%
      \put(5952,716){\makebox(0,0){\strut{}$(0,0)$}}%
      \put(3600,386){\makebox(0,0){\strut{}$\vec{k}$}}%
      \put(998,1262){\makebox(0,0)[r]{\strut{}-4}}%
      \put(998,1781){\makebox(0,0)[r]{\strut{}-2}}%
      \put(998,2300){\makebox(0,0)[r]{\strut{} 0}}%
      \put(998,2819){\makebox(0,0)[r]{\strut{} 2}}%
      \put(998,3338){\makebox(0,0)[r]{\strut{} 4}}%
      \put(668,2300){\rotatebox{90}{\makebox(0,0){\strut{}$\omega$}}}%
      \put(6527,1000){\makebox(0,0)[l]{\strut{} 0.0001}}%
      \put(6527,1433){\makebox(0,0)[l]{\strut{} 0.001}}%
      \put(6527,1866){\makebox(0,0)[l]{\strut{} 0.01}}%
      \put(6527,2299){\makebox(0,0)[l]{\strut{} 0.1}}%
      \put(6527,2732){\makebox(0,0)[l]{\strut{} 1}}%
      \put(6527,3165){\makebox(0,0)[l]{\strut{} 10}}%
      \put(6527,3598){\makebox(0,0)[l]{\strut{} 100}}%
    }%
    \gplbacktext
    \put(0,0){\includegraphics{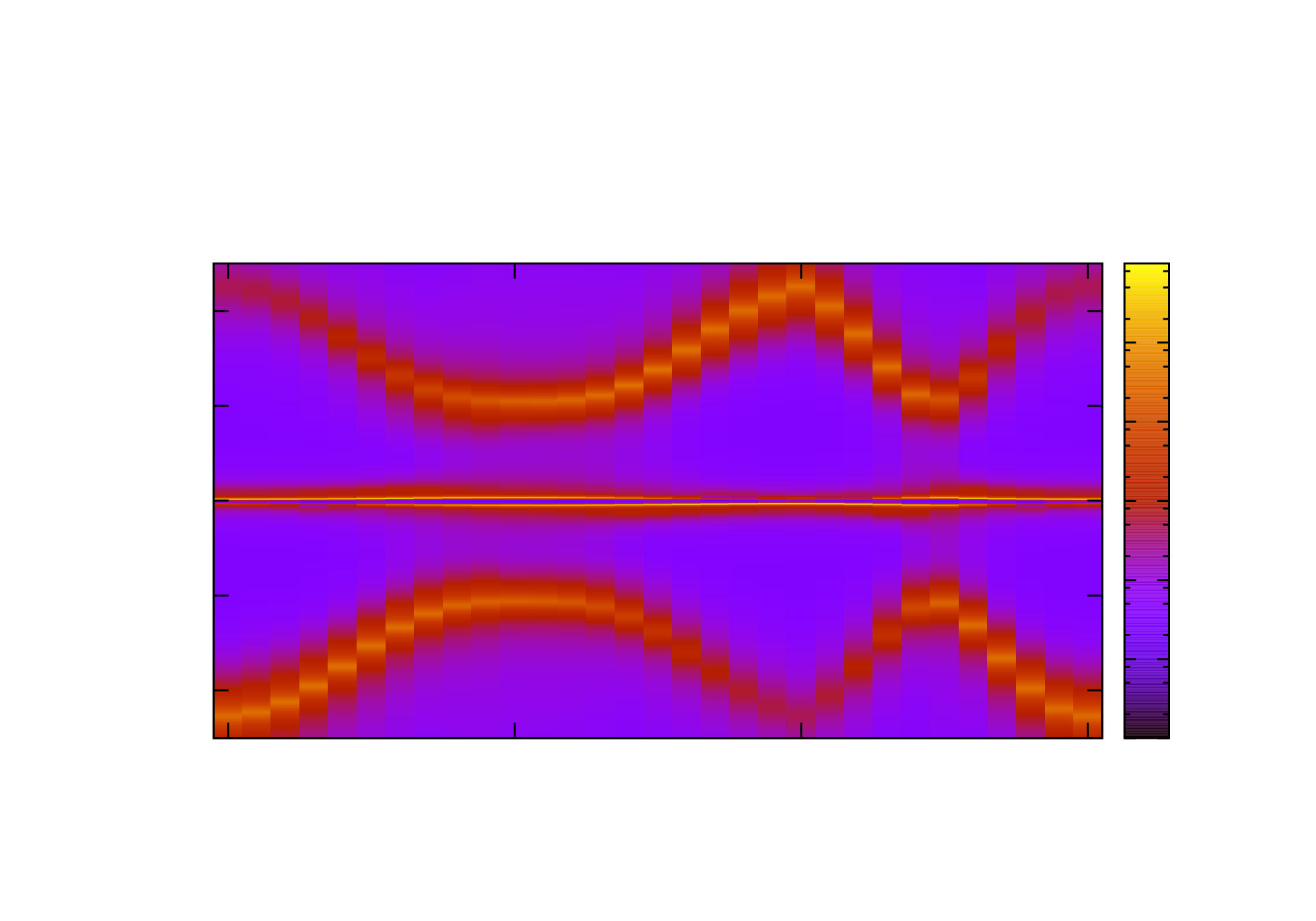}}%
    \gplfronttext
  \end{picture}%
\endgroup

%% file: tex-DMFT-free-spectrum-Trace-zoomed-3d.tex
\begingroup
  \makeatletter
  \providecommand\color[2][]{%
    \GenericError{(gnuplot) \space\space\space\@spaces}{%
      Package color not loaded in conjunction with
      terminal option `colourtext'%
    }{See the gnuplot documentation for explanation.%
    }{Either use 'blacktext' in gnuplot or load the package
      color.sty in LaTeX.}%
    \renewcommand\color[2][]{}%
  }%
  \providecommand\includegraphics[2][]{%
    \GenericError{(gnuplot) \space\space\space\@spaces}{%
      Package graphicx or graphics not loaded%
    }{See the gnuplot documentation for explanation.%
    }{The gnuplot epslatex terminal needs graphicx.sty or graphics.sty.}%
    \renewcommand\includegraphics[2][]{}%
  }%
  \providecommand\rotatebox[2]{#2}%
  \@ifundefined{ifGPcolor}{%
    \newif\ifGPcolor
    \GPcolortrue
  }{}%
  \@ifundefined{ifGPblacktext}{%
    \newif\ifGPblacktext
    \GPblacktextfalse
  }{}%
  \let\gplgaddtomacro\g@addto@macro
  \gdef\gplbacktext{}%
  \gdef\gplfronttext{}%
  \makeatother
  \ifGPblacktext
    \def\colorrgb#1{}%
    \def\colorgray#1{}%
  \else
    \ifGPcolor
      \def\colorrgb#1{\color[rgb]{#1}}%
      \def\colorgray#1{\color[gray]{#1}}%
      \expandafter\def\csname LTw\endcsname{\color{white}}%
      \expandafter\def\csname LTb\endcsname{\color{black}}%
      \expandafter\def\csname LTa\endcsname{\color{black}}%
      \expandafter\def\csname LT0\endcsname{\color[rgb]{1,0,0}}%
      \expandafter\def\csname LT1\endcsname{\color[rgb]{0,1,0}}%
      \expandafter\def\csname LT2\endcsname{\color[rgb]{0,0,1}}%
      \expandafter\def\csname LT3\endcsname{\color[rgb]{1,0,1}}%
      \expandafter\def\csname LT4\endcsname{\color[rgb]{0,1,1}}%
      \expandafter\def\csname LT5\endcsname{\color[rgb]{1,1,0}}%
      \expandafter\def\csname LT6\endcsname{\color[rgb]{0,0,0}}%
      \expandafter\def\csname LT7\endcsname{\color[rgb]{1,0.3,0}}%
      \expandafter\def\csname LT8\endcsname{\color[rgb]{0.5,0.5,0.5}}%
    \else
      \def\colorrgb#1{\color{black}}%
      \def\colorgray#1{\color[gray]{#1}}%
      \expandafter\def\csname LTw\endcsname{\color{white}}%
      \expandafter\def\csname LTb\endcsname{\color{black}}%
      \expandafter\def\csname LTa\endcsname{\color{black}}%
      \expandafter\def\csname LT0\endcsname{\color{black}}%
      \expandafter\def\csname LT1\endcsname{\color{black}}%
      \expandafter\def\csname LT2\endcsname{\color{black}}%
      \expandafter\def\csname LT3\endcsname{\color{black}}%
      \expandafter\def\csname LT4\endcsname{\color{black}}%
      \expandafter\def\csname LT5\endcsname{\color{black}}%
      \expandafter\def\csname LT6\endcsname{\color{black}}%
      \expandafter\def\csname LT7\endcsname{\color{black}}%
      \expandafter\def\csname LT8\endcsname{\color{black}}%
    \fi
  \fi
  \setlength{\unitlength}{0.0500bp}%
  \begin{picture}(7200.00,5040.00)%
    \gplgaddtomacro\gplbacktext{%
      \csname LTb\endcsname%
      \put(3599,3819){\makebox(0,0){\strut{}Trace of the spectral function $A(\vec{k},\omega)$ at $U=0$}}%
    }%
    \gplgaddtomacro\gplfronttext{%
      \csname LTb\endcsname%
      \put(6018,3656){\makebox(0,0)[r]{\strut{}$A(\vec{k},\omega)$}}%
      \csname LTb\endcsname%
      \put(1209,716){\makebox(0,0){\strut{}$(0,0)$}}%
      \put(2803,716){\makebox(0,0){\strut{}$(\pi,0)$}}%
      \put(4397,716){\makebox(0,0){\strut{}$(\pi,\pi)$}}%
      \put(5991,716){\makebox(0,0){\strut{}$(0,0)$}}%
      \put(3600,606){\makebox(0,0){\strut{}$\vec{k}$}}%
      \put(998,1002){\makebox(0,0)[r]{\strut{}-0.2}}%
      \put(998,1326){\makebox(0,0)[r]{\strut{}-0.15}}%
      \put(998,1651){\makebox(0,0)[r]{\strut{}-0.1}}%
      \put(998,1976){\makebox(0,0)[r]{\strut{}-0.05}}%
      \put(998,2300){\makebox(0,0)[r]{\strut{} 0}}%
      \put(998,2624){\makebox(0,0)[r]{\strut{} 0.05}}%
      \put(998,2949){\makebox(0,0)[r]{\strut{} 0.1}}%
      \put(998,3274){\makebox(0,0)[r]{\strut{} 0.15}}%
      \put(998,3598){\makebox(0,0)[r]{\strut{} 0.2}}%
      \put(272,2300){\rotatebox{90}{\makebox(0,0){\strut{}$\omega$}}}%
      \put(6527,1001){\makebox(0,0)[l]{\strut{} 0.01}}%
      \put(6527,1520){\makebox(0,0)[l]{\strut{} 0.1}}%
      \put(6527,2039){\makebox(0,0)[l]{\strut{} 1}}%
      \put(6527,2559){\makebox(0,0)[l]{\strut{} 10}}%
      \put(6527,3078){\makebox(0,0)[l]{\strut{} 100}}%
      \put(6527,3598){\makebox(0,0)[l]{\strut{} 1000}}%
    }%
    \gplbacktext
    \put(0,0){\includegraphics{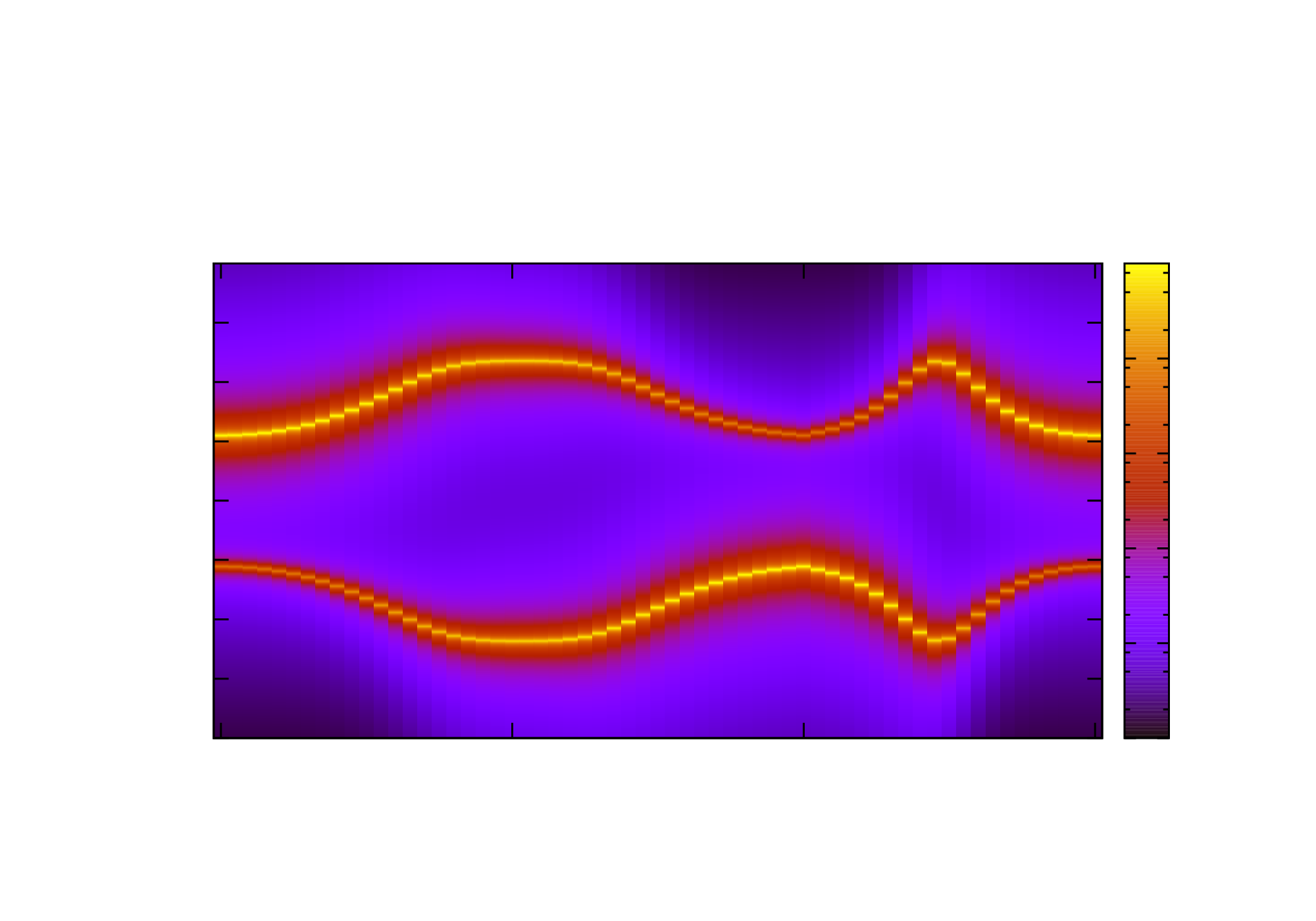}}%
    \gplfronttext
  \end{picture}%
\endgroup

%% file: tex-DMFT-spectrum-Trace-U0125-zoomed-3d.tex
\begingroup
  \makeatletter
  \providecommand\color[2][]{%
    \GenericError{(gnuplot) \space\space\space\@spaces}{%
      Package color not loaded in conjunction with
      terminal option `colourtext'%
    }{See the gnuplot documentation for explanation.%
    }{Either use 'blacktext' in gnuplot or load the package
      color.sty in LaTeX.}%
    \renewcommand\color[2][]{}%
  }%
  \providecommand\includegraphics[2][]{%
    \GenericError{(gnuplot) \space\space\space\@spaces}{%
      Package graphicx or graphics not loaded%
    }{See the gnuplot documentation for explanation.%
    }{The gnuplot epslatex terminal needs graphicx.sty or graphics.sty.}%
    \renewcommand\includegraphics[2][]{}%
  }%
  \providecommand\rotatebox[2]{#2}%
  \@ifundefined{ifGPcolor}{%
    \newif\ifGPcolor
    \GPcolortrue
  }{}%
  \@ifundefined{ifGPblacktext}{%
    \newif\ifGPblacktext
    \GPblacktextfalse
  }{}%
  \let\gplgaddtomacro\g@addto@macro
  \gdef\gplbacktext{}%
  \gdef\gplfronttext{}%
  \makeatother
  \ifGPblacktext
    \def\colorrgb#1{}%
    \def\colorgray#1{}%
  \else
    \ifGPcolor
      \def\colorrgb#1{\color[rgb]{#1}}%
      \def\colorgray#1{\color[gray]{#1}}%
      \expandafter\def\csname LTw\endcsname{\color{white}}%
      \expandafter\def\csname LTb\endcsname{\color{black}}%
      \expandafter\def\csname LTa\endcsname{\color{black}}%
      \expandafter\def\csname LT0\endcsname{\color[rgb]{1,0,0}}%
      \expandafter\def\csname LT1\endcsname{\color[rgb]{0,1,0}}%
      \expandafter\def\csname LT2\endcsname{\color[rgb]{0,0,1}}%
      \expandafter\def\csname LT3\endcsname{\color[rgb]{1,0,1}}%
      \expandafter\def\csname LT4\endcsname{\color[rgb]{0,1,1}}%
      \expandafter\def\csname LT5\endcsname{\color[rgb]{1,1,0}}%
      \expandafter\def\csname LT6\endcsname{\color[rgb]{0,0,0}}%
      \expandafter\def\csname LT7\endcsname{\color[rgb]{1,0.3,0}}%
      \expandafter\def\csname LT8\endcsname{\color[rgb]{0.5,0.5,0.5}}%
    \else
      \def\colorrgb#1{\color{black}}%
      \def\colorgray#1{\color[gray]{#1}}%
      \expandafter\def\csname LTw\endcsname{\color{white}}%
      \expandafter\def\csname LTb\endcsname{\color{black}}%
      \expandafter\def\csname LTa\endcsname{\color{black}}%
      \expandafter\def\csname LT0\endcsname{\color{black}}%
      \expandafter\def\csname LT1\endcsname{\color{black}}%
      \expandafter\def\csname LT2\endcsname{\color{black}}%
      \expandafter\def\csname LT3\endcsname{\color{black}}%
      \expandafter\def\csname LT4\endcsname{\color{black}}%
      \expandafter\def\csname LT5\endcsname{\color{black}}%
      \expandafter\def\csname LT6\endcsname{\color{black}}%
      \expandafter\def\csname LT7\endcsname{\color{black}}%
      \expandafter\def\csname LT8\endcsname{\color{black}}%
    \fi
  \fi
  \setlength{\unitlength}{0.0500bp}%
  \begin{picture}(7200.00,5040.00)%
    \gplgaddtomacro\gplbacktext{%
      \csname LTb\endcsname%
      \put(3599,3929){\makebox(0,0){\strut{}Trace of the spectral function $A(\vec{k},\omega)$ at $U=0.125$}}%
    }%
    \gplgaddtomacro\gplfronttext{%
      \csname LTb\endcsname%
      \put(6018,3656){\makebox(0,0)[r]{\strut{}$A(\vec{k},\omega)$}}%
      \csname LTb\endcsname%
      \put(1248,716){\makebox(0,0){\strut{}$(0,0)$}}%
      \put(2816,716){\makebox(0,0){\strut{}$(\pi,0)$}}%
      \put(4384,716){\makebox(0,0){\strut{}$(\pi,\pi)$}}%
      \put(5952,716){\makebox(0,0){\strut{}$(0,0)$}}%
      \put(3600,606){\makebox(0,0){\strut{}$\vec{k}$}}%
      \put(998,1002){\makebox(0,0)[r]{\strut{}-0.2}}%
      \put(998,1326){\makebox(0,0)[r]{\strut{}-0.15}}%
      \put(998,1651){\makebox(0,0)[r]{\strut{}-0.1}}%
      \put(998,1976){\makebox(0,0)[r]{\strut{}-0.05}}%
      \put(998,2300){\makebox(0,0)[r]{\strut{} 0}}%
      \put(998,2624){\makebox(0,0)[r]{\strut{} 0.05}}%
      \put(998,2949){\makebox(0,0)[r]{\strut{} 0.1}}%
      \put(998,3274){\makebox(0,0)[r]{\strut{} 0.15}}%
      \put(998,3598){\makebox(0,0)[r]{\strut{} 0.2}}%
      \put(272,2300){\rotatebox{90}{\makebox(0,0){\strut{}$\omega$}}}%
      \put(6527,1000){\makebox(0,0)[l]{\strut{} 0.01}}%
      \put(6527,1650){\makebox(0,0)[l]{\strut{} 0.1}}%
      \put(6527,2299){\makebox(0,0)[l]{\strut{} 1}}%
      \put(6527,2948){\makebox(0,0)[l]{\strut{} 10}}%
      \put(6527,3598){\makebox(0,0)[l]{\strut{} 100}}%
    }%
    \gplbacktext
    \put(0,0){\includegraphics{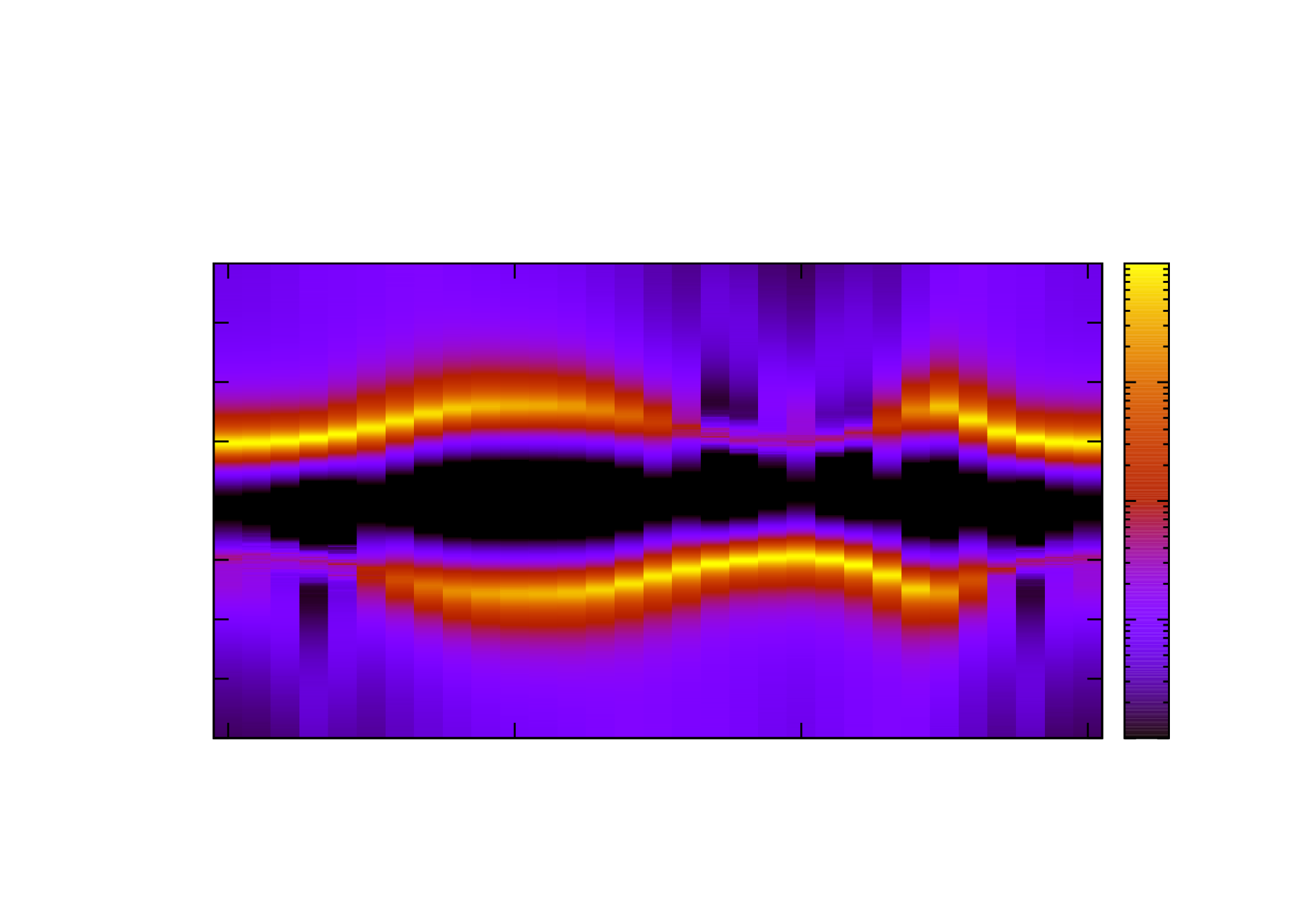}}%
    \gplfronttext
  \end{picture}%
\endgroup

%% file: tex-DMFT-spectrum-ff-U02-3d.tex
\begingroup
  \makeatletter
  \providecommand\color[2][]{%
    \GenericError{(gnuplot) \space\space\space\@spaces}{%
      Package color not loaded in conjunction with
      terminal option `colourtext'%
    }{See the gnuplot documentation for explanation.%
    }{Either use 'blacktext' in gnuplot or load the package
      color.sty in LaTeX.}%
    \renewcommand\color[2][]{}%
  }%
  \providecommand\includegraphics[2][]{%
    \GenericError{(gnuplot) \space\space\space\@spaces}{%
      Package graphicx or graphics not loaded%
    }{See the gnuplot documentation for explanation.%
    }{The gnuplot epslatex terminal needs graphicx.sty or graphics.sty.}%
    \renewcommand\includegraphics[2][]{}%
  }%
  \providecommand\rotatebox[2]{#2}%
  \@ifundefined{ifGPcolor}{%
    \newif\ifGPcolor
    \GPcolortrue
  }{}%
  \@ifundefined{ifGPblacktext}{%
    \newif\ifGPblacktext
    \GPblacktextfalse
  }{}%
  \let\gplgaddtomacro\g@addto@macro
  \gdef\gplbacktext{}%
  \gdef\gplfronttext{}%
  \makeatother
  \ifGPblacktext
    \def\colorrgb#1{}%
    \def\colorgray#1{}%
  \else
    \ifGPcolor
      \def\colorrgb#1{\color[rgb]{#1}}%
      \def\colorgray#1{\color[gray]{#1}}%
      \expandafter\def\csname LTw\endcsname{\color{white}}%
      \expandafter\def\csname LTb\endcsname{\color{black}}%
      \expandafter\def\csname LTa\endcsname{\color{black}}%
      \expandafter\def\csname LT0\endcsname{\color[rgb]{1,0,0}}%
      \expandafter\def\csname LT1\endcsname{\color[rgb]{0,1,0}}%
      \expandafter\def\csname LT2\endcsname{\color[rgb]{0,0,1}}%
      \expandafter\def\csname LT3\endcsname{\color[rgb]{1,0,1}}%
      \expandafter\def\csname LT4\endcsname{\color[rgb]{0,1,1}}%
      \expandafter\def\csname LT5\endcsname{\color[rgb]{1,1,0}}%
      \expandafter\def\csname LT6\endcsname{\color[rgb]{0,0,0}}%
      \expandafter\def\csname LT7\endcsname{\color[rgb]{1,0.3,0}}%
      \expandafter\def\csname LT8\endcsname{\color[rgb]{0.5,0.5,0.5}}%
    \else
      \def\colorrgb#1{\color{black}}%
      \def\colorgray#1{\color[gray]{#1}}%
      \expandafter\def\csname LTw\endcsname{\color{white}}%
      \expandafter\def\csname LTb\endcsname{\color{black}}%
      \expandafter\def\csname LTa\endcsname{\color{black}}%
      \expandafter\def\csname LT0\endcsname{\color{black}}%
      \expandafter\def\csname LT1\endcsname{\color{black}}%
      \expandafter\def\csname LT2\endcsname{\color{black}}%
      \expandafter\def\csname LT3\endcsname{\color{black}}%
      \expandafter\def\csname LT4\endcsname{\color{black}}%
      \expandafter\def\csname LT5\endcsname{\color{black}}%
      \expandafter\def\csname LT6\endcsname{\color{black}}%
      \expandafter\def\csname LT7\endcsname{\color{black}}%
      \expandafter\def\csname LT8\endcsname{\color{black}}%
    \fi
  \fi
  \setlength{\unitlength}{0.0500bp}%
  \begin{picture}(7200.00,5040.00)%
    \gplgaddtomacro\gplbacktext{%
      \csname LTb\endcsname%
      \put(3599,3819){\makebox(0,0){\strut{}Trace of the spectral function $A(\vec{k},\omega)$ at $U=0.2$}}%
    }%
    \gplgaddtomacro\gplfronttext{%
      \csname LTb\endcsname%
      \put(6018,3656){\makebox(0,0)[r]{\strut{}$A(\vec{k},\omega)$}}%
      \csname LTb\endcsname%
      \put(1248,716){\makebox(0,0){\strut{}$(0,0)$}}%
      \put(2816,716){\makebox(0,0){\strut{}$(\pi,0)$}}%
      \put(4384,716){\makebox(0,0){\strut{}$(\pi,\pi)$}}%
      \put(5952,716){\makebox(0,0){\strut{}$(0,0)$}}%
      \put(3600,606){\makebox(0,0){\strut{}$\vec{k}$}}%
      \put(998,1002){\makebox(0,0)[r]{\strut{}-0.2}}%
      \put(998,1326){\makebox(0,0)[r]{\strut{}-0.15}}%
      \put(998,1651){\makebox(0,0)[r]{\strut{}-0.1}}%
      \put(998,1976){\makebox(0,0)[r]{\strut{}-0.05}}%
      \put(998,2300){\makebox(0,0)[r]{\strut{} 0}}%
      \put(998,2624){\makebox(0,0)[r]{\strut{} 0.05}}%
      \put(998,2949){\makebox(0,0)[r]{\strut{} 0.1}}%
      \put(998,3274){\makebox(0,0)[r]{\strut{} 0.15}}%
      \put(998,3598){\makebox(0,0)[r]{\strut{} 0.2}}%
      \put(272,2300){\rotatebox{90}{\makebox(0,0){\strut{}$\omega$}}}%
      \put(6527,1000){\makebox(0,0)[l]{\strut{} 0.01}}%
      \put(6527,1650){\makebox(0,0)[l]{\strut{} 0.1}}%
      \put(6527,2299){\makebox(0,0)[l]{\strut{} 1}}%
      \put(6527,2948){\makebox(0,0)[l]{\strut{} 10}}%
      \put(6527,3598){\makebox(0,0)[l]{\strut{} 100}}%
    }%
    \gplbacktext
    \put(0,0){\includegraphics{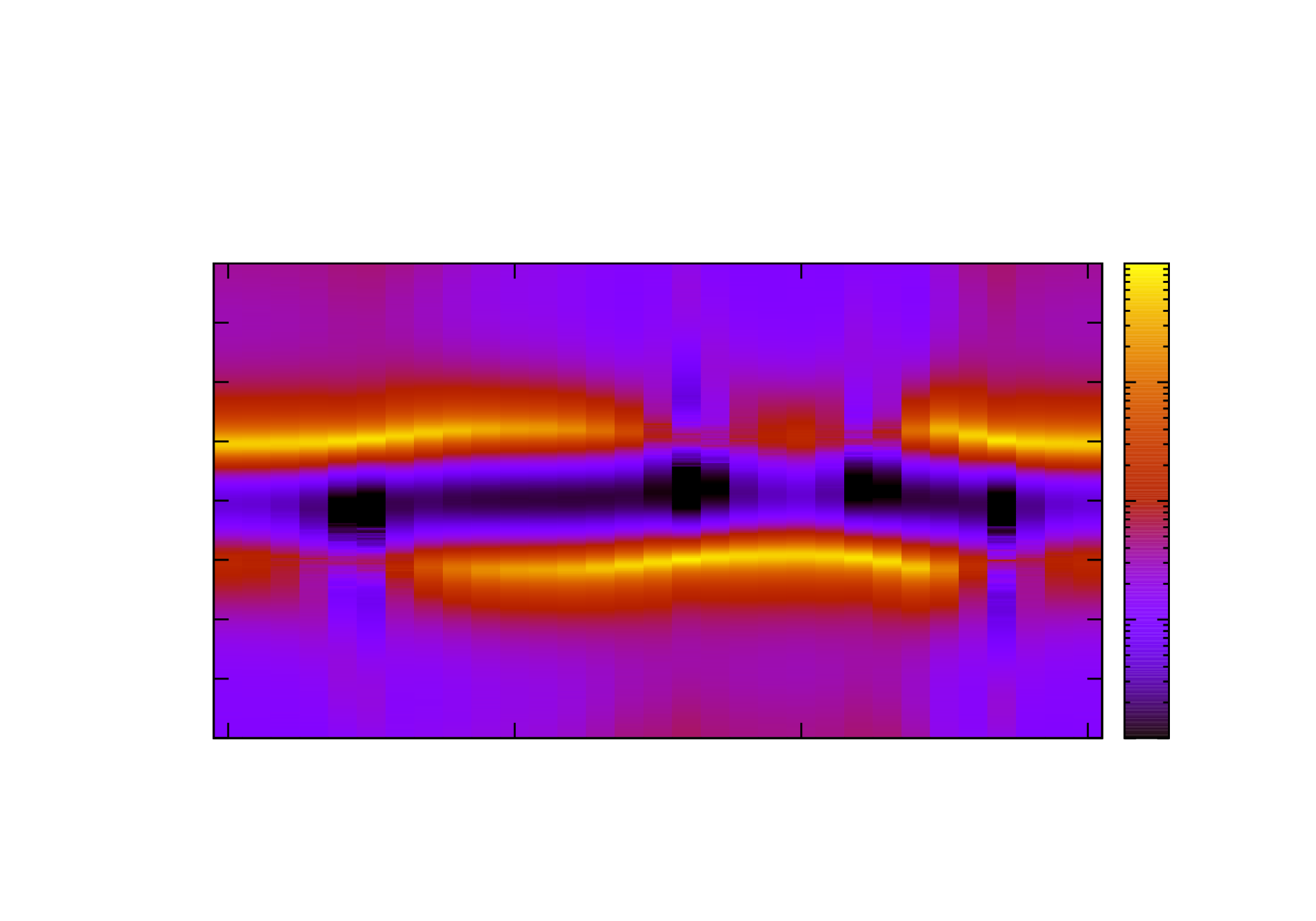}}%
    \gplfronttext
  \end{picture}%
\endgroup

%% file: tex-DMFT-spectrum-Trace-U0275-zoomed-3d.tex
\begingroup
  \makeatletter
  \providecommand\color[2][]{%
    \GenericError{(gnuplot) \space\space\space\@spaces}{%
      Package color not loaded in conjunction with
      terminal option `colourtext'%
    }{See the gnuplot documentation for explanation.%
    }{Either use 'blacktext' in gnuplot or load the package
      color.sty in LaTeX.}%
    \renewcommand\color[2][]{}%
  }%
  \providecommand\includegraphics[2][]{%
    \GenericError{(gnuplot) \space\space\space\@spaces}{%
      Package graphicx or graphics not loaded%
    }{See the gnuplot documentation for explanation.%
    }{The gnuplot epslatex terminal needs graphicx.sty or graphics.sty.}%
    \renewcommand\includegraphics[2][]{}%
  }%
  \providecommand\rotatebox[2]{#2}%
  \@ifundefined{ifGPcolor}{%
    \newif\ifGPcolor
    \GPcolortrue
  }{}%
  \@ifundefined{ifGPblacktext}{%
    \newif\ifGPblacktext
    \GPblacktextfalse
  }{}%
  \let\gplgaddtomacro\g@addto@macro
  \gdef\gplbacktext{}%
  \gdef\gplfronttext{}%
  \makeatother
  \ifGPblacktext
    \def\colorrgb#1{}%
    \def\colorgray#1{}%
  \else
    \ifGPcolor
      \def\colorrgb#1{\color[rgb]{#1}}%
      \def\colorgray#1{\color[gray]{#1}}%
      \expandafter\def\csname LTw\endcsname{\color{white}}%
      \expandafter\def\csname LTb\endcsname{\color{black}}%
      \expandafter\def\csname LTa\endcsname{\color{black}}%
      \expandafter\def\csname LT0\endcsname{\color[rgb]{1,0,0}}%
      \expandafter\def\csname LT1\endcsname{\color[rgb]{0,1,0}}%
      \expandafter\def\csname LT2\endcsname{\color[rgb]{0,0,1}}%
      \expandafter\def\csname LT3\endcsname{\color[rgb]{1,0,1}}%
      \expandafter\def\csname LT4\endcsname{\color[rgb]{0,1,1}}%
      \expandafter\def\csname LT5\endcsname{\color[rgb]{1,1,0}}%
      \expandafter\def\csname LT6\endcsname{\color[rgb]{0,0,0}}%
      \expandafter\def\csname LT7\endcsname{\color[rgb]{1,0.3,0}}%
      \expandafter\def\csname LT8\endcsname{\color[rgb]{0.5,0.5,0.5}}%
    \else
      \def\colorrgb#1{\color{black}}%
      \def\colorgray#1{\color[gray]{#1}}%
      \expandafter\def\csname LTw\endcsname{\color{white}}%
      \expandafter\def\csname LTb\endcsname{\color{black}}%
      \expandafter\def\csname LTa\endcsname{\color{black}}%
      \expandafter\def\csname LT0\endcsname{\color{black}}%
      \expandafter\def\csname LT1\endcsname{\color{black}}%
      \expandafter\def\csname LT2\endcsname{\color{black}}%
      \expandafter\def\csname LT3\endcsname{\color{black}}%
      \expandafter\def\csname LT4\endcsname{\color{black}}%
      \expandafter\def\csname LT5\endcsname{\color{black}}%
      \expandafter\def\csname LT6\endcsname{\color{black}}%
      \expandafter\def\csname LT7\endcsname{\color{black}}%
      \expandafter\def\csname LT8\endcsname{\color{black}}%
    \fi
  \fi
  \setlength{\unitlength}{0.0500bp}%
  \begin{picture}(7200.00,5040.00)%
    \gplgaddtomacro\gplbacktext{%
      \csname LTb\endcsname%
      \put(3599,3819){\makebox(0,0){\strut{}Trace of the spectral function $A(\vec{k},\omega)$ at $U=0.275$}}%
    }%
    \gplgaddtomacro\gplfronttext{%
      \csname LTb\endcsname%
      \put(6018,3656){\makebox(0,0)[r]{\strut{}$A(\vec{k},\omega)$}}%
      \csname LTb\endcsname%
      \put(1248,716){\makebox(0,0){\strut{}$(0,0)$}}%
      \put(2816,716){\makebox(0,0){\strut{}$(\pi,0)$}}%
      \put(4384,716){\makebox(0,0){\strut{}$(\pi,\pi)$}}%
      \put(5952,716){\makebox(0,0){\strut{}$(0,0)$}}%
      \put(3600,606){\makebox(0,0){\strut{}$\vec{k}$}}%
      \put(998,1002){\makebox(0,0)[r]{\strut{}-0.2}}%
      \put(998,1326){\makebox(0,0)[r]{\strut{}-0.15}}%
      \put(998,1651){\makebox(0,0)[r]{\strut{}-0.1}}%
      \put(998,1976){\makebox(0,0)[r]{\strut{}-0.05}}%
      \put(998,2300){\makebox(0,0)[r]{\strut{} 0}}%
      \put(998,2624){\makebox(0,0)[r]{\strut{} 0.05}}%
      \put(998,2949){\makebox(0,0)[r]{\strut{} 0.1}}%
      \put(998,3274){\makebox(0,0)[r]{\strut{} 0.15}}%
      \put(998,3598){\makebox(0,0)[r]{\strut{} 0.2}}%
      \put(272,2300){\rotatebox{90}{\makebox(0,0){\strut{}$\omega$}}}%
      \put(6527,1001){\makebox(0,0)[l]{\strut{} 0.01}}%
      \put(6527,1650){\makebox(0,0)[l]{\strut{} 0.1}}%
      \put(6527,2299){\makebox(0,0)[l]{\strut{} 1}}%
      \put(6527,2948){\makebox(0,0)[l]{\strut{} 10}}%
      \put(6527,3598){\makebox(0,0)[l]{\strut{} 100}}%
    }%
    \gplbacktext
    \put(0,0){\includegraphics{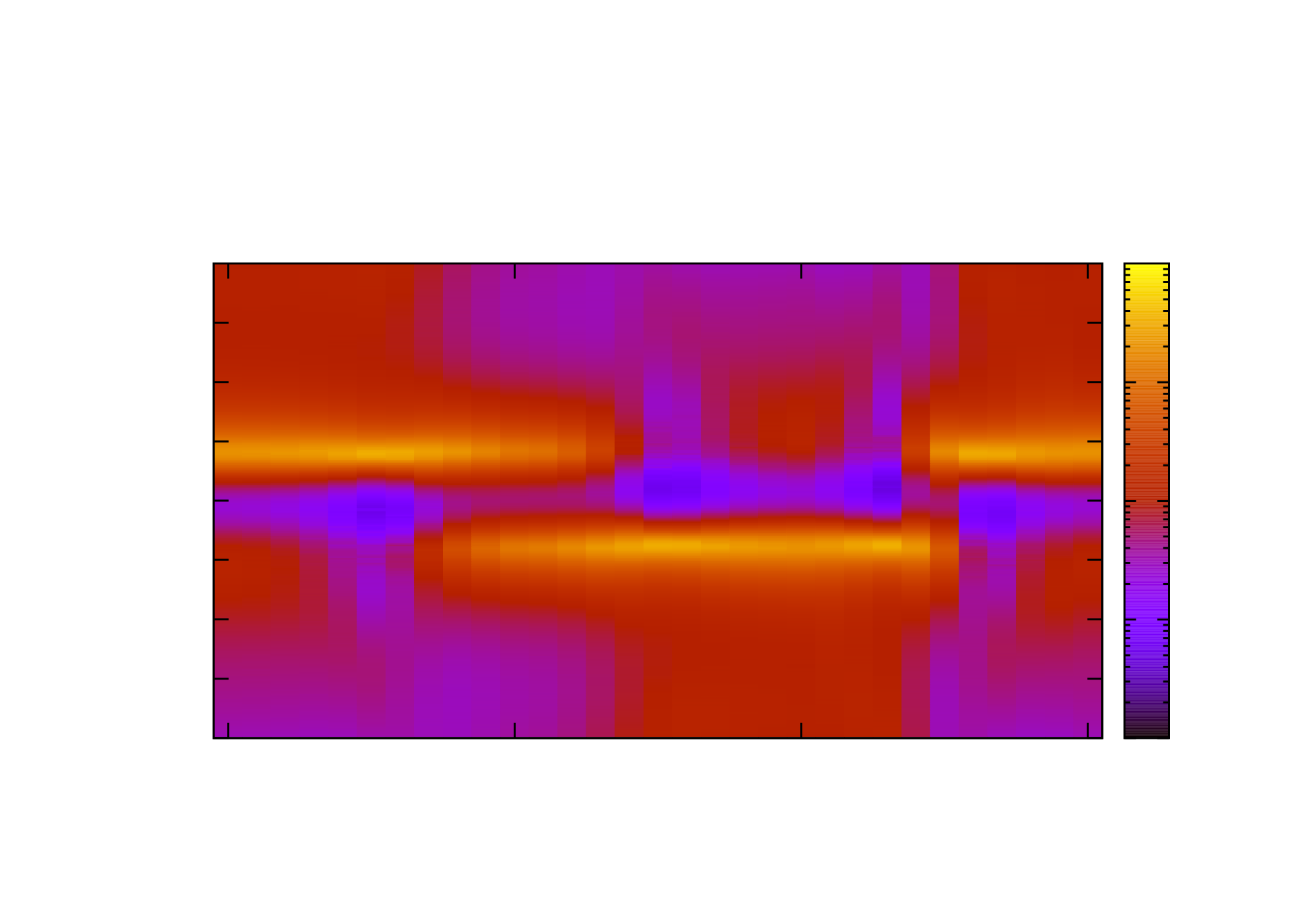}}%
    \gplfronttext
  \end{picture}%
\endgroup

%% file: tex-DMFT-spectrum-Trace-U05-zoomed-3d.tex
\begingroup
  \makeatletter
  \providecommand\color[2][]{%
    \GenericError{(gnuplot) \space\space\space\@spaces}{%
      Package color not loaded in conjunction with
      terminal option `colourtext'%
    }{See the gnuplot documentation for explanation.%
    }{Either use 'blacktext' in gnuplot or load the package
      color.sty in LaTeX.}%
    \renewcommand\color[2][]{}%
  }%
  \providecommand\includegraphics[2][]{%
    \GenericError{(gnuplot) \space\space\space\@spaces}{%
      Package graphicx or graphics not loaded%
    }{See the gnuplot documentation for explanation.%
    }{The gnuplot epslatex terminal needs graphicx.sty or graphics.sty.}%
    \renewcommand\includegraphics[2][]{}%
  }%
  \providecommand\rotatebox[2]{#2}%
  \@ifundefined{ifGPcolor}{%
    \newif\ifGPcolor
    \GPcolortrue
  }{}%
  \@ifundefined{ifGPblacktext}{%
    \newif\ifGPblacktext
    \GPblacktexttrue
  }{}%
  \let\gplgaddtomacro\g@addto@macro
  \gdef\gplbacktext{}%
  \gdef\gplfronttext{}%
  \makeatother
  \ifGPblacktext
    \def\colorrgb#1{}%
    \def\colorgray#1{}%
  \else
    \ifGPcolor
      \def\colorrgb#1{\color[rgb]{#1}}%
      \def\colorgray#1{\color[gray]{#1}}%
      \expandafter\def\csname LTw\endcsname{\color{white}}%
      \expandafter\def\csname LTb\endcsname{\color{black}}%
      \expandafter\def\csname LTa\endcsname{\color{black}}%
      \expandafter\def\csname LT0\endcsname{\color[rgb]{1,0,0}}%
      \expandafter\def\csname LT1\endcsname{\color[rgb]{0,1,0}}%
      \expandafter\def\csname LT2\endcsname{\color[rgb]{0,0,1}}%
      \expandafter\def\csname LT3\endcsname{\color[rgb]{1,0,1}}%
      \expandafter\def\csname LT4\endcsname{\color[rgb]{0,1,1}}%
      \expandafter\def\csname LT5\endcsname{\color[rgb]{1,1,0}}%
      \expandafter\def\csname LT6\endcsname{\color[rgb]{0,0,0}}%
      \expandafter\def\csname LT7\endcsname{\color[rgb]{1,0.3,0}}%
      \expandafter\def\csname LT8\endcsname{\color[rgb]{0.5,0.5,0.5}}%
    \else
      \def\colorrgb#1{\color{black}}%
      \def\colorgray#1{\color[gray]{#1}}%
      \expandafter\def\csname LTw\endcsname{\color{white}}%
      \expandafter\def\csname LTb\endcsname{\color{black}}%
      \expandafter\def\csname LTa\endcsname{\color{black}}%
      \expandafter\def\csname LT0\endcsname{\color{black}}%
      \expandafter\def\csname LT1\endcsname{\color{black}}%
      \expandafter\def\csname LT2\endcsname{\color{black}}%
      \expandafter\def\csname LT3\endcsname{\color{black}}%
      \expandafter\def\csname LT4\endcsname{\color{black}}%
      \expandafter\def\csname LT5\endcsname{\color{black}}%
      \expandafter\def\csname LT6\endcsname{\color{black}}%
      \expandafter\def\csname LT7\endcsname{\color{black}}%
      \expandafter\def\csname LT8\endcsname{\color{black}}%
    \fi
  \fi
  \setlength{\unitlength}{0.0500bp}%
  \begin{picture}(7200.00,5040.00)%
    \gplgaddtomacro\gplbacktext{%
      \csname LTb\endcsname%
      \put(3599,3819){\makebox(0,0){\strut{}Trace of the spectral function $A(\vec{k},\omega)$ at $U=0.5$}}%
    }%
    \gplgaddtomacro\gplfronttext{%
      \csname LTb\endcsname%
      \put(6018,3656){\makebox(0,0)[r]{\strut{}$A(\vec{k},\omega)$}}%
      \csname LTb\endcsname%
      \put(1248,716){\makebox(0,0){\strut{}$(0,0)$}}%
      \put(2816,716){\makebox(0,0){\strut{}$(\pi,0)$}}%
      \put(4384,716){\makebox(0,0){\strut{}$(\pi,\pi)$}}%
      \put(5952,716){\makebox(0,0){\strut{}$(0,0)$}}%
      \put(3600,386){\makebox(0,0){\strut{}$\vec{k}$}}%
      \put(998,1002){\makebox(0,0)[r]{\strut{}-0.2}}%
      \put(998,1326){\makebox(0,0)[r]{\strut{}-0.15}}%
      \put(998,1651){\makebox(0,0)[r]{\strut{}-0.1}}%
      \put(998,1976){\makebox(0,0)[r]{\strut{}-0.05}}%
      \put(998,2300){\makebox(0,0)[r]{\strut{} 0}}%
      \put(998,2624){\makebox(0,0)[r]{\strut{} 0.05}}%
      \put(998,2949){\makebox(0,0)[r]{\strut{} 0.1}}%
      \put(998,3274){\makebox(0,0)[r]{\strut{} 0.15}}%
      \put(998,3598){\makebox(0,0)[r]{\strut{} 0.2}}%
      \put(272,2300){\rotatebox{90}{\makebox(0,0){\strut{}$\omega$}}}%
      \put(6527,1001){\makebox(0,0)[l]{\strut{} 0.01}}%
      \put(6527,1650){\makebox(0,0)[l]{\strut{} 0.1}}%
      \put(6527,2299){\makebox(0,0)[l]{\strut{} 1}}%
      \put(6527,2948){\makebox(0,0)[l]{\strut{} 10}}%
      \put(6527,3598){\makebox(0,0)[l]{\strut{} 100}}%
    }%
    \gplbacktext
    \put(0,0){\includegraphics{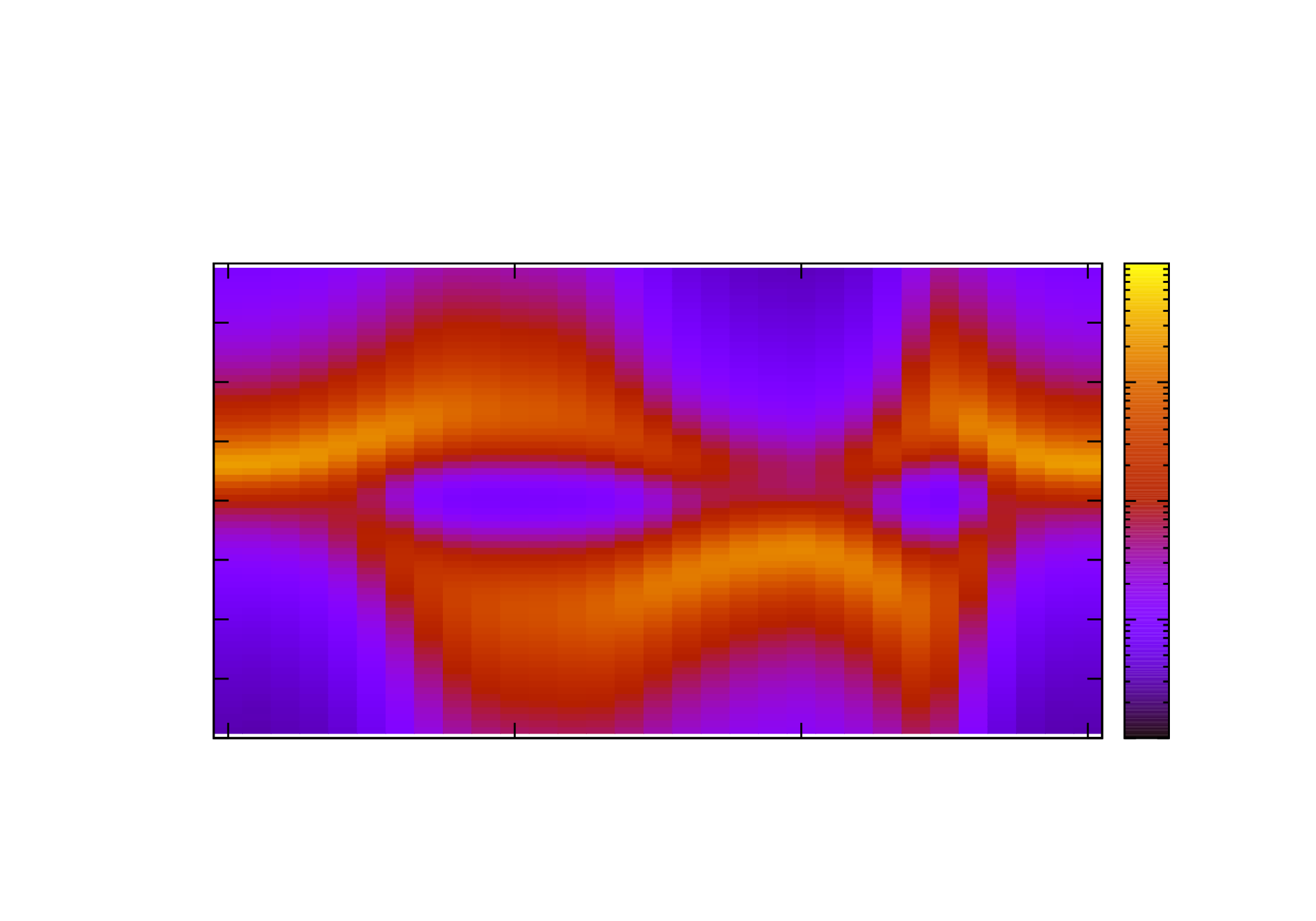}}%
    \gplfronttext
  \end{picture}%
\endgroup

%% file: tex-DMFT-spectrum-Trace-U055-zoomed-3d.tex
\begingroup
  \makeatletter
  \providecommand\color[2][]{%
    \GenericError{(gnuplot) \space\space\space\@spaces}{%
      Package color not loaded in conjunction with
      terminal option `colourtext'%
    }{See the gnuplot documentation for explanation.%
    }{Either use 'blacktext' in gnuplot or load the package
      color.sty in LaTeX.}%
    \renewcommand\color[2][]{}%
  }%
  \providecommand\includegraphics[2][]{%
    \GenericError{(gnuplot) \space\space\space\@spaces}{%
      Package graphicx or graphics not loaded%
    }{See the gnuplot documentation for explanation.%
    }{The gnuplot epslatex terminal needs graphicx.sty or graphics.sty.}%
    \renewcommand\includegraphics[2][]{}%
  }%
  \providecommand\rotatebox[2]{#2}%
  \@ifundefined{ifGPcolor}{%
    \newif\ifGPcolor
    \GPcolortrue
  }{}%
  \@ifundefined{ifGPblacktext}{%
    \newif\ifGPblacktext
    \GPblacktextfalse
  }{}%
  \let\gplgaddtomacro\g@addto@macro
  \gdef\gplbacktext{}%
  \gdef\gplfronttext{}%
  \makeatother
  \ifGPblacktext
    \def\colorrgb#1{}%
    \def\colorgray#1{}%
  \else
    \ifGPcolor
      \def\colorrgb#1{\color[rgb]{#1}}%
      \def\colorgray#1{\color[gray]{#1}}%
      \expandafter\def\csname LTw\endcsname{\color{white}}%
      \expandafter\def\csname LTb\endcsname{\color{black}}%
      \expandafter\def\csname LTa\endcsname{\color{black}}%
      \expandafter\def\csname LT0\endcsname{\color[rgb]{1,0,0}}%
      \expandafter\def\csname LT1\endcsname{\color[rgb]{0,1,0}}%
      \expandafter\def\csname LT2\endcsname{\color[rgb]{0,0,1}}%
      \expandafter\def\csname LT3\endcsname{\color[rgb]{1,0,1}}%
      \expandafter\def\csname LT4\endcsname{\color[rgb]{0,1,1}}%
      \expandafter\def\csname LT5\endcsname{\color[rgb]{1,1,0}}%
      \expandafter\def\csname LT6\endcsname{\color[rgb]{0,0,0}}%
      \expandafter\def\csname LT7\endcsname{\color[rgb]{1,0.3,0}}%
      \expandafter\def\csname LT8\endcsname{\color[rgb]{0.5,0.5,0.5}}%
    \else
      \def\colorrgb#1{\color{black}}%
      \def\colorgray#1{\color[gray]{#1}}%
      \expandafter\def\csname LTw\endcsname{\color{white}}%
      \expandafter\def\csname LTb\endcsname{\color{black}}%
      \expandafter\def\csname LTa\endcsname{\color{black}}%
      \expandafter\def\csname LT0\endcsname{\color{black}}%
      \expandafter\def\csname LT1\endcsname{\color{black}}%
      \expandafter\def\csname LT2\endcsname{\color{black}}%
      \expandafter\def\csname LT3\endcsname{\color{black}}%
      \expandafter\def\csname LT4\endcsname{\color{black}}%
      \expandafter\def\csname LT5\endcsname{\color{black}}%
      \expandafter\def\csname LT6\endcsname{\color{black}}%
      \expandafter\def\csname LT7\endcsname{\color{black}}%
      \expandafter\def\csname LT8\endcsname{\color{black}}%
    \fi
  \fi
  \setlength{\unitlength}{0.0500bp}%
  \begin{picture}(7200.00,5040.00)%
    \gplgaddtomacro\gplbacktext{%
      \csname LTb\endcsname%
      \put(3599,3819){\makebox(0,0){\strut{}Trace of the spectral function $A(\vec{k},\omega)$ at $U=0.55$}}%
    }%
    \gplgaddtomacro\gplfronttext{%
      \csname LTb\endcsname%
      \put(6018,3656){\makebox(0,0)[r]{\strut{}$A(\vec{k},\omega)$}}%
      \csname LTb\endcsname%
      \put(1248,716){\makebox(0,0){\strut{}$(0,0)$}}%
      \put(2816,716){\makebox(0,0){\strut{}$(\pi,0)$}}%
      \put(4384,716){\makebox(0,0){\strut{}$(\pi,\pi)$}}%
      \put(5952,716){\makebox(0,0){\strut{}$(0,0)$}}%
      \put(3600,606){\makebox(0,0){\strut{}$\vec{k}$}}%
      \put(998,1002){\makebox(0,0)[r]{\strut{}-0.2}}%
      \put(998,1326){\makebox(0,0)[r]{\strut{}-0.15}}%
      \put(998,1651){\makebox(0,0)[r]{\strut{}-0.1}}%
      \put(998,1976){\makebox(0,0)[r]{\strut{}-0.05}}%
      \put(998,2300){\makebox(0,0)[r]{\strut{} 0}}%
      \put(998,2624){\makebox(0,0)[r]{\strut{} 0.05}}%
      \put(998,2949){\makebox(0,0)[r]{\strut{} 0.1}}%
      \put(998,3274){\makebox(0,0)[r]{\strut{} 0.15}}%
      \put(998,3598){\makebox(0,0)[r]{\strut{} 0.2}}%
      \put(272,2300){\rotatebox{90}{\makebox(0,0){\strut{}$\omega$}}}%
      \put(6527,1001){\makebox(0,0)[l]{\strut{} 0.01}}%
      \put(6527,1650){\makebox(0,0)[l]{\strut{} 0.1}}%
      \put(6527,2299){\makebox(0,0)[l]{\strut{} 1}}%
      \put(6527,2948){\makebox(0,0)[l]{\strut{} 10}}%
      \put(6527,3598){\makebox(0,0)[l]{\strut{} 100}}%
    }%
    \gplbacktext
    \put(0,0){\includegraphics{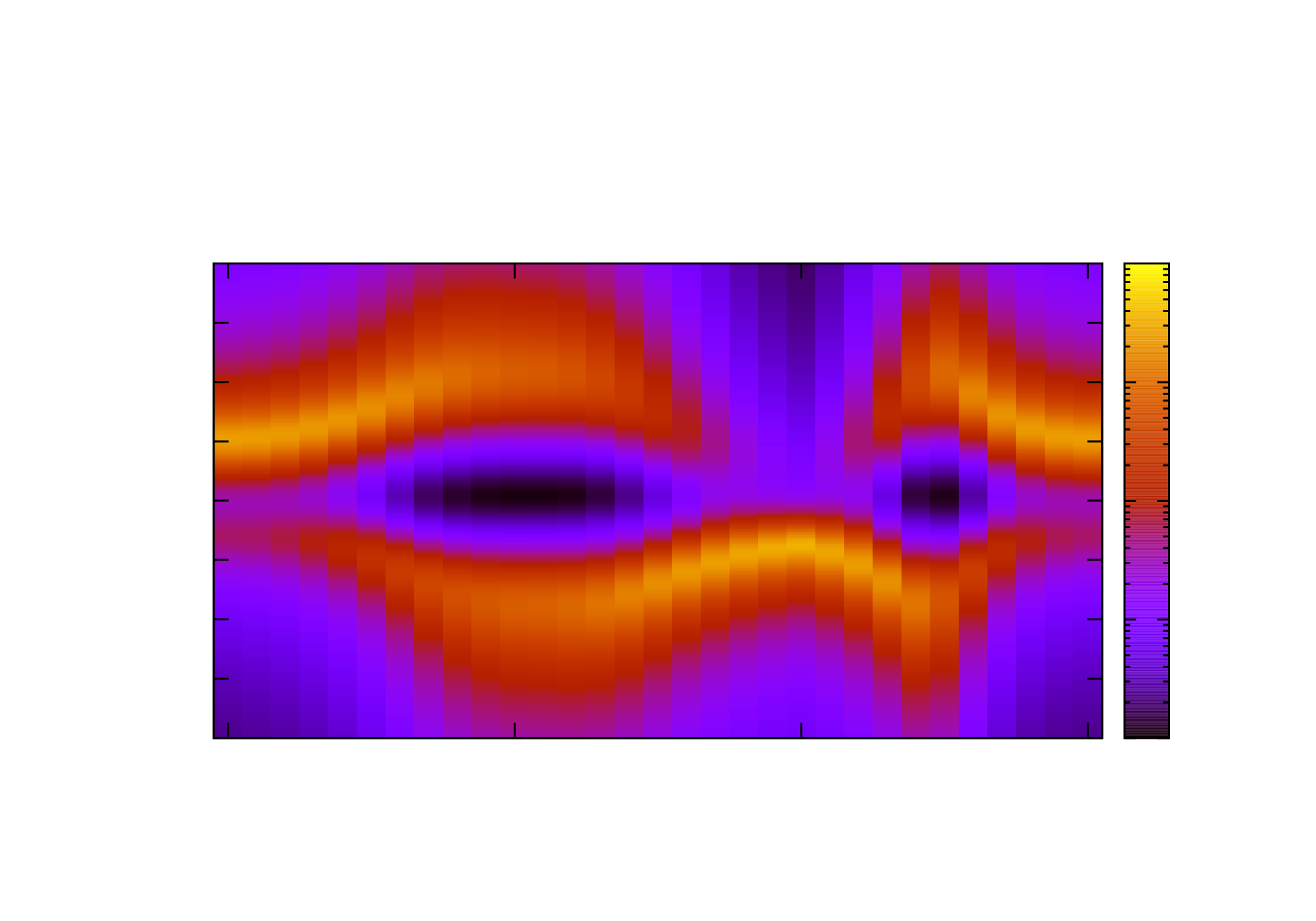}}%
    \gplfronttext
  \end{picture}%
\endgroup

%% file: tex-CPA-spec-3d.tex
\begingroup
  \makeatletter
  \providecommand\color[2][]{%
    \GenericError{(gnuplot) \space\space\space\@spaces}{%
      Package color not loaded in conjunction with
      terminal option `colourtext'%
    }{See the gnuplot documentation for explanation.%
    }{Either use 'blacktext' in gnuplot or load the package
      color.sty in LaTeX.}%
    \renewcommand\color[2][]{}%
  }%
  \providecommand\includegraphics[2][]{%
    \GenericError{(gnuplot) \space\space\space\@spaces}{%
      Package graphicx or graphics not loaded%
    }{See the gnuplot documentation for explanation.%
    }{The gnuplot epslatex terminal needs graphicx.sty or graphics.sty.}%
    \renewcommand\includegraphics[2][]{}%
  }%
  \providecommand\rotatebox[2]{#2}%
  \@ifundefined{ifGPcolor}{%
    \newif\ifGPcolor
    \GPcolortrue
  }{}%
  \@ifundefined{ifGPblacktext}{%
    \newif\ifGPblacktext
    \GPblacktexttrue
  }{}%
  \let\gplgaddtomacro\g@addto@macro
  \gdef\gplbacktext{}%
  \gdef\gplfronttext{}%
  \makeatother
  \ifGPblacktext
    \def\colorrgb#1{}%
    \def\colorgray#1{}%
  \else
    \ifGPcolor
      \def\colorrgb#1{\color[rgb]{#1}}%
      \def\colorgray#1{\color[gray]{#1}}%
      \expandafter\def\csname LTw\endcsname{\color{white}}%
      \expandafter\def\csname LTb\endcsname{\color{black}}%
      \expandafter\def\csname LTa\endcsname{\color{black}}%
      \expandafter\def\csname LT0\endcsname{\color[rgb]{1,0,0}}%
      \expandafter\def\csname LT1\endcsname{\color[rgb]{0,1,0}}%
      \expandafter\def\csname LT2\endcsname{\color[rgb]{0,0,1}}%
      \expandafter\def\csname LT3\endcsname{\color[rgb]{1,0,1}}%
      \expandafter\def\csname LT4\endcsname{\color[rgb]{0,1,1}}%
      \expandafter\def\csname LT5\endcsname{\color[rgb]{1,1,0}}%
      \expandafter\def\csname LT6\endcsname{\color[rgb]{0,0,0}}%
      \expandafter\def\csname LT7\endcsname{\color[rgb]{1,0.3,0}}%
      \expandafter\def\csname LT8\endcsname{\color[rgb]{0.5,0.5,0.5}}%
    \else
      \def\colorrgb#1{\color{black}}%
      \def\colorgray#1{\color[gray]{#1}}%
      \expandafter\def\csname LTw\endcsname{\color{white}}%
      \expandafter\def\csname LTb\endcsname{\color{black}}%
      \expandafter\def\csname LTa\endcsname{\color{black}}%
      \expandafter\def\csname LT0\endcsname{\color{black}}%
      \expandafter\def\csname LT1\endcsname{\color{black}}%
      \expandafter\def\csname LT2\endcsname{\color{black}}%
      \expandafter\def\csname LT3\endcsname{\color{black}}%
      \expandafter\def\csname LT4\endcsname{\color{black}}%
      \expandafter\def\csname LT5\endcsname{\color{black}}%
      \expandafter\def\csname LT6\endcsname{\color{black}}%
      \expandafter\def\csname LT7\endcsname{\color{black}}%
      \expandafter\def\csname LT8\endcsname{\color{black}}%
    \fi
  \fi
  \setlength{\unitlength}{0.0500bp}%
  \begin{picture}(7200.00,5040.00)%
    \gplgaddtomacro\gplbacktext{%
      \csname LTb\endcsname%
      \put(3599,3929){\makebox(0,0){\strut{}Disorder average of the trace of the spectral function $A(\vec{k},\omega)$}}%
    }%
    \gplgaddtomacro\gplfronttext{%
      \csname LTb\endcsname%
      \put(6018,3656){\makebox(0,0)[r]{\strut{}$A(\vec{k},\omega)$}}%
      \color{black}%
      \color{black}%
      \color{black}%
      \color{black}%
      \color{black}%
      \color{black}%
      \color{black}%
      \color{black}%
      \color{black}%
      \color{black}%
      \color{black}%
      \color{black}%
      \color{black}%
      \color{black}%
      \color{black}%
      \color{black}%
      \color{black}%
      \color{black}%
      \color{black}%
      \color{black}%
      \color{black}%
      \color{black}%
      \color{black}%
      \color{black}%
      \color{black}%
      \color{black}%
      \color{black}%
      \color{black}%
      \color{black}%
      \color{black}%
      \color{black}%
      \color{black}%
      \color{black}%
      \color{black}%
      \color{black}%
      \color{black}%
      \color{black}%
      \color{black}%
      \color{black}%
      \color{black}%
      \color{black}%
      \color{black}%
      \color{black}%
      \color{black}%
      \color{black}%
      \color{black}%
      \color{black}%
      \color{black}%
      \color{black}%
      \color{black}%
      \color{black}%
      \color{black}%
      \color{black}%
      \color{black}%
      \color{black}%
      \color{black}%
      \color{black}%
      \color{black}%
      \color{black}%
      \color{black}%
      \color{black}%
      \color{black}%
      \color{black}%
      \color{black}%
      \color{black}%
      \color{black}%
      \color{black}%
      \color{black}%
      \color{black}%
      \color{black}%
      \color{black}%
      \color{black}%
      \color{black}%
      \color{black}%
      \color{black}%
      \color{black}%
      \color{black}%
      \color{black}%
      \color{black}%
      \color{black}%
      \color{black}%
      \color{black}%
      \color{black}%
      \color{black}%
      \color{black}%
      \color{black}%
      \color{black}%
      \color{black}%
      \color{black}%
      \color{black}%
      \color{black}%
      \color{black}%
      \color{black}%
      \color{black}%
      \color{black}%
      \color{black}%
      \color{black}%
      \color{black}%
      \color{black}%
      \csname LTb\endcsname%
      \put(1170,716){\makebox(0,0){\strut{}(0,0) }}%
      \put(2655,716){\makebox(0,0){\strut{}($\pi$,0) }}%
      \put(4275,716){\makebox(0,0){\strut{}($\pi$,$\pi$)}}%
      \put(6030,716){\makebox(0,0){\strut{}($0$,$0$)}}%
      \put(3600,386){\makebox(0,0){\strut{}$\vec{k}$}}%
      \put(998,1002){\makebox(0,0)[r]{\strut{}-0.2}}%
      \put(998,1326){\makebox(0,0)[r]{\strut{}-0.15}}%
      \put(998,1651){\makebox(0,0)[r]{\strut{}-0.1}}%
      \put(998,1976){\makebox(0,0)[r]{\strut{}-0.05}}%
      \put(998,2300){\makebox(0,0)[r]{\strut{} 0}}%
      \put(998,2624){\makebox(0,0)[r]{\strut{} 0.05}}%
      \put(998,2949){\makebox(0,0)[r]{\strut{} 0.1}}%
      \put(998,3274){\makebox(0,0)[r]{\strut{} 0.15}}%
      \put(998,3598){\makebox(0,0)[r]{\strut{} 0.2}}%
      \put(272,2300){\rotatebox{90}{\makebox(0,0){\strut{}$\omega$}}}%
      \put(6527,1000){\makebox(0,0)[l]{\strut{} 0.01}}%
      \put(6527,1650){\makebox(0,0)[l]{\strut{} 0.1}}%
      \put(6527,2299){\makebox(0,0)[l]{\strut{} 1}}%
      \put(6527,2948){\makebox(0,0)[l]{\strut{} 10}}%
      \put(6527,3598){\makebox(0,0)[l]{\strut{} 100}}%
    }%
    \gplbacktext
    \put(0,0){\includegraphics{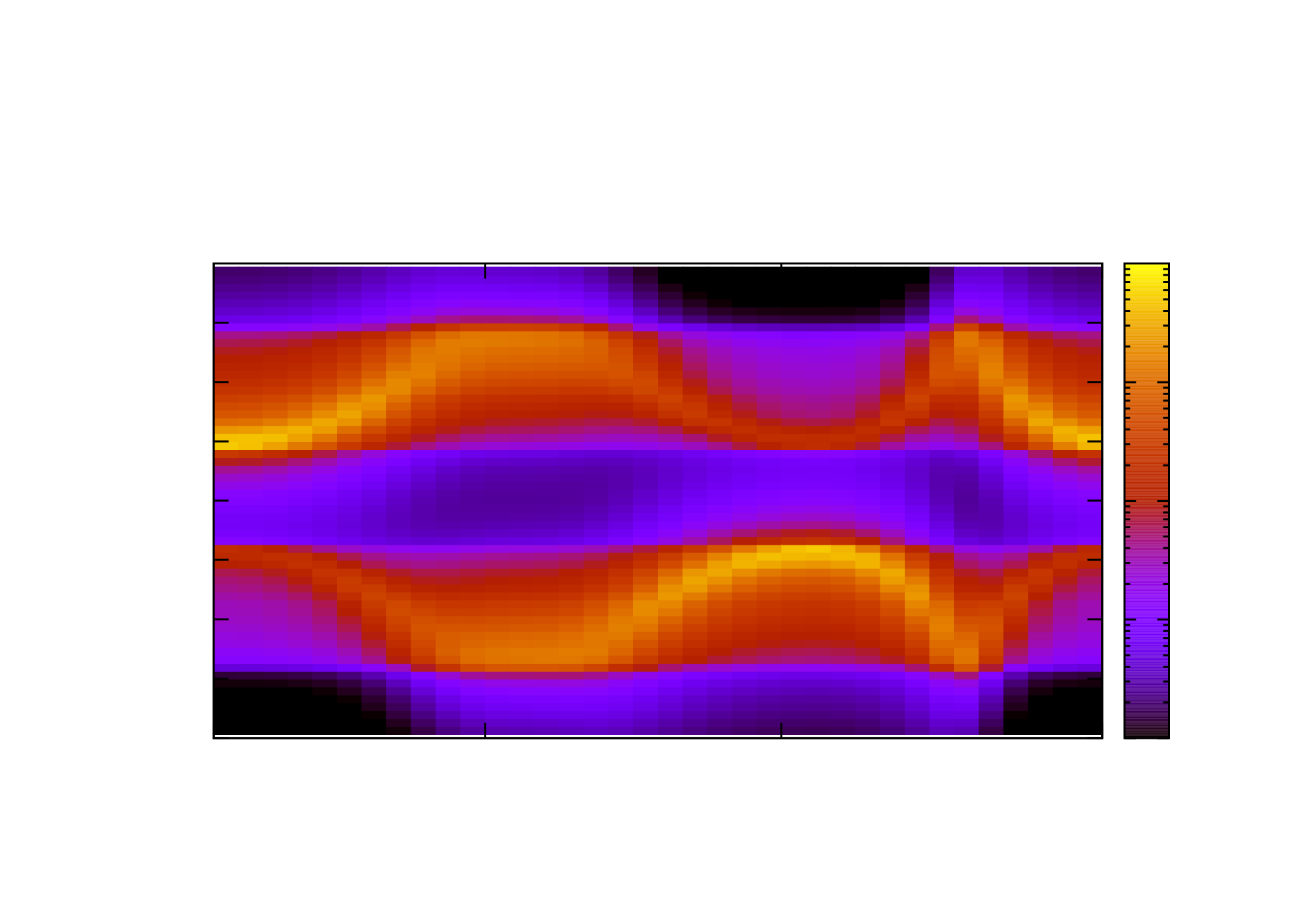}}%
    \gplfronttext
  \end{picture}%
\endgroup